\documentclass{aa}
\usepackage{graphicx}
\usepackage{txfonts}
\bibpunct{(}{)}{;}{a}{}{,}

\def\llm{\textsc{LLmodels}}

\def\synth3{\textsc{Synth3}}

\def\logg{\log(g)}
\def\teff{T_{\rm eff}}

\def\kms{km~s$^{-1}$}
\def\ms{m~s$^{-1}$}
\def\vsini{\upsilon\sin i}

\def\lamlam{\lambda\lambda}

\newcommand{\abn}[1]{\alpha(\mathrm{#1})}

\ifdefined\ion
\renewcommand{\ion}[2]{\textup{#1\,\textsc{\uppercase{#2}}}}
\else
\newcommand{\ion}[2]{\textup{#1\,\textsc{\uppercase{#2}}}}
\fi

\def\bs{\langle B \rangle}

\def\b{|\mathrm{\mathbf{B}}|}

\def\btimesfi{\sum|\mathbf{B}|_if_i}
\def\geff{g_{\rm eff}}
\def\stokesi{Stokes-\ensuremath{\rm I}}

\begin{document}

\title{Magnetic fields in M dwarfs from the CARMENES survey}

\author{
D.~Shulyak\inst{1,2}
\and
A.~Reiners\inst{1}
\and
E.~Nagel\inst{3}
\and
L.~Tal-Or\inst{1,4}
\and
J.\,A.~Caballero\inst{9}
\and
M.~Zechmeister\inst{2}
\and
V.\,J.\,S.~B\'ejar\inst{11,15}
\and
M.~Cort\'es-Contreras\inst{9}
\and
E.\,L.~Martin\inst{9}
\and
A.~Kaminski\inst{7}
\and
I.~Ribas\inst{5,6}
\and
A.~Quirrenbach\inst{7}
\and
P.\,J.~Amado\inst{8}
\and
G.~Anglada-Escud\'e\inst{8,10}
\and
F.\,F.~Bauer\inst{8}
\and
S.~Dreizler\inst{2}
\and
E.\,W.~Guenther\inst{12}
\and
T.~Henning\inst{13}
\and
S.\,V.~Jeffers\inst{2}
\and
M.~K\"urster\inst{12}
\and
M.~Lafarga\inst{5,6}
\and
J.\,C.~Morales\inst{5,6}
\and
S.~Pedraz\inst{14}
}

\institute{
Max-Planck-Institute f\"ur Sonnensystemforschung, Justus-von-Liebig-Weg 3, D-37075 G\"ottingen, Germany\\
\email{shulyak@mps.mpg.de}
\and
Institut f\"ur Astrophysik, G\"ottingen Universit\"at, Friedrich-Hund-Platz 1, D-37075 G\"ottingen, Germany
\and
Hamburger Sternwarte, Gojenbergsweg 112, D-21029 Hamburg, Germany
\and
Department of Geophysics, Raymond and Beverly Sackler Faculty of Exact Sciences, Tel Aviv University, Tel Aviv, 6997801, Israel
\and
Institut de Ci\'encies de l’Espai (ICE, CSIC), Campus UAB, c/ de Can Magrans s/n, E-08193, Bellaterra, Barcelona, Spain
\and
Institut d’Estudis Espacials de Catalunya (IEEC), E-08034, Barcelona, Spain
\and
Landessternwarte, Zentrum f\"ur Astronomie der Universit\"at Heidelberg, K\"onigstuhl 12, D-69117, Heidelberg, Germany
\and
Instituto de Astrofisica de Andalucia (IAA-CSIC), Glorieta de la Astronomia s/n, E-18008, Granada, Spain
\and
Centro de Astrobiologia (CSIC-INTA), ESAC, Camino Bajo del Castillo s/n, E-28692, Villanueva de la Ca\~nada, Madrid, Spain
\and
Queen Mary University of London, Mile End Road, London E1 4NS, UK
\and
Universidad de La Laguna, Departamento de Astrofisica, C/ Via L\'actea s/n, La Laguna, Tenerife, E-38206, Spain
\and
Th\"uringer Landessternwarte Tautenburg, Sternwarte 5, D-07778 Tautenburg, Germany
\and
Max-Planck-Institut f\"ur Astronomie, K\"onigstuhl 17, D-69117, Heidelberg, Germany
\and
Centro Astron\'omico Hispano-Alem\'an (CSIC-MPG), Observatorio Astron\'omico de Calar Alto, Sierra de los Filabres, E-04550, G\'ergal, Almeria, Spain
\and
Instituto de Astrof\'{\i}sica de Canarias, V\'{\i}a  L\'actea, s/n E38205, La Laguna, Tenerife, Spain
}

\date{Received ; accepted}

 
  \abstract
   {M dwarfs are known to generate the strongest magnetic fields among main-sequence stars with convective envelopes, 
but the link between the magnetic fields and underlying dynamo mechanisms, rotation, and activity
still lacks a consistent picture.}
   {In this work we measure magnetic fields from the high-resolution near-infrared spectra taken with 
the CARMENES radial-velocity planet survey in a sample of $29$ active M dwarfs and compare our results against 
stellar parameters.}
   {We use the state-of-the-art radiative transfer code to measure total magnetic flux densities
from the Zeeman broadening of spectral lines and filling factors.}
   {We detect strong kG magnetic fields in all our targets.
In $16$ stars the magnetic fields were measured for the first time.
Our measurements are consistent with the magnetic field saturation in stars with rotation periods $P<4$~d. 
The analysis of the magnetic filling factors reveal two different patterns 
of either very smooth distribution or a more patchy one, which
can be connected to the dynamo state of the stars and/or stellar mass.}
   {Our measurements extend the list of M dwarfs with strong surface magnetic fields.
   They also allow us to better constrain the interplay between
the magnetic energy, stellar rotation, and underlying dynamo action. The high spectral resolution
and observations at near-infrared wavelengths are the beneficial capabilities of the CARMENES instrument
that allow us to address important questions about the stellar magnetism.}

\keywords{stars: low-mass -- stars: magnetic field -- stars: rotation -- stars: atmospheres}

   \maketitle
%

\section{Introduction}

Magnetic fields are found in all type of stars throughout the Herzsprung-Russel diagram \citep{2001ASPC..248.....M}.
Among main-sequence stars with outer convective envelopes, M dwarfs are know to generate
strong kG magnetic fields \citep{1985ApJ...299L..47S,1996ApJ...459L..95J,2007ApJ...656.1121R,
2009ApJ...692..538R,2010ApJ...710..924R}. These fields are generated
by convective dynamos that operate in stellar interiors and are powered by stellar rotation
\citep{1981ApJ...248..279P,2003A&A...397..147P,2011ApJ...743...48W, 2014ApJ...794..144R}. 
Dynamo-generated magnetic fields reach the surface of a star and initiate a number
of phenomena that we call stellar activity: stellar spots, plages, hot chromospheres and
coronae, etc., which are observed indirectly via the analysis of, e.g., emission lines, X-rays,
photometric variability \citep[e.g.,]{2011ApJ...727...56I,2017ApJ...834...85N} and radio
emission \citep[e.g.,][]{2012ApJ...746...23M}.

Perhaps, one of the most remarkable finding was establishing of the so-called rotation-activity 
relation \citep{1984ApJ...279..763N,2007AcA....57..149K,2014ApJ...794..144R,2017ApJ...834...85N}. 
A key feature of this relation is the existence of two distinct branches (or groups) 
of stars that show very different behaviour of X-ray fluxes with stellar rotation.
On the first branch the amount of X-ray flux decreases as rotation periods of stars increase
as they spin down due to the magnetic braking.
This is a direct evidence that stellar rotation powers dynamo action in these stars.
To the contrary, when the rotation period reaches a certain value of about $4$ days 
(the exact value is actually a function of the stellar mass),
the X-ray fluxes show no (or very marginal) dependence on rotation \citep{2014ApJ...794..144R}.
In this case the stellar dynamo saturates and the amount of non-thermal
energy released by a star reaches a certain limit. There might be
several explanations for the observed phenomena,
but it is generally believed that they are connected to the underlying dynamo
\citep[see discussion in][]{2014ApJ...794..144R}.

Similar to the X-ray emission, \citet{2009ApJ...692..538R} found the same two-branch
behaviour of magnetic flux densities with stellar rotation. The value of the saturated magnetic field
was not well constrained because of limitation of available analysis methods, but it was 
believed to be somewhere equal or slightly above $4$~kG.

As analysis methods improved, it became possible to measure magnetic fields in M dwarfs
from a direct spectrum synthesis in atomic and molecular lines 
\citep{2002A&A...385..701B,2008A&A...482..387A,2010A&A...523A..37S,2014A&A...563A..35S}. 
In their investigation \citet[][Sh2017 hereafter]{2017NatAs...1E.184S}
reported the detection of the magnetic fields well above the presumed saturated value
in few M dwarf stars. The authors used rich spectropolarimetric observations of a 
sample of stars obtained with ESPaDOnS (Canada-France-Hawaii Telescope) and NARVAL (Telescope Bernard Lyot)
instruments and carried out
magnetic field measurements from atomic titanium (Ti) and molecular iron hydride (FeH) lines. In particular,
to date the strongest average magnetic field of about $\bs\approx7$~kG was reported
for the fully convective star WX~UMa, which questioned the concept of the magnetic field saturation. 
Furthermore, Sh2017 were able
to measure magnetic fields in stars with large projected rotational velocities ($\vsini$) from the effect
of magnetic intensification of the well separated Ti lines located in the $Z$-band.
Because many M dwarfs with short periods also have large $\vsini$ values, these stars were previously
excluded from the measurements of total magnetic fields thus biasing results towards
stars with relatively small $\vsini$ values.

Combining their results of magnetic field measurements from \stokesi 
with the results from the polarimetric studies of global magnetic field geometries
presented in \citet{2010MNRAS.407.2269M}, Sh2017 found that the stars that generate strongest fields
are also those with very simple, dipole-dominant magnetic field geometries, while
stars with multipole-dominant global fields do not generate fields stronger than $\approx4$~kG.
They also pointed out that from their limited sample ($25$ stars) it could be seen that the magnetic field in
stars with dipole-dominated geometry does not show an obvious saturation effect. 
These findings provided first observational evidence for the existence 
of two distinct dynamo states in M dwarfs that differ
not only in the geometry of the large-scale magnetic field, which is known since the extensive
polarimetric campaign reported in \citet{2010MNRAS.407.2269M}, but also in total magnetic energy generated.
To explain the observed dichotomy of dipole- vs. multipole-dominant magnetic fields, 
an idea of bi-stable dynamo models was proposed and confirmed with dedicated simulations \citep{2013A&A...549L...5G}.
The finding of Sh2017 that the magnetic fields may not saturate in particular stars or that they saturate to a much
larger values compared to what was thought before needs to be confirmed with additional observations and theoretical
models.

Motivated by our recent findings and availability of the new CARMENES 
instrument, in this work we carry out first magnetic field measurements from the 
high-resolution near-infrared spectroscopy in a subsample of stars with short rotation periods.
Because CARMENES does not have polarimetry, we perform
measurements of the total magnetic flux density but we are not able to constrain dynamo states
in our targets for which measurements of their fields are done for the first time. 
Therefore, our main goal is to constrain the relation between the magnetic field and stellar rotation
and to mark targets with strong magnetic fields for future spectropolarimetric
follow up campaigns.

\section{Observations}

The detailed description of the CARMENES instrument and the overview of the survey
can be found in \citet{2014SPIE.9147E..1FQ,2015A&A...577A.128A,2018A&A...612A..49R}. 
CARMENES is the first high-resolution spectrograph that simultaneously covers the optical and near-infrared wavelength 
range between $520$~nm and $1710$~nm. In two channels (VIS and NIR), the instrument provides data at a resolution 
higher than $R=80\,000$. In particular, the NIR channel covers the wavelength range $960$-$1000$~nm that is essential for our analysis.

In this work we concentrate on the subsample
of $31$ stars presented in \citet{2018A&A...614A.122T} which we called the RV-loud sample.
All these stars had at least
$11$ measurements over the last two years ($2016$--$2017$), projected rotational velocity of $\vsini>2$~\kms, 
and radial velocity scatter amplitude $>10$~\ms\ (as measured from the visual arm of the instrument).
Further details about RV-loud sample are summarized in Table~1 in \citet{2018A&A...614A.122T}.
In this work we make use of the near-infrared arm (NIR) of CARMENES
because this is where our magnetically sensitive spectral features are located.
The data reduction and wavelength calibration for the CARMENES spectra is done with the dedicated tool
CARACAL \citep{2016SPIE.9910E..0EC}, and the high-precision radial velocities are then computed 
using SERVAL package \citep{2018A&A...609A..12Z}. Additionally to radial velocities, SERVAL also
produces a final co-added template spectrum of each star which is built from all available
individual exposures. These templates represent a smoothed and oversampled version
of the stellar spectrum with rejected obvious outliers (cosmic rays, night sky emission lines, etc.)
and boosted signal-to-noise. We also use these templates for some of our magnetic field measurements.

The NIR spectra from CARMENES are often affected by strong telluric absorption from atmospheric water. 
This is a challenge for our analysis because several of the Ti lines are at wavelengths close to telluric water lines. 
We rejected observations of Ti lines when contamination by telluric lines systematically affected our profile analysis.

In Table~\ref{tab:obs} we provide information on the number of individual exposures
taken per star, number of used spectra after rejecting those with very low SNR
and/or telluric removal artifacts, and the SNR of the finally co-added spectra that were used
in the analysis of the magnetic fields.

\begin{table}[!ht]
\small
\caption{Summary of observations.\label{tab:obs}}
\begin{center}
\begin{tabular}{lccc}
\hline
\hline
                 &                      &                 &                      \\
  Karmn          &    No. of taken      & No. of used     &  SNR of coadded      \\
                 &    spectra           & spectra         &    spectra           \\
                 &                      & Ti / FeH          &    Ti / FeH        \\
\hline           
J01033+623       &          19          &    7 / 7          &  190 / 257         \\
J01352-072       &          11          &    5 / 9          &  145 / 248         \\
J02088+494       &          16          &    2 / 13         &   80 / 370         \\
J03473-019       &          11          &    7 / 9          &  195 / 274         \\
J04472+206       &          12          &    2 / 6          &   82 / 182         \\
J05365+113       &          55          &   13 / 17         &  280 / 448         \\
J06000+027       &          14          &    5 / 10         &  156 / 285         \\
J07446+035       &          34          &   26 / 26         &  428 / 585         \\
J08298+267       &          11          &    5 / 5          &  165 / 218         \\
J09449-123       &          11          &    2 / 9          &   93 / 260         \\
J12156+526       &          13          &    4 / 10         &  146 / 300         \\
J12189+111       &          12          &   10 / 10         &  210 / 287         \\
J14173+454       &          11          &    6 / 9          &  113 / 186         \\
J15218+209       &          38          &   30 / 30         &  420 / 607         \\
J15499+796       &          16          &    3 / 6          &   65 / 181         \\
J16313+408       &          11          &    7 / 8          &  118 / 167         \\
J16555-083       &          105         &   43 / 79         &  164 / 348         \\
J16570-043       &          13          &    4 / 6          &  176 / 378         \\
J17338+169       &          14          &    5 / 9          &  133 / 223         \\
J18022+642       &          16          &    7 / 13         &  157 / 280         \\
J18189+661       &          13          &    6 / 12         &  164 / 288         \\
J18356+329$^a$   &          11          &    1 / 5          &   30 / 87          \\
J18498-238       &          41          &    2 / 31         &  149 / 755         \\
J19169+051S      &          33          &   17 / 24         &  121 / 187         \\
J19255+096$^a$   &          26          &    2 / 12         &   24 / 74          \\
J19511+464       &          13          &    6 / 9          &  164 / 286         \\
J20093-012       &          14          &   10 / 10         &  133 / 189         \\
J22012+283       &          12          &   10 / 10         &  219 / 296         \\
J22468+443       &          93          &   26 / 79         &  404 / 1071        \\
J22518+317       &          11          &    4 / 9          &  146 / 282         \\
J23548+385       &          11          &    5 / 9          &  143 / 247         \\
\hline
\end{tabular}
\end{center}
\textbf{Note:} $^a$The two stars marked with were excluded from the analysis of stellar magnetic fields due to very low SNR of the finally co-added spectra.
\end{table}

\section{Methods}

Our magnetic field measurements are based on the direct spectral synthesis 
of spectroscopic features subject to the photospheric magnetic fields.
In order to predict observed line profiles we utilize the \textsc{Magnesyn} spectrum synthesis code,
which is the part of the \llm\ stellar atmosphere package \citep{2004A&A...428..993S}. 
The radiative transfer equation is solved
in all four Stokes parameters for a given configuration of the magnetic field.
Our approach to measure magnetic fields is mostly the same as in our early study
of active M dwarfs presented in Sh2017.

There are two main problems that limit our analysis of cool star spectra. First, despite obvious improvements
in molecular opacity, most of the spectra of M dwarfs are still far from being satisfactory fit to the accuracy
required by modern high-resolution spectrograpths. Much of the background opacity is still missing or not
accurate enough. This becomes gradually worse in late-type M dwarfs where molecules dominate opacity in all spectral
domains. Second, the line formation itself is not well understood because of uncertainties in model atmospheres and
line broadening mechanisms. Therefore it is very difficult to choose a proper set of spectral lines that would
accurately separate effects of the magnetic field and atmospheric parameters (e.g., effective temperature, metallicity, rotation, etc.).
Similar to Sh2017, we used lines of the FeH molecule at $\lamlam990-995$~nm of the Wing--Ford $F^4\,\Delta-X^4\,\Delta$ transitions, 
as well as Ti lines located at $\lamlam960-980$~nm. We choose FeH lines because they contain magnetically very sensitive 
and weakly sensitive lines, and the whole Wing--Ford band is virtually free from telluric contamination. 
However, the fundamental problem with FeH lines is the absence of an accurate theory
for the calculation of Land\'e $g$-factors \citep{2006ApJ...636..548A,2002A&A...385..701B}. 
Apart from a few examples \citep{2008ApJ...679..854H,2012EAS....58...63C,2014IAUS..302..164C}, the laboratory measurements are not available
for the lines that are most important for magnetic field analysis. On another hand, the $g$-factors for Ti lines are accurately known 
and can be computed assuming standard Russell–Saunders coupling scheme. 
Unfortunately, these Ti lines do suffer from strong telluric contamination. In particular, the \ion{Ti}{I}~$\lambda$974.3~nm line 
is essential for disentangling magnetic broadening from the effects of metallicity and rotation because
this line has zero magnetic sensitivity (i.e., effective g-factor $\geff=0$), but very often the profile of this line is severely distorted by a 
nearby telluric feature.
A few more Ti lines are also affected by direct telluric hits. Therefore, it is essential to extract maximum information by
using both FeH and Ti lines and cross check the consistency between the results. For that, we utilize semi-empirical Land\'e g-factors for FeH lines
as described in \citet{2010A&A...523A..37S}, and remove telluric contribution from Ti lines by direct modeling of telluric absorption
with the \textsc{MolecFit}\footnote{\tt https://www.eso.org/sci/software/pipelines/skytools/molecfit} tool \citep{2015A&A...576A..77S,2015A&A...576A..78K}.

Removing telluric features with \textsc{MolecFit} is done by fitting the atmospheric transmission model to the observed
spectrum and adjusting mixing ratios of corresponding atmospheric gases, such as H$_{\rm 2}$0, CO$_{\rm 2}$, etc. 
The atmospheric temperature and pressure as a function of altitude for a given observing time were extracted from
the Global Data Assimilation System (GDAS) database\footnote{\tt https://ready.arl.noaa.gov/gdas1.php}.
The other essential parameters such as, e.g., surface temperature and pressure, local humidity, airmass, are taken from the fits files
generated by the instrument software. Current version of \textsc{MolecFit} utilizes molecular line lists from the HITRAN 
database\footnote{https://www.cfa.harvard.edu/hitran}.
After the state of the Earth's atmosphere has been defined, the code runs Levenberg-Marquardt $\chi^2$-minimization
algorithm to find best fit molecular number densities. The code can also fit polynomials of different orders for the normalization
of the transmission spectrum. Once the best fit parameters have been found, the observed spectrum is divided by the predicted transmission model.
This procedure was applied to each individual CARMENES exposure. The accuracy of the telluric removal depends strongly
on the SNR of the data and atmospheric humidity. For instance, if telluric lines are deep then even small mismatch between
model and observations in line cores may give rise to strong artefacts after spectra division.
Therefore, the application of the same procedure to spectra obtained 
at different seasons and/or different observatories can lead to deviations the spectra corrected for atmospheric transmission.

In stars with large projected rotational velocities, the rotational broadening dominates the shape of line profiles, which means that
narrow and dense FeH lines become strongly blended and it is difficult or even
impossible to accurately constrain the strength of surface magnetic fields from FeH lines alone. 
To the contrary, Ti lines are well separated from each other and remain unblended even with
largest $\vsini$'s found in M dwarf stars. The only reason why these lines were never used in magnetic field
studies is exactly because of strong telluric contamination. 
However, if telluric can be accurately removed, these lines
become perfect probes of magnetic fields in fast rotating stars because one can use the effect 
of magnetic intensification to measure magnetic flux density. For instance, in the case 
of a saturated spectral line, its equivalent width will be proportional to the intensity of the magnetic field
and the number of individual Zeeman components, so-called Zeeman pattern \citep{2004ASSL..307.....L}.
Therefore, at large $\vsini$ the depths of individual spectral lines subject to strong magnetic field 
will depend on their Zeeman patterns.
The lines of Ti in $\lamlam960-980$~nm region all have very different Zeeman patterns and thus
one can fit their depths to measure intensity of the magnetic field (provided that other stellar parameters are known).
The obvious caveats in this analysis is that the magnetic intensification (to a certain degree)
can be mimicked by other effects, such as, e.g., metallicity or temperature effects: 
changing either of them will make lines look deeper/shallower,
so that one can always satisfactorily fit a magnetically sensitive line with a non-magnetic model by adjusting 
atmospheric parameters. Therefore, using the magnetic insensitive \ion{Ti}{I}~$\lambda$974.3~nm
line is the only way to break this degeneracy because it helps to constrain stellar parameters
separately from the magnetic field.
From the magnetic intensification we can only measure magnetic fields that are strong enough 
to noticeably affect the equivalent widths of spectroscopic lines.
Nevertheless, this effect opened a way to measure magnetic fields in stars with extreme $\vsini$ values
that was for many years believed impossible \citep[][Sh2017]{2017ApJ...835L...4K}.

Our approach is to measure magnetic flux density from unpolarized (\stokesi) light.
This way we can capture the magnetic fields that are organized at both large and small scales.
However, the geometry of the magnetic field remains unconstrained, i.e., we are blind
to the sign of the field that we are looking at. To address this question one would need 
additional polarimetric observations with another instrument(s). Still, strong surface magnetic fields
in stars with convective envelopes originate from spots and other active regions that differ in size, number, etc.
which distorts profiles of spectral lines differently from a case where the magnetic field
was homogeneously distributed over the stellar surface. 
This gives us a possibility to measure what we call
the complexity of surface field by assuming a distribution of magnetic field components $\b_{i}$ each
covering a particular portion of the star represented by filling factor $\mathrm{f}_i$.
The total magnetic field is then a simple sum $\bs=\btimesfi$.
We use the Levenberg-Marquardt minimization algorithm to find
best fit values of filling factors for a given combination of atmospheric parameters.
We treat simultaneously rotational velocity $\vsini$, atmospheric abundance of a given element,
continuum scaling factor for each spectral region that we fit,
and $11$ filling factors distributed between $0$~kG and $20$~kG as free parameters.
The continuum scaling factors are needed to account for possible mismatch
between observed and predicted spectra that may originate from, e.g., artefacts left after
flat fielding and/or order merging, missing molecular opacity,
uncertainties in model atmospheres, etc. Note that because the stellar continuum cannot be accurately defined, especially
in late type M dwarfs with strong absorption cause my numerous photospheric water lines in the vicinity of our Ti lines,
we fit a low order polynomial to each CARMENES echelle order prior to merging them into single 1D spectrum.
This spectrum is thus normalized to stellar pseudo-continuum, but small offsets could still be present
and we account for them in our fitting approach with additional scaling factors.

Similar to Sh2017, we compute theoretical line profiles using MARCS model atmospheres \citep{2008A&A...486..951G}.
The effective temperatures  of our sample stars were computed from their spectral types employing
dedicated calibrations \citep{1995ApJS..101..117K,2004AJ....127.3516G}.
We took spectral types from the CARMENCITA database \citep{2015A&A...577A.128A,2016csss.confE.148C}.
The other parameter~--~surface gravity ($\logg$)~--~was calculated from the stellar radii
and mass assuming $R\propto M^{0.9}$ relation which provides surface gravities close to those predicted
by stellar evolution models \citep{2008ApJS..178...89D,2012MNRAS.427..127B}.
We did not use effective temperatures and 
surface gravities from \citet{2018A&A...615A...6P}
because they were not able to derive such parameters for the 
most active and fastest rotating CARMENES stars, which just comprise our sample.

The transition parameters of Ti lines were extracted from the Vienna Atomic Line Database (VALD) \citep{2015PhyS...90e4005R},
and for FeH lines we used molecular constants taken from \citet{2003ApJ...594..651D}\footnote{\tt http://bernath.uwaterloo.ca/FeH}.
Transition probabilities for some of these lines were corrected according to \citet{2010A&A...523A..58W}.

Also, similar to Sh2017, we treat pressure broadening by including contributions from hydrogen and helium only
and ignore contribution from molecular hydrogen H$_{\rm 2}$ because it grossly overestimates
the observed widths of atomic lines especially at late spectral types \citep[see, e.g.,][]{2007ARep...51..282P}.
Another source of line broadening is the velocity field caused by convective motions.
However, as shown by 3D hydrodynamic simulations \citep{2009A&A...508.1429W}, 
the velocity fields in atmospheres of M dwarfs with temperatures $\teff<3500$~K
are well below $1$~\kms. This has a weak impact on line profiles, leaving the Zeeman effect and rotation
to be the dominating broadening mechanisms. Therefore, we assumed zero micro- and macroturbulent velocities
in all calculations.

\begin{figure}
\centerline{
\includegraphics[width=0.49\hsize]{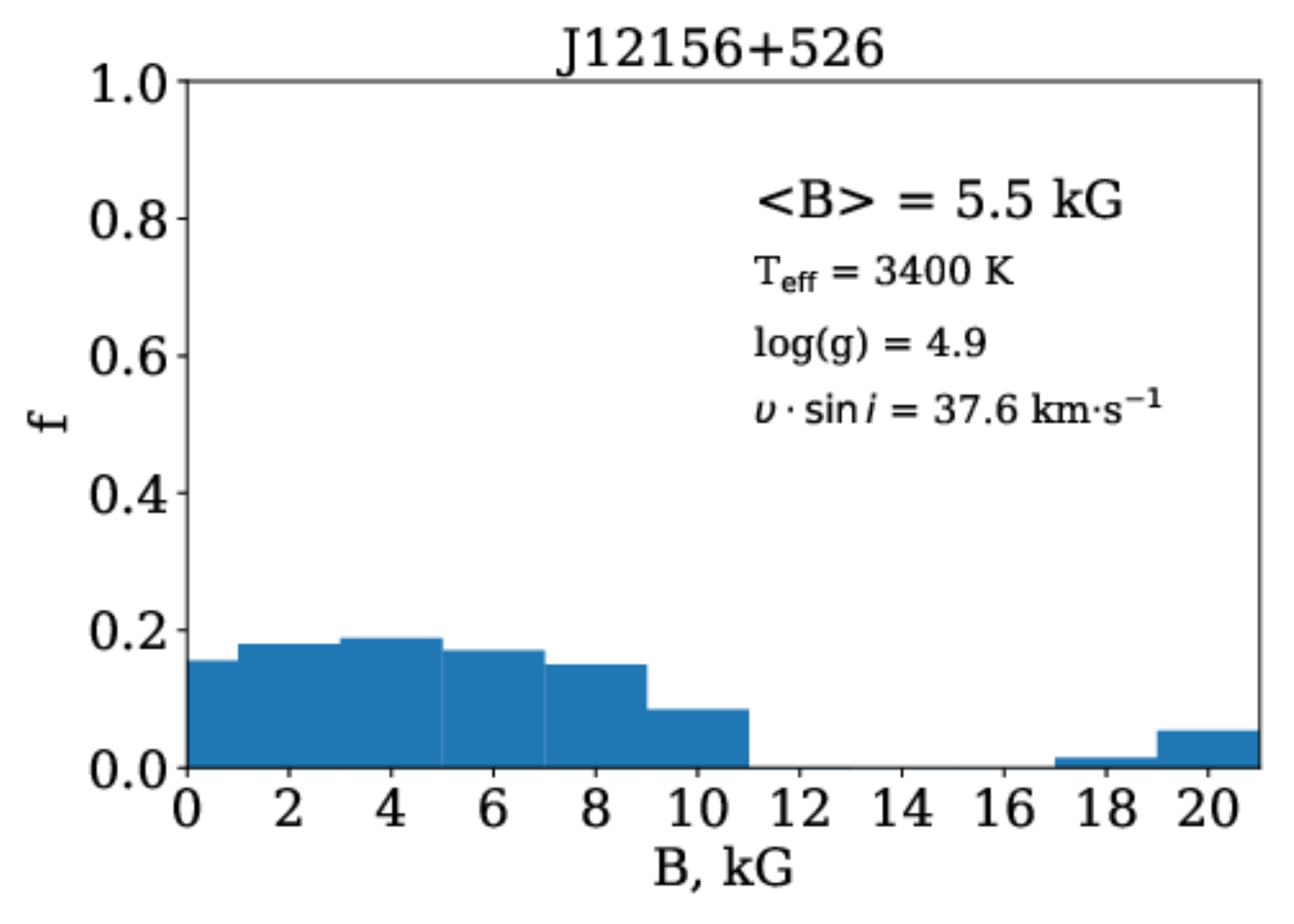}
\includegraphics[width=0.49\hsize]{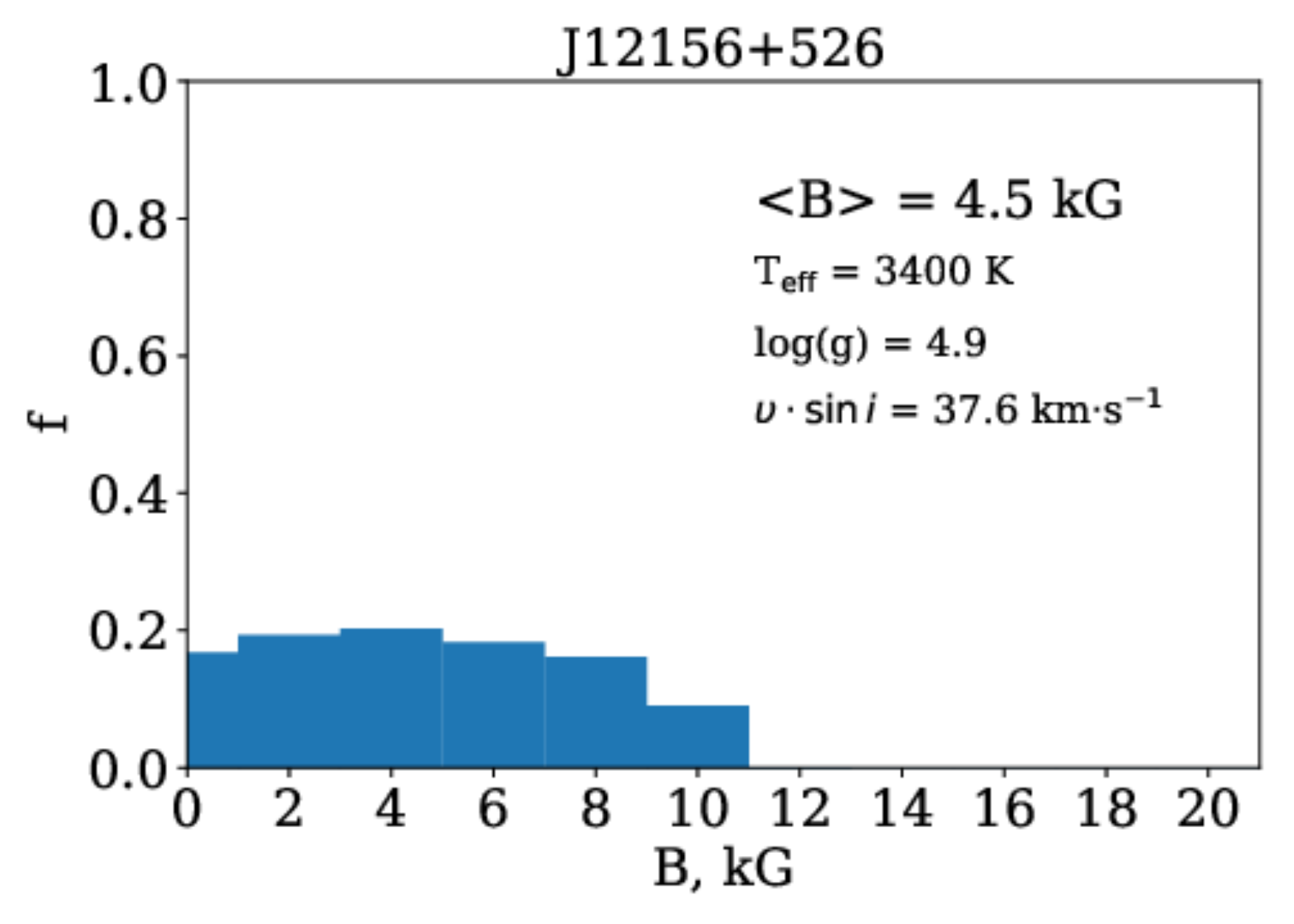}
}
\caption{\label{fig:high-bf-example}
Example of magnetic filling factors of J12156+526 as derived from \ion{Ti}{I} lines. 
Original distribution contains high field components (left panel), and ignoring them with subsequent
re-scaling of the rest of filling factors drops the average magnetic field by $1$~kG (right panel).}
\end{figure}

The filling factors that we derive with our method sometimes show high-field magnetic components. 
A characteristic example
is illustrated in Fig.~\ref{fig:high-bf-example}, where we show the filling factors of J12156+526 as derived from the Ti lines.
Here we derive a very smooth distribution of filling factors between $0$~kG and $10$~kG,
but then we also find two stand-alone magnetic field components at the tailed $18$~kG and $20$~kG, respectively.
As mentioned in Sh2017, these components result from the code trying to fit details in observed line profiles
originated from the poor data quality and/or bad understanding of background molecular absorption. 
The filling factors of these components are small, but they contribute a significant
fraction to the final result: ignoring them drops the average magnetic field from $5.5$~kG to $4.5$~kG, respectively.
Note that the magnetic components with strongest field appear always irregardless of the choice of the limiting
magnetic field. In other words, it does not matter if one truncates the last magnetic component to, say, $15$~kG
instead of $20$~kG because the $15$~kG magnetic component will then simply appear to have large filling factor, etc.
We also cannot use too few magnetic field components because we would under-fit profiles of spectral lines.
Because we have no physically motivated reason to believe that these high field components truly exist,
we ignore them in our estimates of stellar magnetic fields.

While the initial guess for abundances and $\vsini$ does not have a strong impact on the final results, 
this is not always the case for filling factors.
Again, we find that this is more often an issue for stars with poor data quality. Interestingly, even if filling factor
distributions look different, the total magnetic field strength is usually not affected much (with a scatter
on the order of a few hundred Gauss in our sample). The stars for which the initial guess of filling factors also affects
the resulting field strength are those with $\vsini>20$~\kms\ and only when we derive fields from FeH lines.
This is expected because strong blending of FeH lines and uncertainties in their transition and
magnetic parameters act together contributing to the uncertainty of the results.

\section{Results}

Our present sample comprises of $29$ active stars 
with short rotation periods ranging between $10$~d and $0.1$~d. Among them,
there are $16$ objects without previous magnetic field measurements,
and $5$ objects with $\vsini>30$~\kms. Two stars, J18356+329 and J19255+096,
were excluded from the analysis because of poor data quality (see Table~\ref{tab:obs}).
We summarize our magnetic field measurements
in Table~\ref{tab:magnetic} and describe them in detail below.
In addition, we provide our model fits to individual spectral lines in Online materials.
Note that because filling factors are correlated parameters, it was not possible to estimate robust error bars 
on the values of the magnetic field strength and filling factors from our approach.
Besides, the formal errors from the chi-square fit on the parameters are likely underestimating the true uncertainties
because of additional sources of uncertainties like, e.g., choice of the spectral lines, accuracy of the telluric
correction, etc. It appears very difficult to quantify all possible sources because we would need
to run a lot more individual measurements for each star which is very computationally demanding.
Therefore, similar to our previous works we put a conservative 
$1$~kG error bars on measured magnetic fields in stars with $\vsini>20$~\kms.
For the rest of the sample we provide uncertainties on the derived magnetic fields
as a difference between measurements from FeH and Ti lines, similar to Sh2017.

\begin{sidewaystable*}[!ht]
\caption{Magnetic field measurements.\label{tab:magnetic}}
\begin{center}
\begin{tabular}{llrccc|ccc|ccc|c|c}
\hline
\hline
             &                                    &        &       &          &          &                   &                     &        &                        &                   &         &            &                \\
Karmn        &  Name                              & Gl/GJ  & SpT   & $\teff$  & $\logg$  &  \multicolumn{3}{c|}{\ion{Ti}{i} lines}          &  \multicolumn{3}{c|}{FeH lines}                      & $P$        & $\bs$          \\
             &                                    & number &       & K        & dex      &   $\abn{Ti}$      & $\vsini$            &  $\bs$ &   $\abn{Fe}$           & $\vsini$          &  $\bs$  & d          & adopted        \\
             &                                    &        &       &          &          &   dex             &  \kms               &   kG   &   dex                  &  \kms             &   kG    &            & kG             \\
\hline  
J01033+623   &   \object{V388 Cas}                &   51   & M5.0  &  3200    &    5.1   & -6.98 $\pm$ 0.02  & 11.7 $\pm$ 0.2      & 5.0    & -4.42 $\pm$  0.01      &  11.3 $\pm$ 0.1   & 4.7     &    1.06    & 4.8 $\pm$ 0.3  \\
J01352-072   &   \object{Barta 161 12}            &        & M4.0  &  3400    &    4.9   & -6.99 $\pm$ 0.03  & 47.8 $\pm$ 0.7      & 5.8    & -4.73 $\pm$  0.02      &  41.8 $\pm$ 0.5   & 4.6     &    0.7031  & 5.8 $\pm$ 1.0  \\
J02088+494   &   \object{G 173-039}               & 3136   & M3.5  &  3400    &    4.9   & -6.91 $\pm$ 0.02  & 22.9 $\pm$ 0.4      & 4.9    & -4.55 $\pm$  0.01      &  21.3 $\pm$ 0.2   & 4.5     &    0.748   & 4.9 $\pm$ 1.0  \\
J03473-019   &   \object{G 080-021}               &        & M3.0  &  3500    &    4.9   & -6.93 $\pm$ 0.01  &  5.7 $\pm$ 0.1      & 3.2    & -4.46 $\pm$  0.01      &   6.4 $\pm$ 0.1   & 3.1     &    3.88    & 3.2 $\pm$ 0.1  \\
J04472+206   &   \object{RX J0447.2+2038}         &        & M5.0  &  3200    &    5.1   & -7.01 $\pm$ 0.03  & 46.5 $\pm$ 0.6      & 5.7    & -4.93 $\pm$  0.03      &  37.5 $\pm$ 0.7   & 5.2     &    0.342   & 5.7 $\pm$ 1.0  \\
J05365+113   &   \object{V2689 Ori}               &  208   & M0.0  &  3800    &    4.7   & -6.99 $\pm$ 0.01  &  0.4 $\pm$ 1.1      & 1.0    & -4.73 $\pm$  0.01      &   2.9 $\pm$ 0.1   & 1.3     &   12.04    & 1.2 $\pm$ 0.3  \\
J06000+027   &   \object{G 099-049}               & 3379   & M4.0  &  3400    &    4.9   & -6.82 $\pm$ 0.02  &  6.5 $\pm$ 0.2      & 2.0    & -4.28 $\pm$  0.01      &   5.6 $\pm$ 0.1   & 3.0     &    1.81    & 2.5 $\pm$ 1.0  \\
J07446+035   &   \object{YZ CMi}                  &  285   & M4.5  &  3300    &    5.0   & -6.93 $\pm$ 0.02  &  4.5 $\pm$ 0.1      & 4.7    & -4.41 $\pm$  0.01      &   5.4 $\pm$ 0.1   & 4.4     &    2.78    & 4.6 $\pm$ 0.3  \\
J08298+267   &   \object{DX Cnc}                  & 1111   & M6.5  &  2700    &    5.3   & -7.19 $\pm$ 0.03  & 12.4 $\pm$ 0.4      & 3.0    & -5.02 $\pm$  0.01      &  10.4 $\pm$ 0.1   & 3.6     &    0.459   & 3.3 $\pm$ 0.6  \\
J09449-123   &   \object{G 161-071}               &        & M5.0  &  3200    &    5.1   & -7.08 $\pm$ 0.03  & 30.1 $\pm$ 0.6      & 5.3    & -4.70 $\pm$  0.01      &  27.3 $\pm$ 0.3   & 4.5     &    0.306   & 5.3 $\pm$ 1.0  \\
J12156+526   &   \object{StKM 2-809}              &        & M4.0  &  3400    &    4.9   & -6.90 $\pm$ 0.02  & 37.6 $\pm$ 0.4      & 4.5    & -4.83 $\pm$  0.01      &  28.0 $\pm$ 0.3   & 4.2     & $<$0.4     & 4.5 $\pm$ 1.0  \\
J12189+111   &   \object{GL Vir}                  & 1156   & M5.0  &  3200    &    5.1   & -6.89 $\pm$ 0.02  & 15.4 $\pm$ 0.3      & 3.3    & -4.22 $\pm$  0.01      &  15.3 $\pm$ 0.1   & 4.0     &    0.491   & 3.6 $\pm$ 0.7  \\
J14173+454   &   \object{RX J1417.3+4525}         &        & M5.0  &  3200    &    5.1   & -7.05 $\pm$ 0.03  & 13.8 $\pm$ 0.4      & 5.1    & -4.56 $\pm$  0.01      &  15.3 $\pm$ 0.1   & 5.3     & $<$0.7     & 5.2 $\pm$ 0.2  \\
J15218+209   &   \object{OT Ser}                  & 9520   & M1.5  &  3600    &    4.8   & -6.95 $\pm$ 0.02  &  3.9 $\pm$ 0.2      & 3.3    & -4.49 $\pm$  0.01      &   5.0 $\pm$ 0.1   & 3.1     &    3.372   & 3.2 $\pm$ 0.2  \\
J15499+796   &   \object{LP 022-420}              &        & M5.0  &  3200    &    5.1   & -6.86 $\pm$ 0.02  & 27.7 $\pm$ 0.5      & 3.4    & -4.73 $\pm$  0.02      &  22.9 $\pm$ 0.5   & 4.2     & $<$0.3     & 3.4 $\pm$ 1.0  \\
J16313+408   &   \object{G 180-060}               & 3959   & M5.0  &  3200    &    5.1   & -6.92 $\pm$ 0.02  &  7.5 $\pm$ 0.2      & 4.1    & -4.31 $\pm$  0.01      &   8.1 $\pm$ 0.1   & 4.0     &    0.51    & 4.1 $\pm$ 0.1  \\
J16555-083   &   \object{vB 8}                    & 644 C  & M7.0  &  2600    &    5.3   & -7.21 $\pm$ 0.04  &  5.8 $\pm$ 0.5      & 2.6    & -5.07 $\pm$  0.01      &   7.0 $\pm$ 0.1   & 3.0     & $<$1.0     & 2.8 $\pm$ 0.4  \\
J16570-043   &   \object{LP 686-027}              & 1207   & M3.5  &  3400    &    4.9   & -6.93 $\pm$ 0.02  & 10.5 $\pm$ 0.2      & 3.1    & -4.31 $\pm$  0.01      &  10.7 $\pm$ 0.1   & 3.5     &    1.21    & 3.3 $\pm$ 0.4  \\
J17338+169   &   \object{1RXS J173353.5+165515}   &        & M5.5  &  3000    &    5.2   & -7.15 $\pm$ 0.04  & 36.6 $\pm$ 1.0      & 6.9    & -5.18 $\pm$  0.03      &  29.1 $\pm$ 0.6   & 4.5     &    0.27    & 6.9 $\pm$ 1.0  \\
J18022+642   &   \object{LP 071-082}              &        & M5.0  &  3200    &    5.1   & -6.90 $\pm$ 0.02  & 10.6 $\pm$ 0.2      & 3.8    & -4.37 $\pm$  0.01      &  10.9 $\pm$ 0.1   & 4.7     &    0.28    & 4.3 $\pm$ 0.9  \\
J18189+661   &   \object{LP 071-165}              & 4053   & M4.5  &  3300    &    5.0   & -6.59 $\pm$ 0.04  & 15.6 $\pm$ 0.3      & 1.8    & -4.30 $\pm$  0.02      &  14.4 $\pm$ 0.2   & 3.4     & $<$0.7     & 2.6 $\pm$ 1.6  \\
J18498-238   &   \object{V1216 Sgr}               &  729   & M3.5  &  3400    &    4.9   & -6.73 $\pm$ 0.03  &  5.4 $\pm$ 0.2      & 1.8    & -4.27 $\pm$  0.01      &   4.1 $\pm$ 0.1   & 2.6     &    2.87    & 2.2 $\pm$ 0.8  \\
J19169+051S  &   \object{V1298 Aql (vB 10)}       & 752 B  & M8.0  &  2500    &    5.3   & -7.55 $\pm$ 0.10  &  5.1 $\pm$ 0.5      & 2.2    & -5.11 $\pm$  0.01      &   5.4 $\pm$ 0.1   & 2.4     &    0.8     & 2.3 $\pm$ 0.2  \\
J19511+464   &   \object{G 208-042}               & 1243   & M4.0  &  3400    &    4.9   & -6.87 $\pm$ 0.02  & 23.3 $\pm$ 0.3      & 3.2    & -4.53 $\pm$  0.01      &  21.1 $\pm$ 0.2   & 3.3     &    0.594   & 3.2 $\pm$ 1.0  \\
J20093-012   &   \object{SCR J2009-0113}          &        & M5.0  &  3200    &    5.1   & -6.95 $\pm$ 0.03  &  5.1 $\pm$ 0.2      & 3.1    & -4.31 $\pm$  0.01      &   5.9 $\pm$ 0.1   & 3.4     & $<$1.8     & 3.2 $\pm$ 0.3  \\
J22012+283   &   \object{V374 Peg}                & 4247   & M4.0  &  3400    &    4.9   & -6.84 $\pm$ 0.01  & 36.3 $\pm$ 0.3      & 4.4    & -4.53 $\pm$  0.01      &  31.5 $\pm$ 0.3   & 4.4     &    0.45    & 4.4 $\pm$ 1.0  \\
J22468+443   &   \object{EV Lac}                  &  873   & M3.5  &  3400    &    4.9   & -6.90 $\pm$ 0.02  &  4.5 $\pm$ 0.2      & 4.2    & -4.44 $\pm$  0.01      &   5.1 $\pm$ 0.1   & 4.0     &    4.379   & 4.1 $\pm$ 0.2  \\
J22518+317   &   \object{GT Peg}                  &  875.1 & M3.0  &  3500    &    4.9   & -6.90 $\pm$ 0.02  & 14.3 $\pm$ 0.2      & 3.1    & -4.49 $\pm$  0.01      &  13.2 $\pm$ 0.1   & 3.7     &    1.64    & 3.4 $\pm$ 0.6  \\
J23548+385   &   \object{RX J2354.8+3831}         &        & M4.0  &  3400    &    5.0   & -6.92 $\pm$ 0.02  &  4.3 $\pm$ 0.2      & 4.8    & -4.34 $\pm$  0.01      &   4.7 $\pm$ 0.1   & 4.4     &    4.76    & 4.6 $\pm$ 0.4  \\
\hline
\end{tabular}
\end{center}
For stars without measured rotational periods we esimate them from $\vsini$ values and stellar radii estimated from spectral types.
The last column lists adopted values of the surface magnetic fields. For stars with $\vsini<20$~\kms\ they were taken as a mean
between meaurements from Ti and FeH lines, respectively, and the error bars were assumed to be the corresponding difference between these measurements.
For stars with $\vsini>20$~\kms\ we adopt measurements from the Ti lines and similar to \citet{2017NatAs...1E.184S} assume conservative uncertainties of $1$~kG.
The rotational velocities derived from Ti lines should be preferred over those derived from FeH lines.
\end{sidewaystable*}

\subsection{Magnetic fields of sample stars}

\begin{figure}
\includegraphics[width=\hsize]{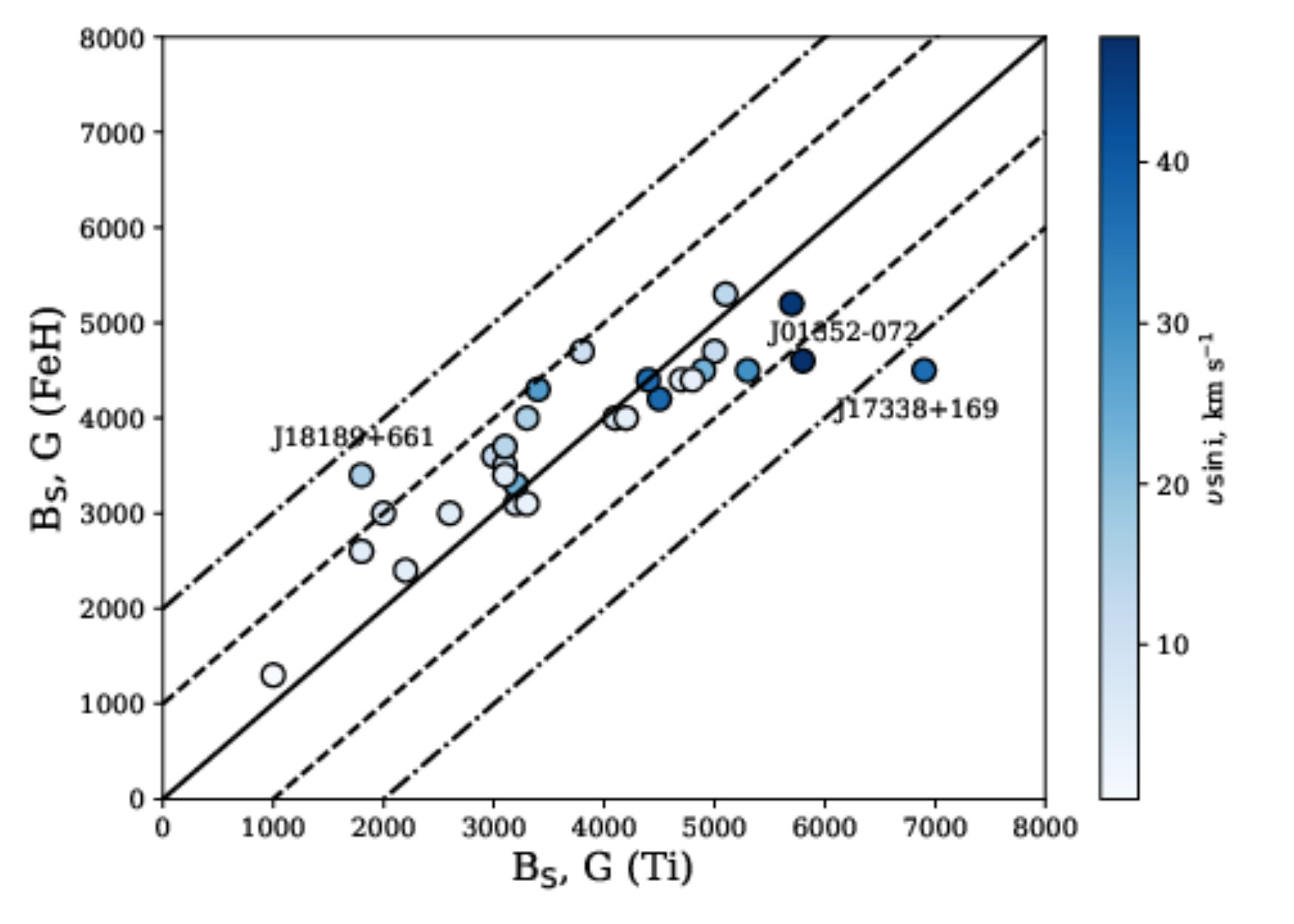}
\caption{\label{fig:bf-ti-feh}
Comparison between the magnetic field measurements from Ti and FeH lines with $\vsini$ color coded.
Dashed and dash-dotted lines represent $1$~kG and $2$~kG deviations from the central line, respectively.}
\end{figure}

Figure~\ref{fig:bf-ti-feh} compares magnetic fields measured from Ti and FeH lines, respectively. 
In general, we obtain very consistent estimates of the magnetic fields from Ti and FeH lines in stars with
good data quality and relatively small $\vsini$ values, such as, e.g., J03473-019, J05365+113, J07446+035, J15218+209, and J22468+443
(see Table~\ref{tab:magnetic}).
Thus, despite of telluric removal problem and uncertainties in magnetic $g$-factors, both Ti and FeH lines
can be successfully used for the magnetic field measurements.

At the same time, for stars J06000+027 and J18498-238 we obtain very different estimates even though
the data for these stars look relatively good and their $\vsini$ values are relatively small, too (see Table~\ref{tab:magnetic}). 
Note that in these stars we measure systematically higher $\vsini$
from Ti lines compared to FeH lines and hence lower values of magnetic fields, 
which can partly be due to the distortion of the magnetic insensitive \ion{Ti}{i}~$\lambda$974.3~nm line
by telluric removal procedure. If we fix $\vsini$ in these stars to the lower values 
(i.e., to the values derived from the fit to FeH lines) we always obtain estimates close to that derived from FeH lines.
It is thus possible that inaccuracies in $\vsini$ values can explain the observed discrepancy in derived magnetic fields
in these objects.

In three stars we find more than $1$~kG deviation between measurements from Ti and FeH lines, and we marked them with
their names in Fig.~\ref{fig:bf-ti-feh}. The largest deviations of $2.4$~kG 
is found in J17338+169.
In this star the red wing of the Ti~$\lambda$974.3~nm line
is affected by telluric removal artifact (see Fig.~\ref{fig:J17338+169-fit}). 
Alternatively, we used different individual spectra to fit this line and 
obtained a weaker field of $\bs=6$~kG with higher $\vsini=39$~\kms instead of $\bs=6.9$~kG,
but the large deviation between measurements from Ti and FeH  still remained.
In J01352-072 the fit to FeH spectrum is very inaccurate. In addition, we could use only a half of the region
covered by FeH lines. At the same time, our fit to Ti lines is much more accurate.
In J18189+661 we could not accurately remove telluric feature from the \ion{Ti}{i}~$974.3$~nm line that
made this line look deeper. As a result, we measured overestimated Ti abundance and hence weaker magnetic field
from Ti lines compared to that from FeH lines.

In stars with rotation velocities $\vsini>20$~\kms\ we often (but not always) measure weaker fields from FeH lines
compared to Ti lines. The same effect was observed and explained in Sh2017 as being likely caused by strong line blending which leads
to the degeneracy in fitting parameters (i.e., temperature, abundance, rotation).
The results of these measurements depend strongly on data quality and set of FeH lines used for the analysis.
Especially in fast rotating stars the result of our measurements from Ti lines must be preferred
over values derived from FeH lines.

\subsection{Comparison with previous measurements}

In our sample there are $12$ objects with previously measured magnetic fields, and in $9$
we find a good agreement between literature values and our measurements. 
Stars for which our new measurements disagree with
previous results are J01033+623, J15218+209, and J19169+051.

In J01033+623, Sh2017 measured a field of $\bs=6.1$~kG from the ESPaDOnS data,
while from CARMENES data we derive much weaker $\bs=4.8$~kG. 
This is because now we derive more smooth distribution of filling factors with a reduced $12$~kG component.
If we exclude this component from our fit to ESPaDOnS data we get very close $\bs=5$~kG average field,
but then the fit to the data looks worse. The same discrepancy is found for FeH lines, and
at this point it is difficult to decide which of the fit should be preferred. 

In J15218+209 we find a $0.5$~kG stronger field from
CARMENES data both from Ti and FeH lines, but our FeH spectrum is very noisy. The difference in
measured magnetic field from Ti lines could be explained by quite different $\vsini$, i.e. we measure
$\vsini=3.9$~\kms\ and $\vsini=4.9$~\kms\ from CARMENES and ESPaDOnS spectra, respectively. We found out that
the higher $\vsini$ measured from the ESPaDOnS data is likely due to the telluric removal artifact that affected
the profile of the magnetic insensitive \ion{Ti}{i}~$\lambda974.3$~nm line which made it broader in ESPaDOnS data.
 
In J19169+051S we measure a stronger field of $\bs=2.2$~kG
compared to previous estimate of $\bs=1.3$~kG from \cite{2007ApJ...656.1121R}. 
Our fit to Ti lines in this star was not very accurate
because we could use only four lines and the observed line profiles were affected by surrounding strong molecular features
prominent at these cool temperatures. On the other hand, FeH spectrum looks better and from them we derive still field of $\bs=2.4$~kG.
Note that \citet{2007ApJ...656.1121R} measured field in J19169+051S indirectly, i.e. without employing
radiative transfer models, which could explain the difference in the magnetic fields between their
and our measurements. On the other hand, our understanding of line formation in cool temperatures of late type
M dwarfs is far from been satisfactorily understood. Uncertainties in, e.g., molecular broadening 
can bias our results. This question needs to be addressed on a larger sample of stars.

\subsection{Complexity of surface magnetic fields}

\begin{figure*}
\centering
\includegraphics[width=0.3\hsize]{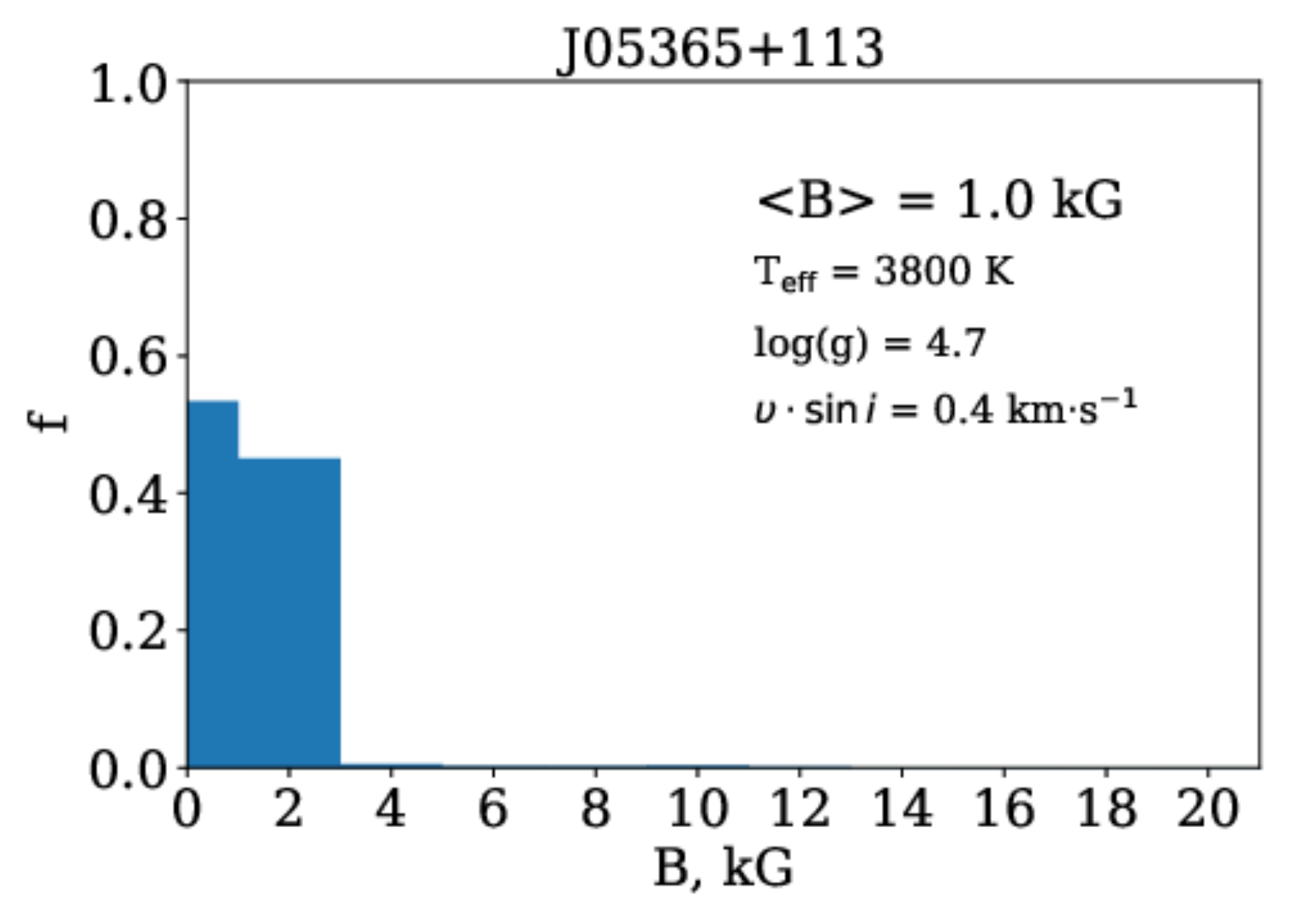}
\includegraphics[width=0.3\hsize]{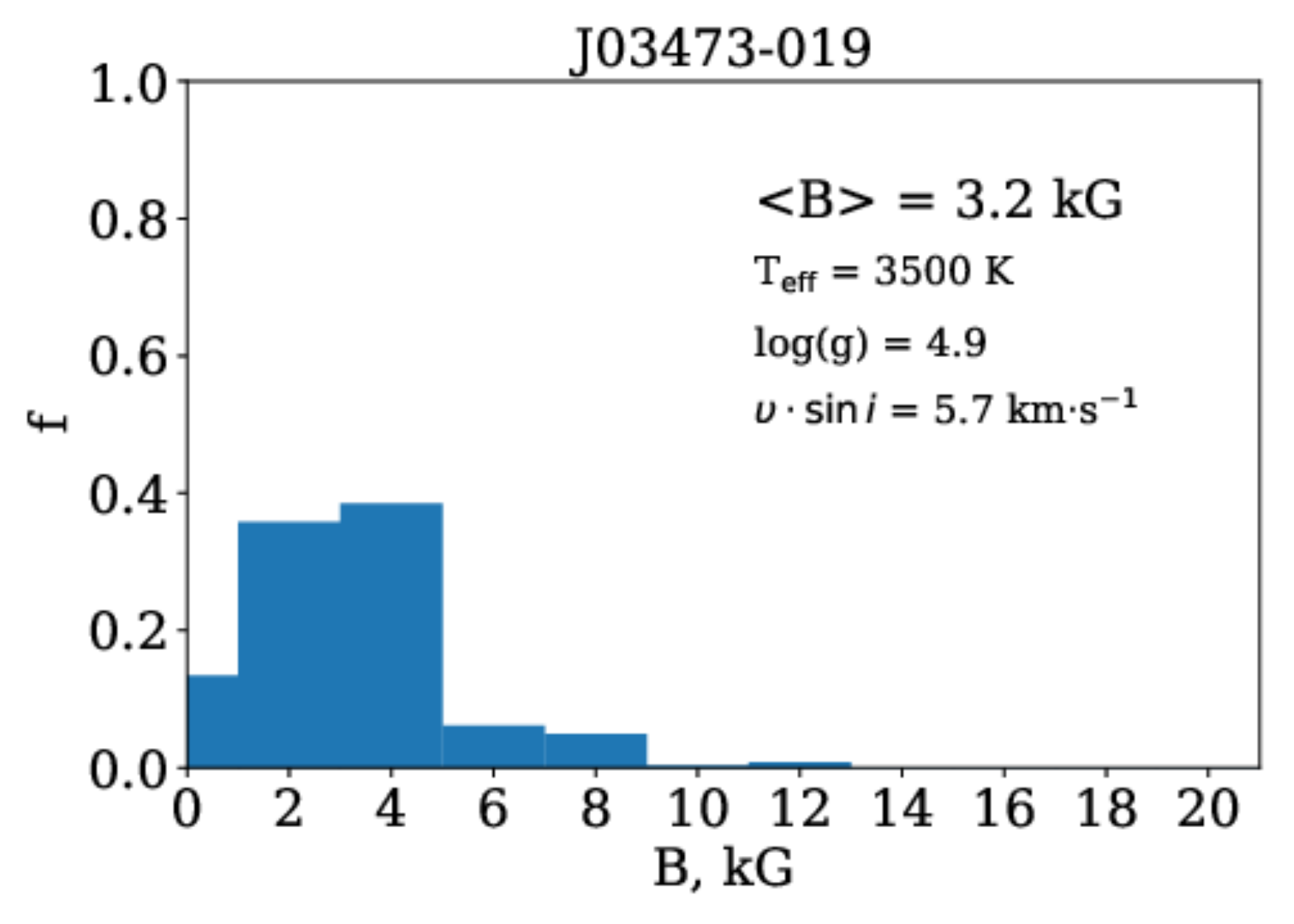}
\includegraphics[width=0.3\hsize]{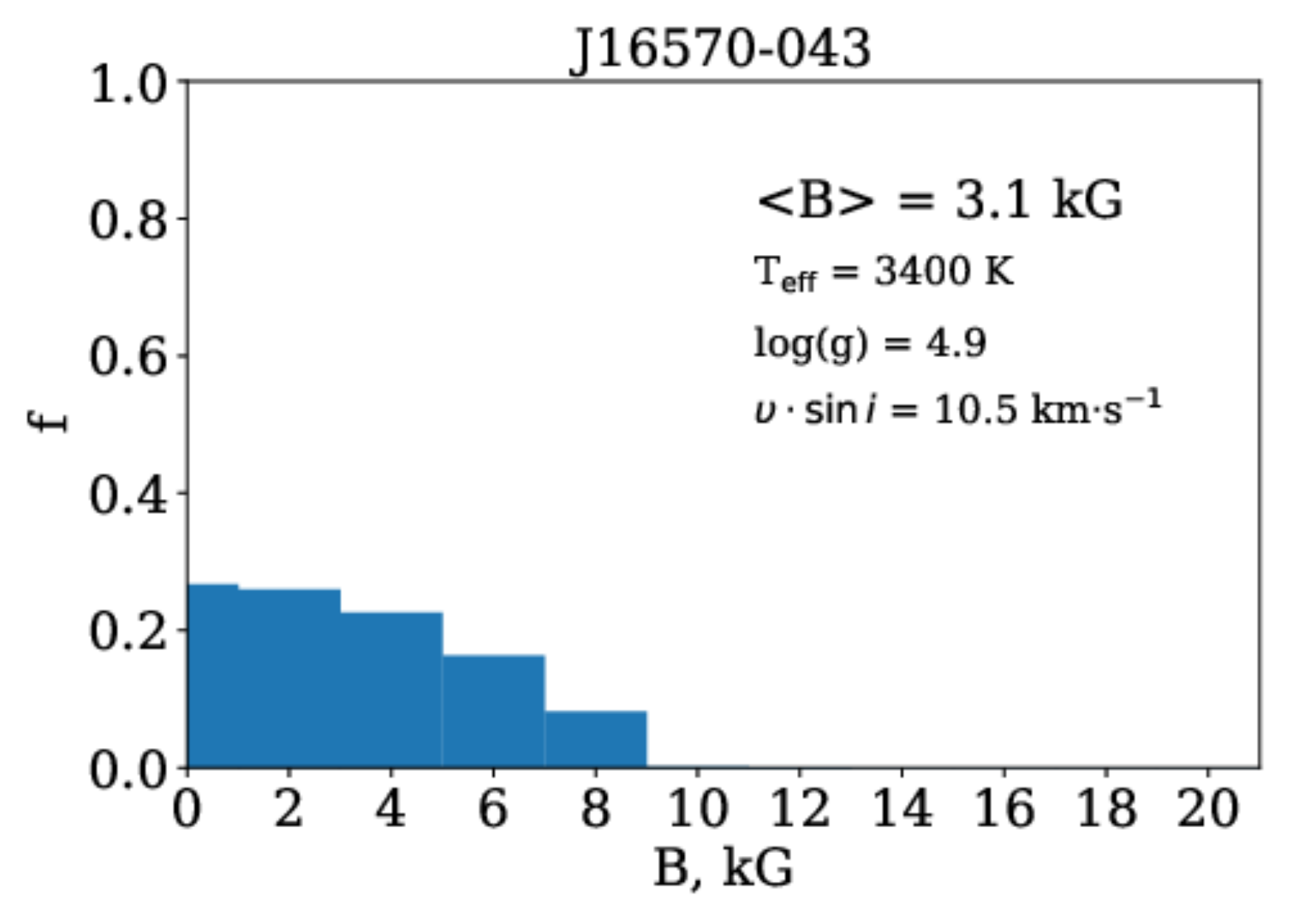}
\includegraphics[width=0.3\hsize]{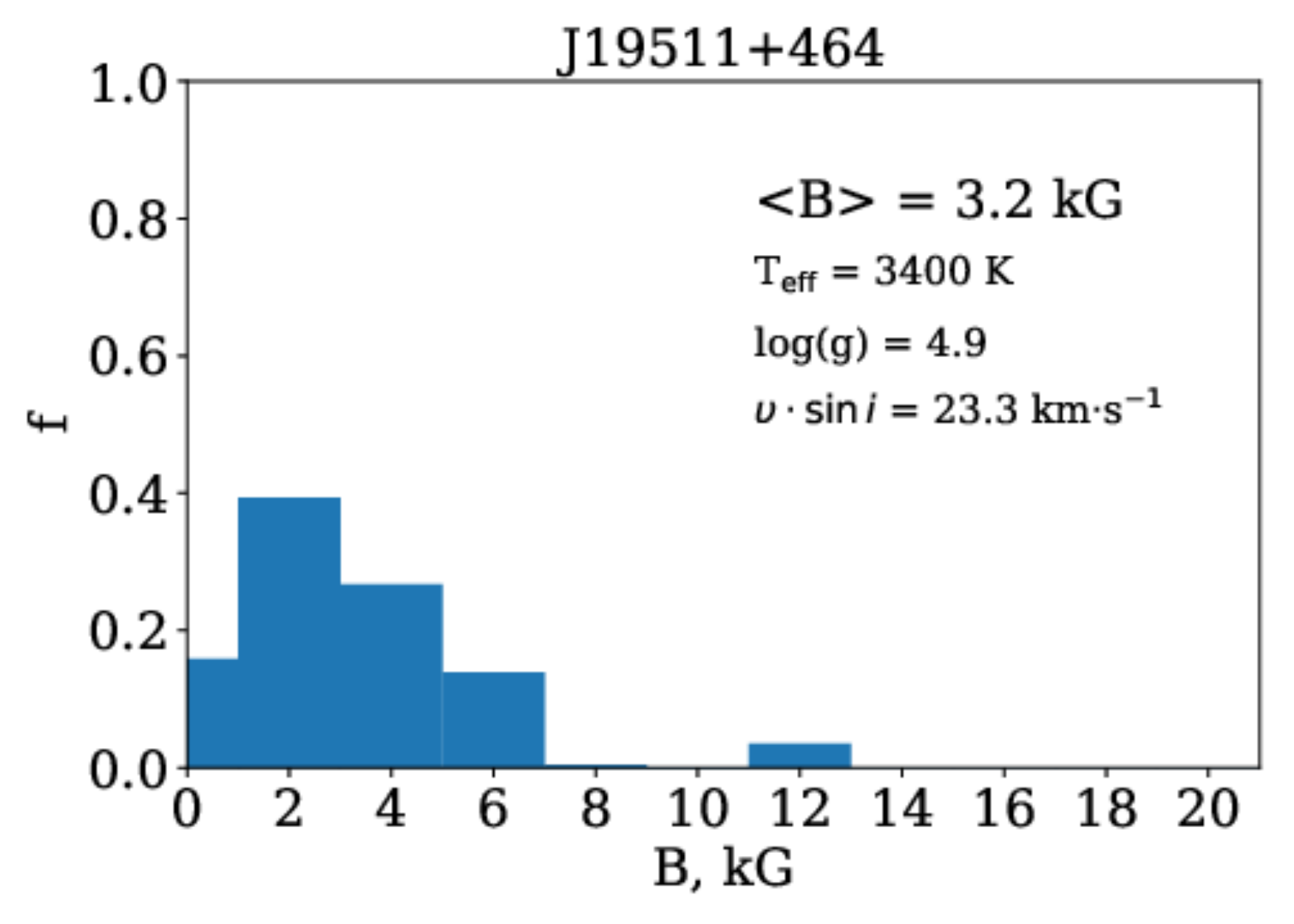}
\includegraphics[width=0.3\hsize]{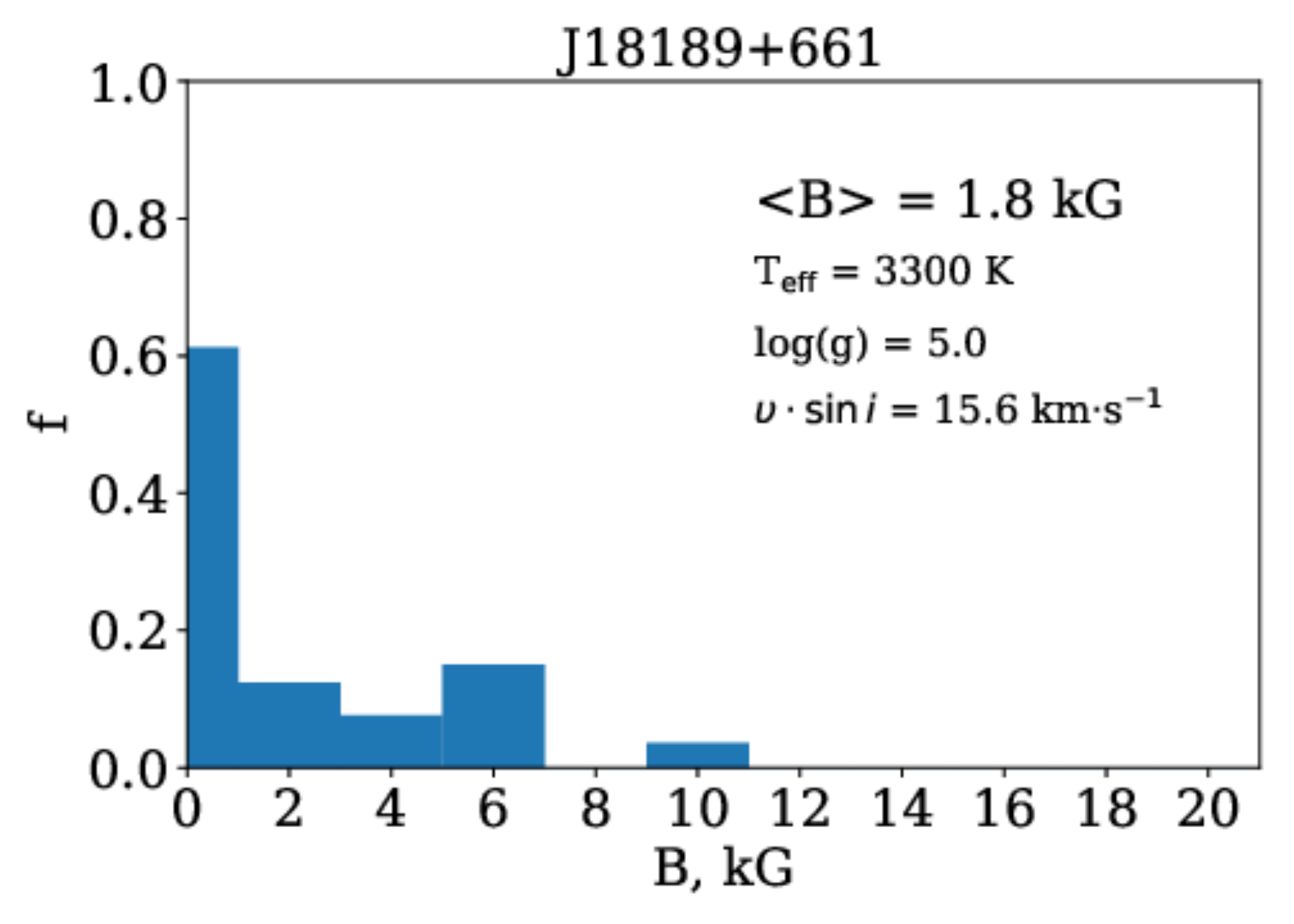}
\includegraphics[width=0.3\hsize]{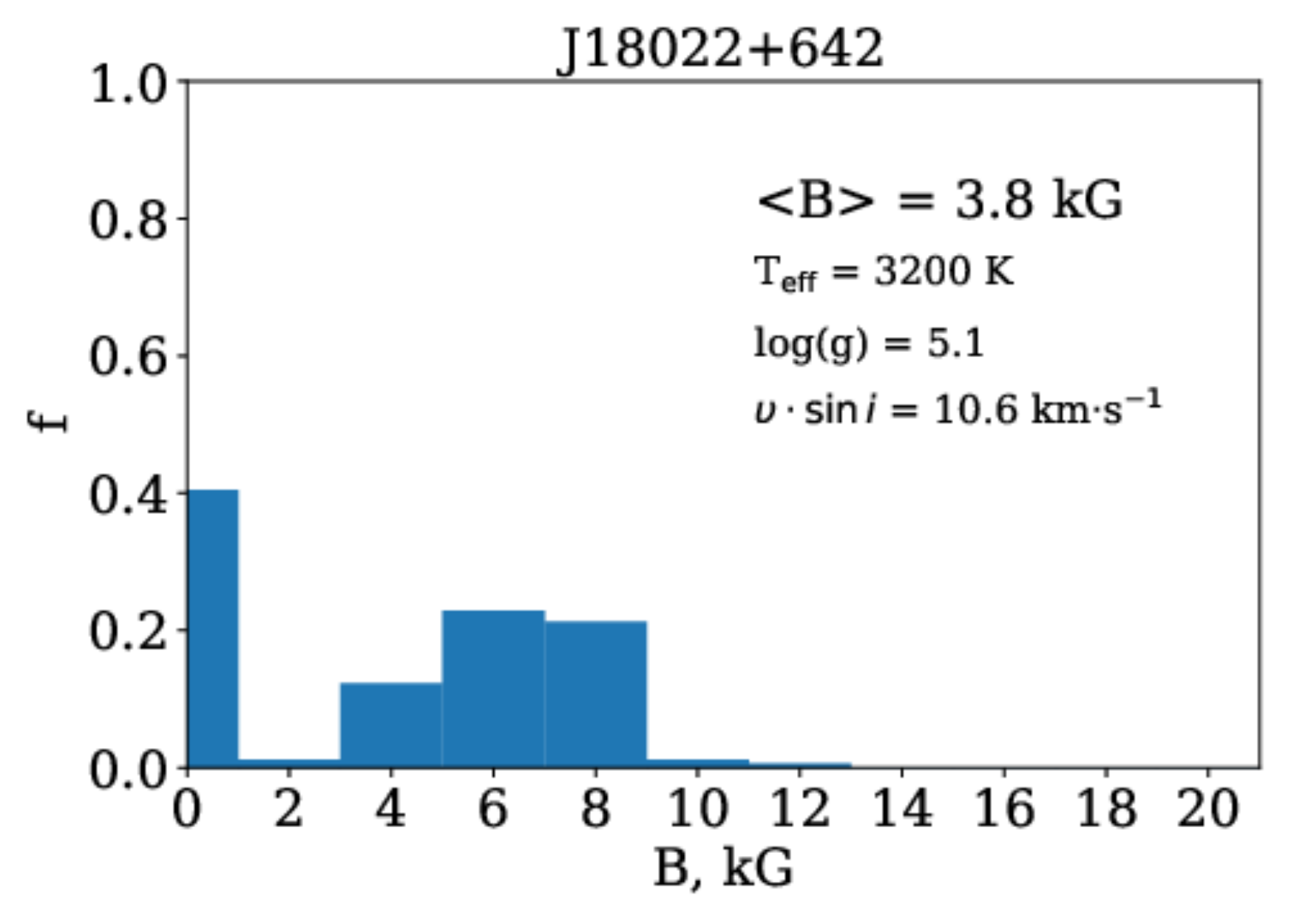}
\includegraphics[width=0.3\hsize]{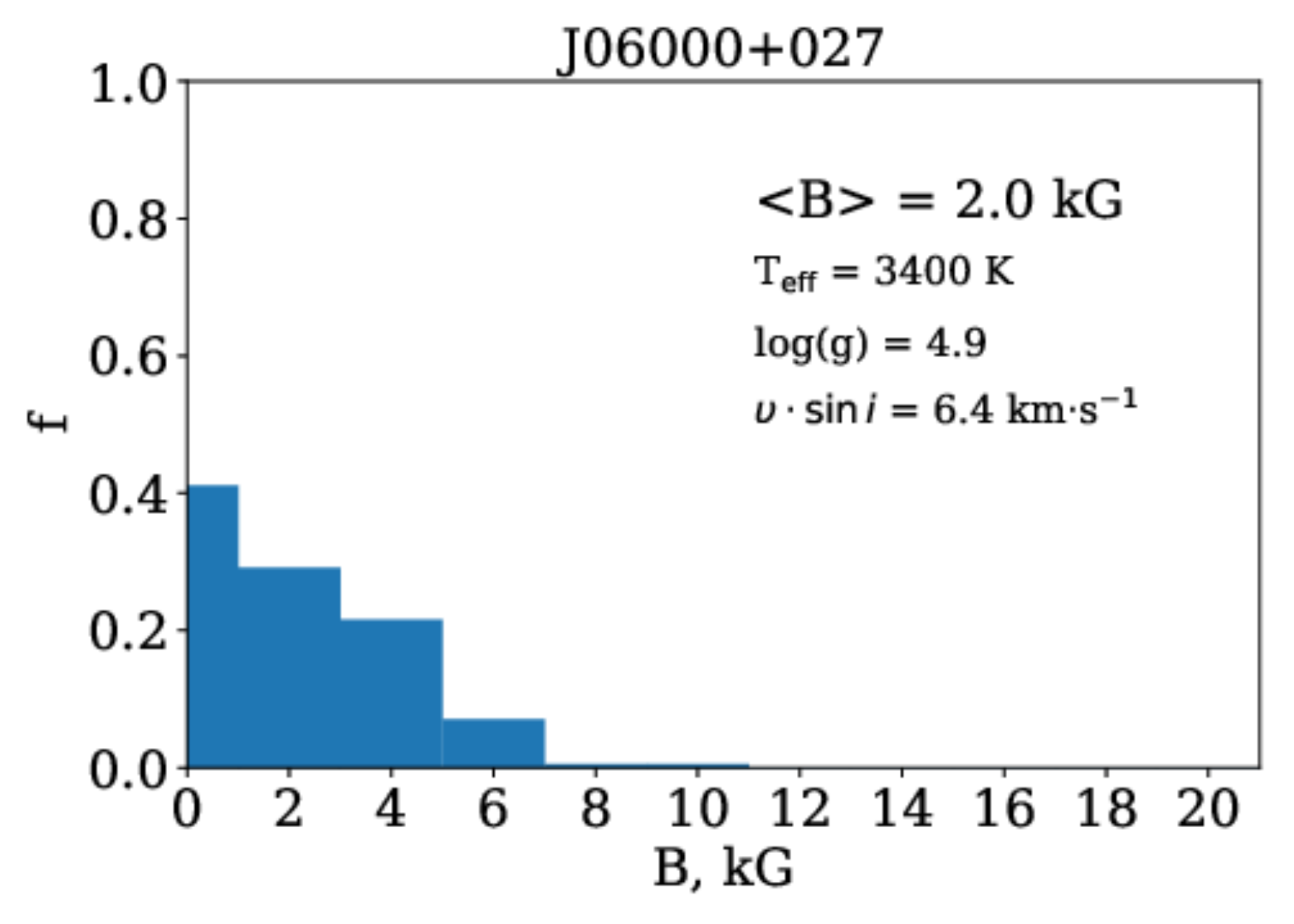}
\includegraphics[width=0.3\hsize]{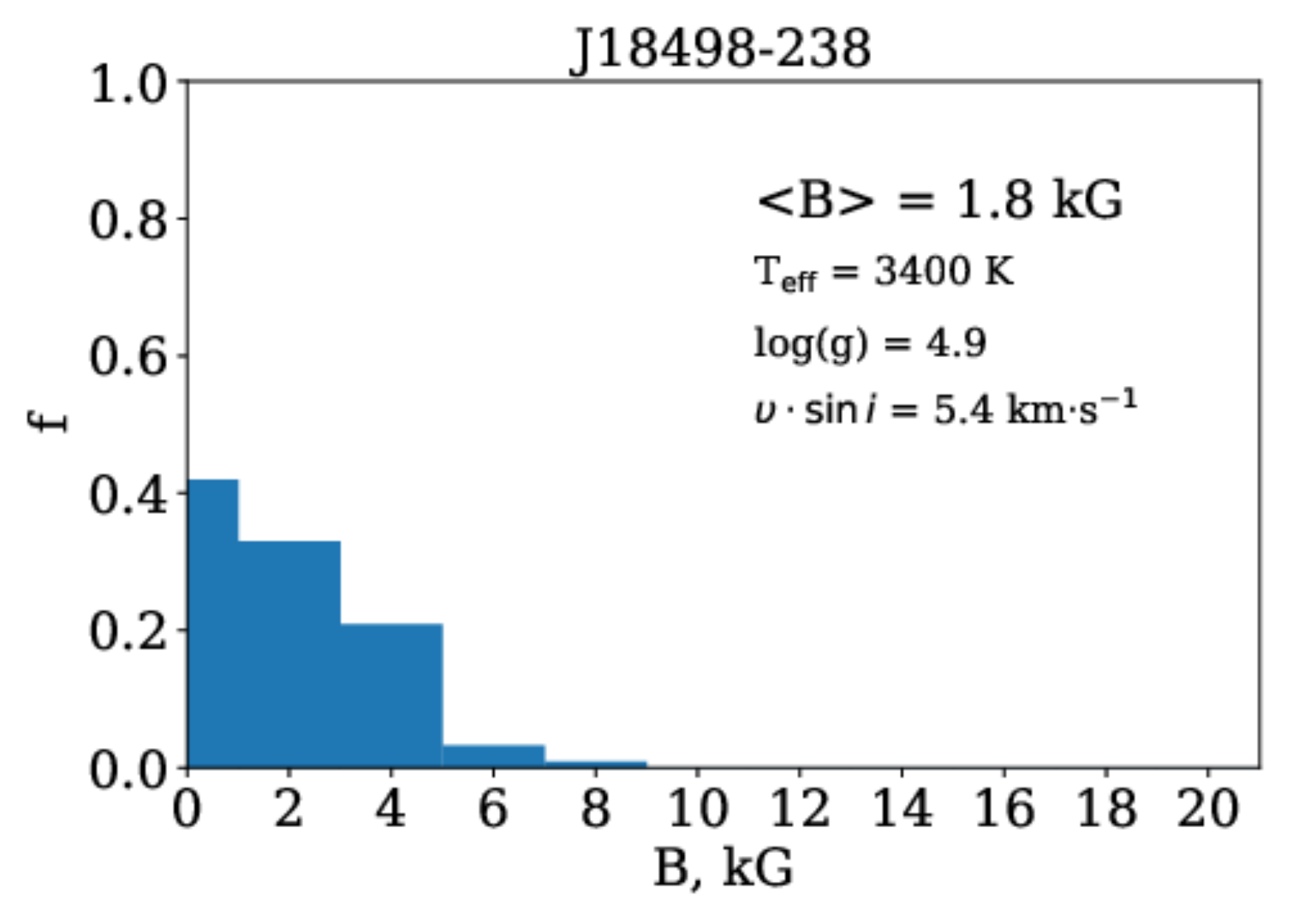}
\includegraphics[width=0.3\hsize]{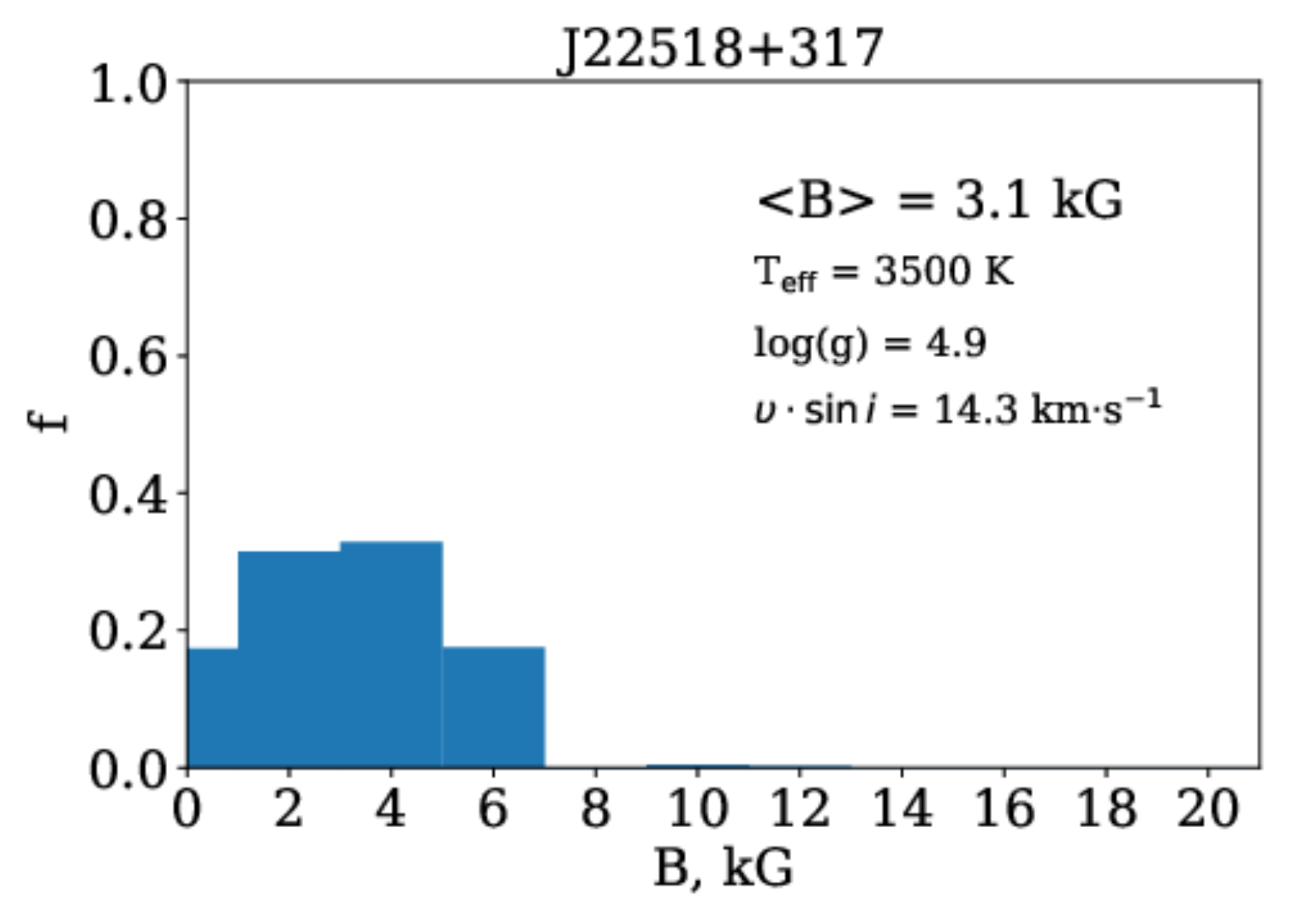}
\includegraphics[width=0.3\hsize]{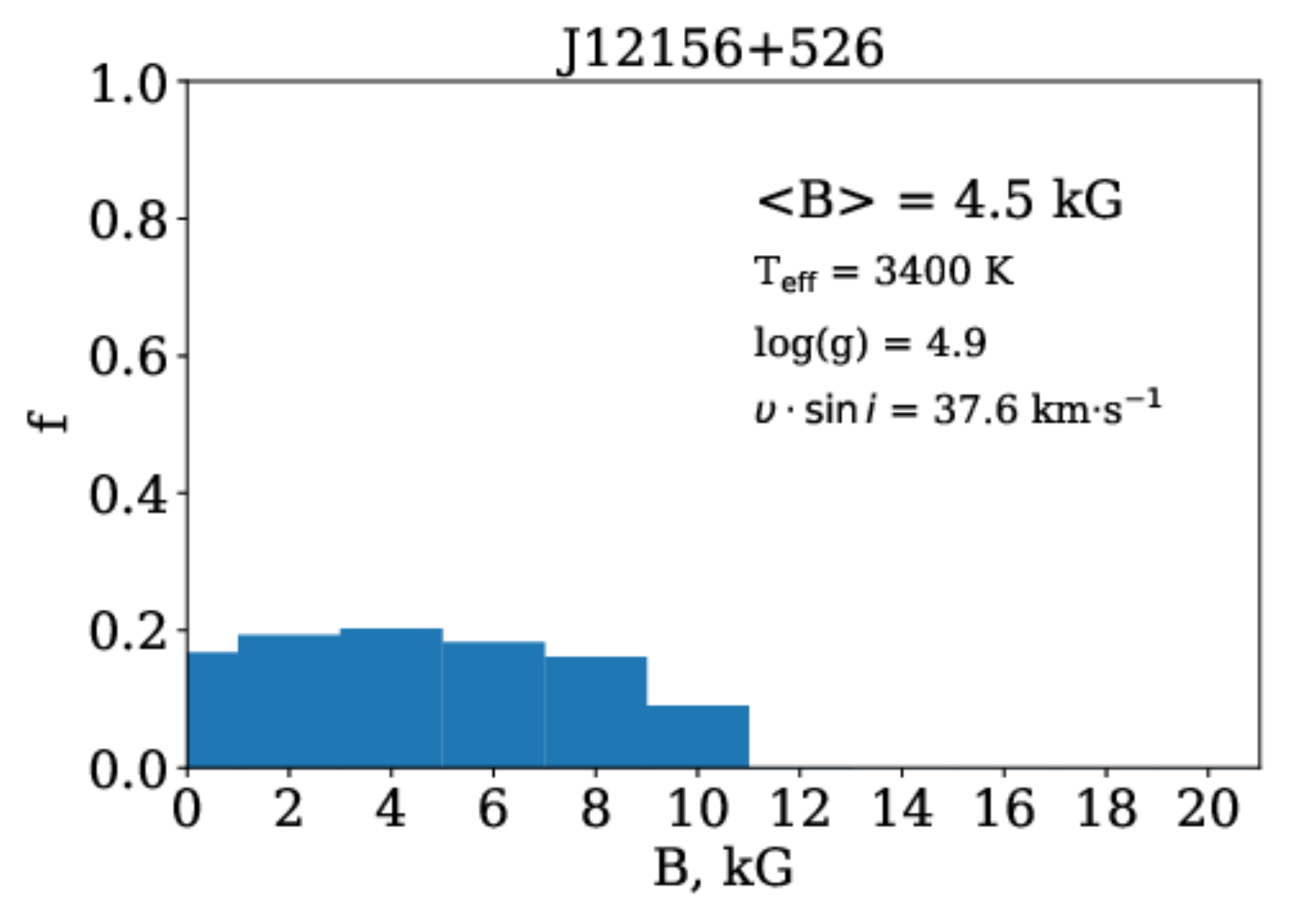}
\includegraphics[width=0.3\hsize]{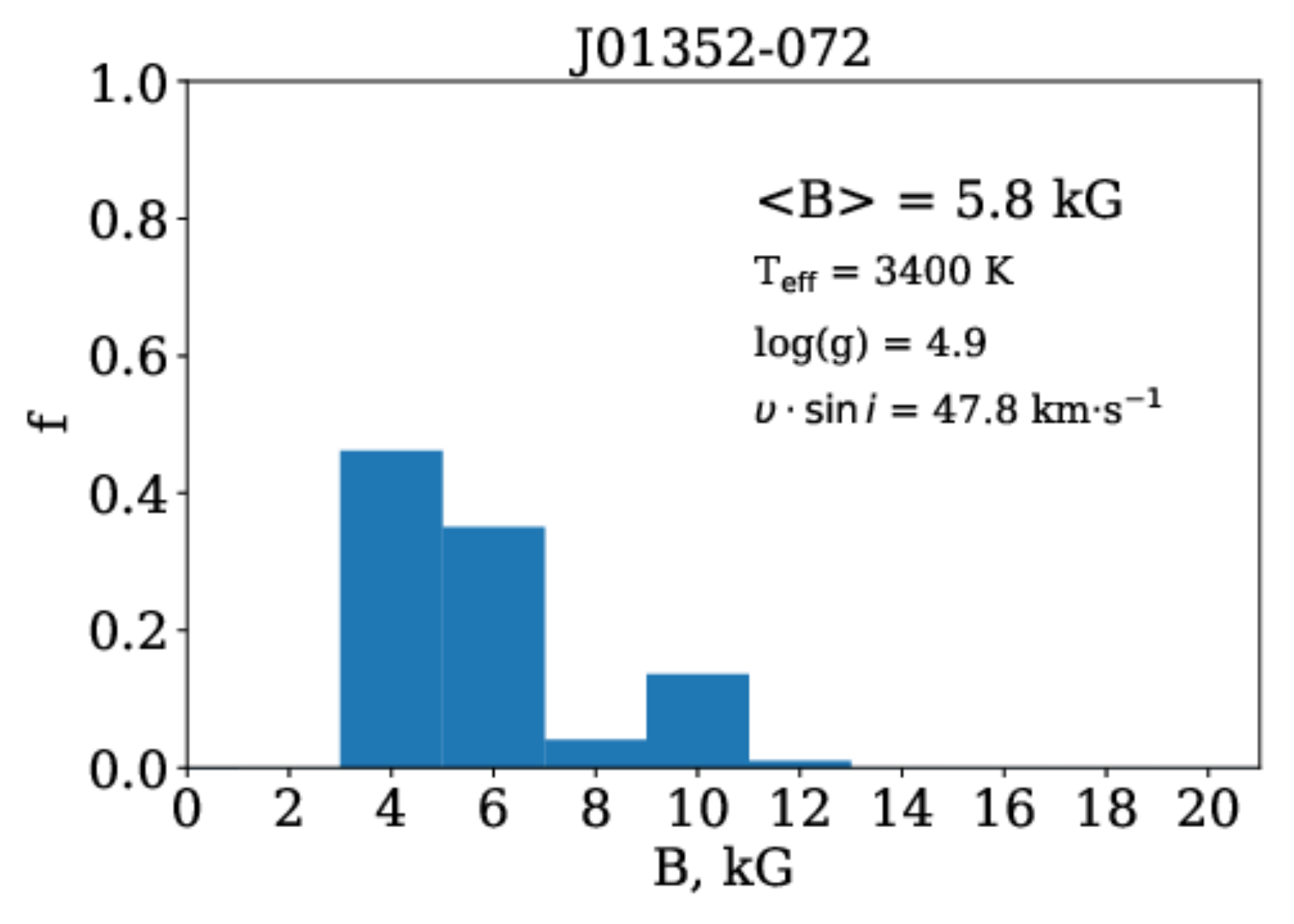}
\includegraphics[width=0.3\hsize]{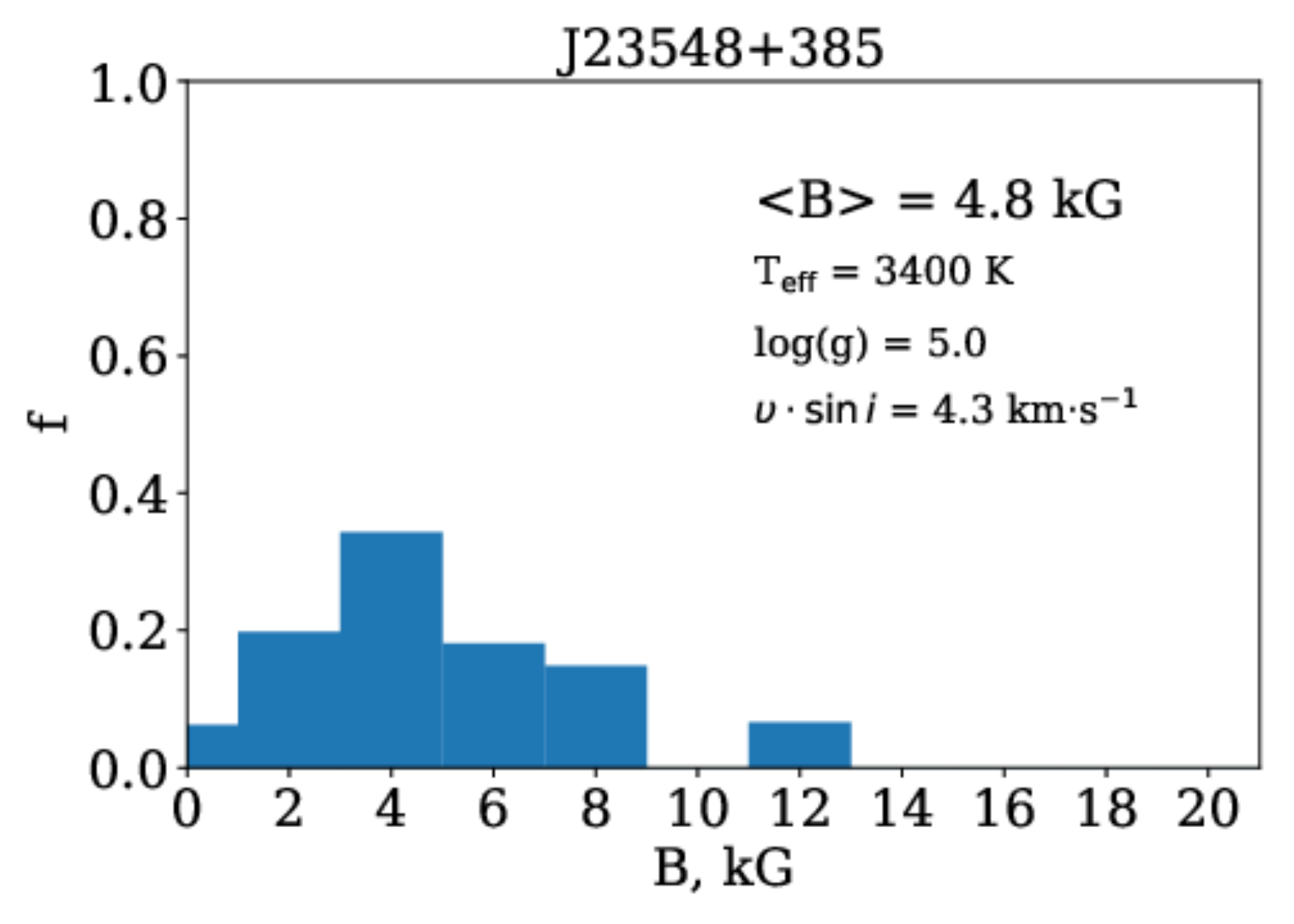}
\includegraphics[width=0.3\hsize]{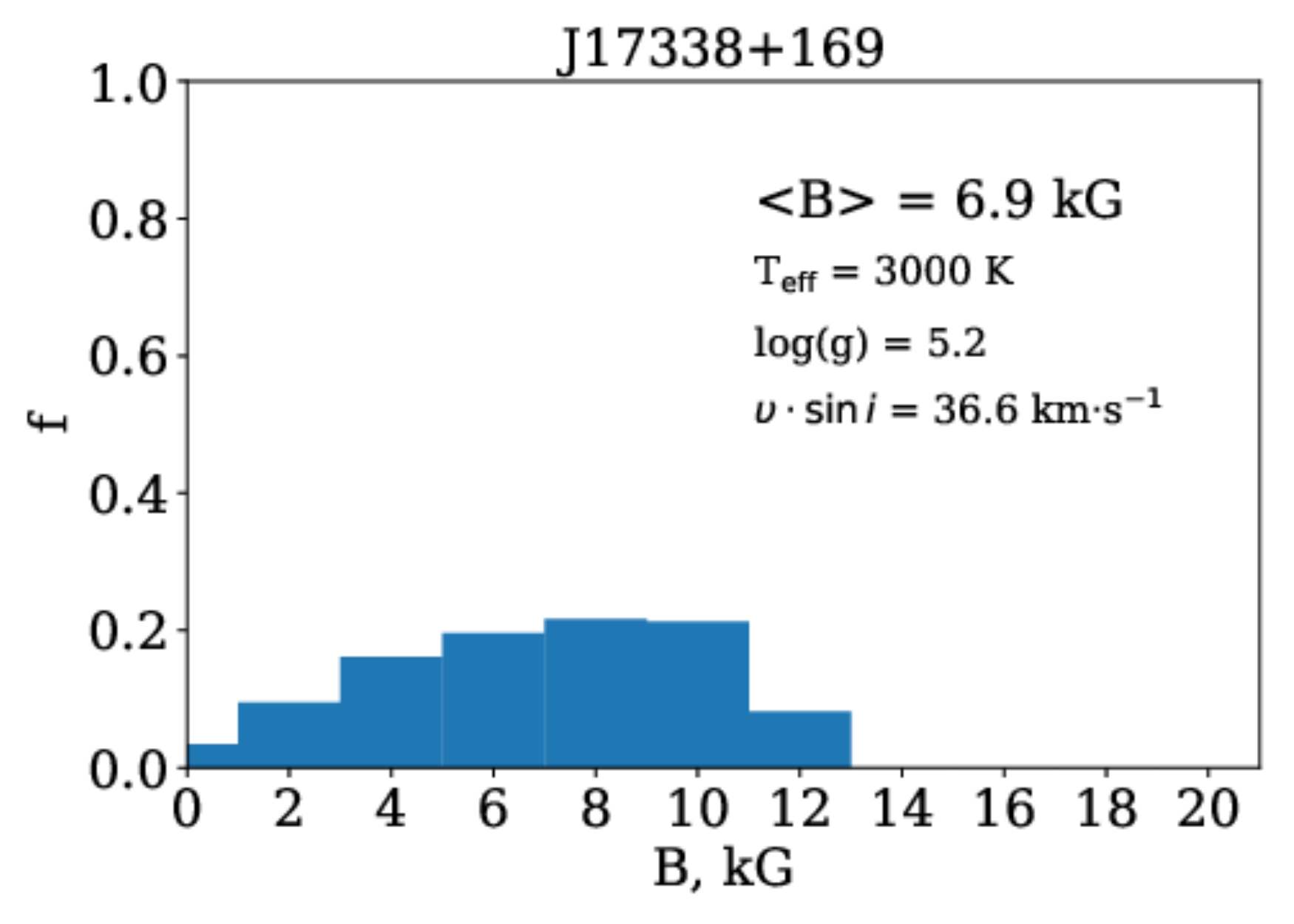}
\includegraphics[width=0.3\hsize]{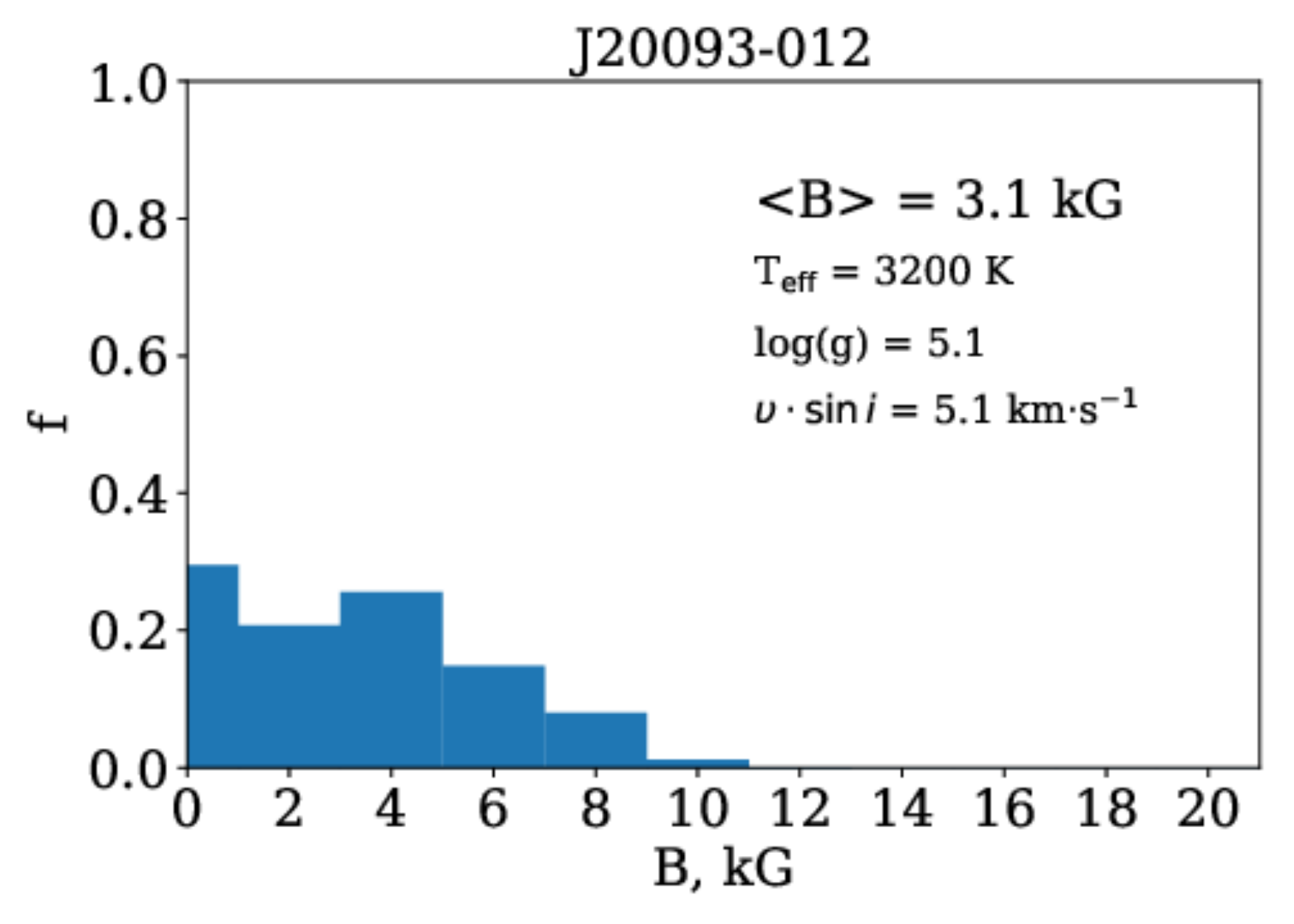}
\includegraphics[width=0.3\hsize]{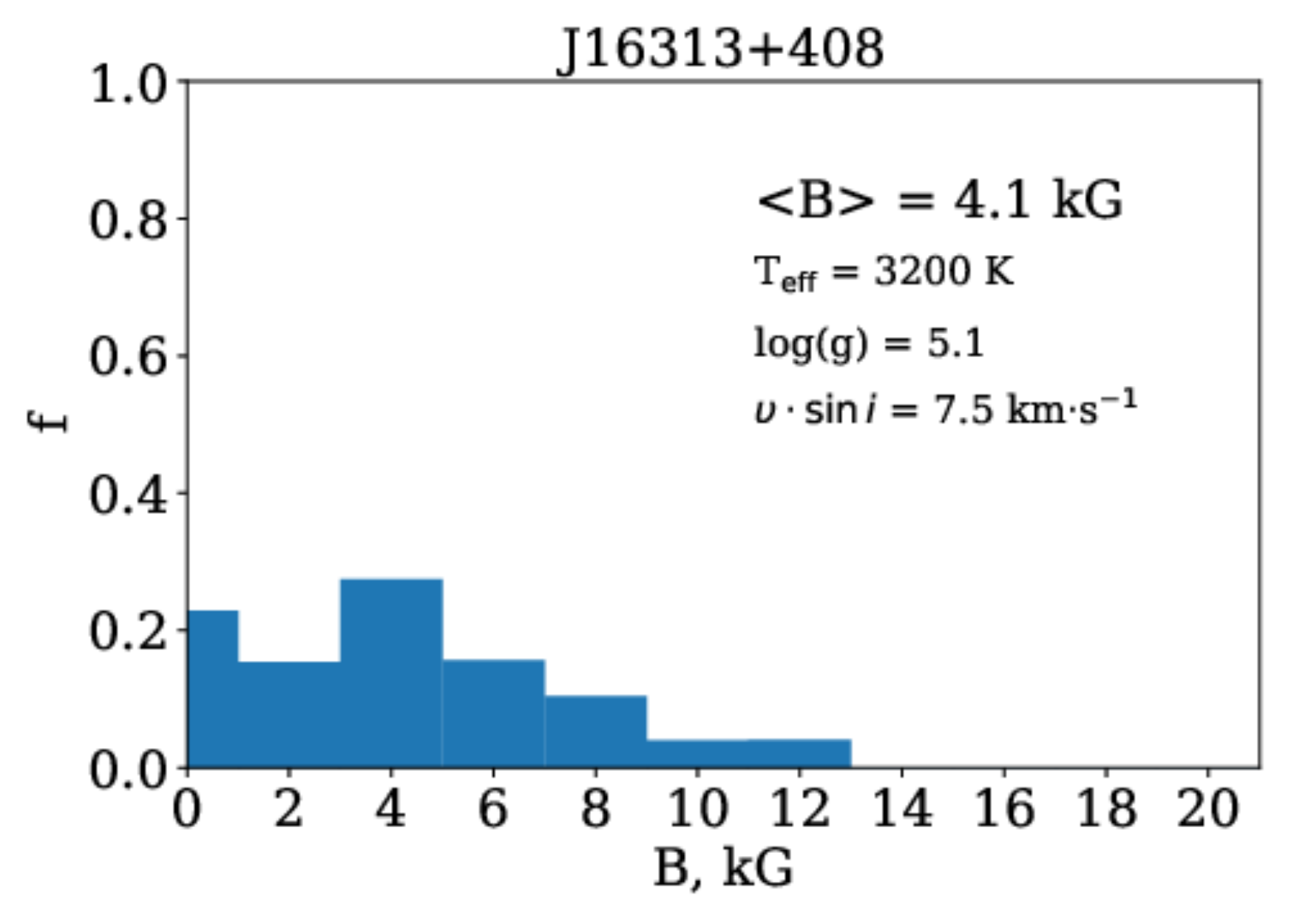}
\caption{\label{fig:f-factors-1}
Distribution of filling factors as derived from Ti lines. 
The stars are sorted according to the SNR of their spectra (SNR decreases from left to the right and top to bottom).}
\end{figure*}

\begin{figure*}
\centering
\includegraphics[width=0.3\hsize]{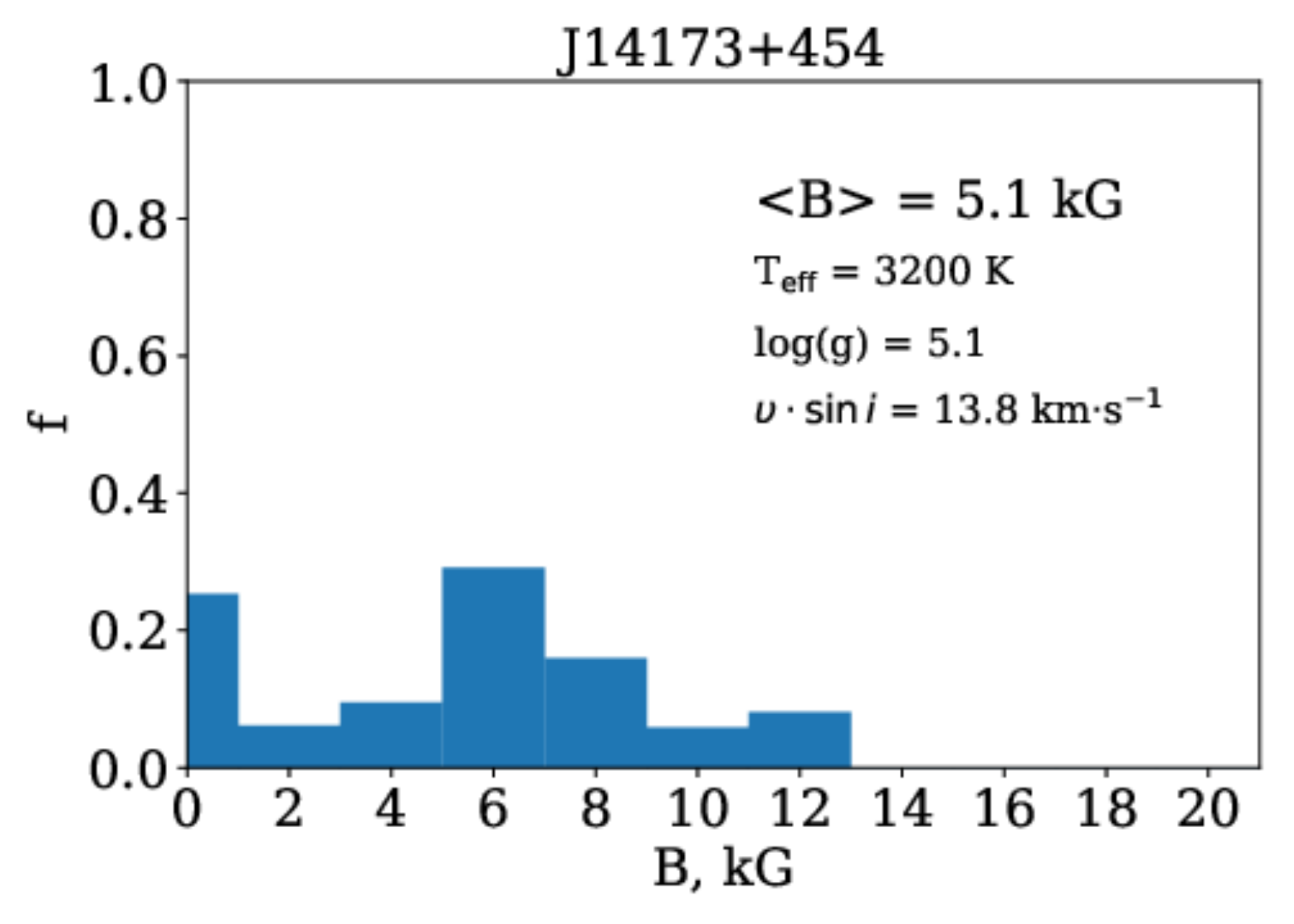}
\includegraphics[width=0.3\hsize]{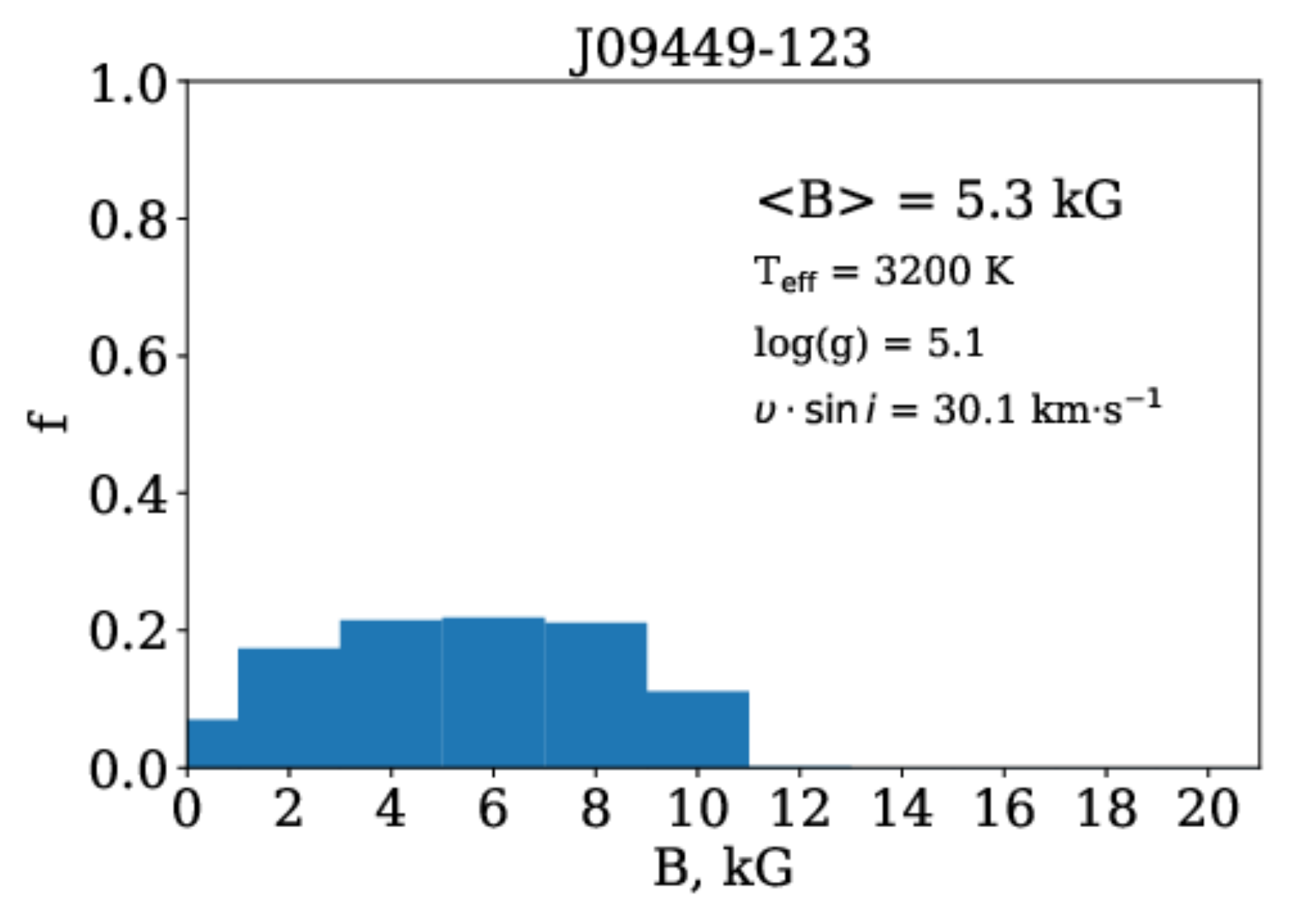}
\includegraphics[width=0.3\hsize]{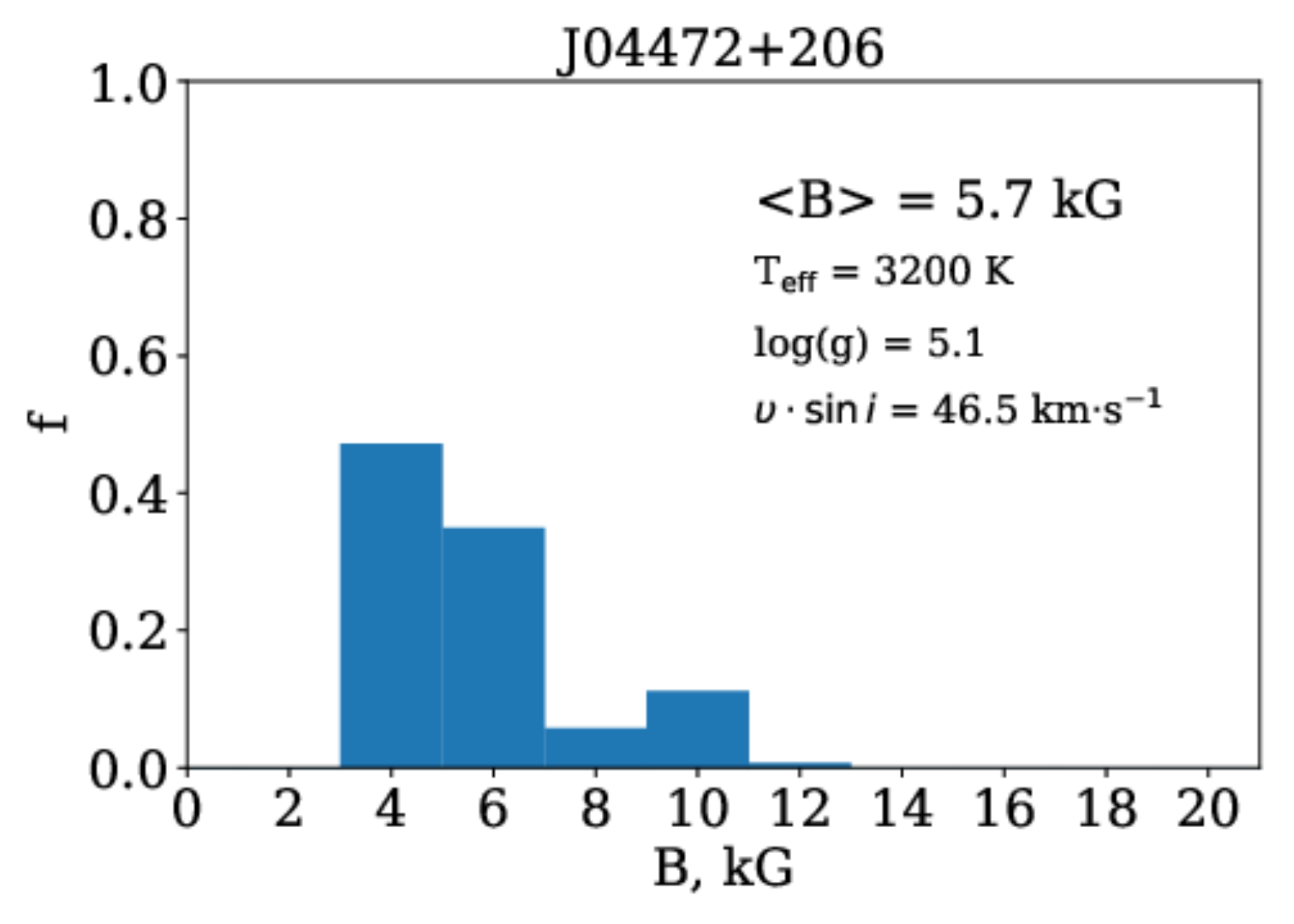}
\includegraphics[width=0.3\hsize]{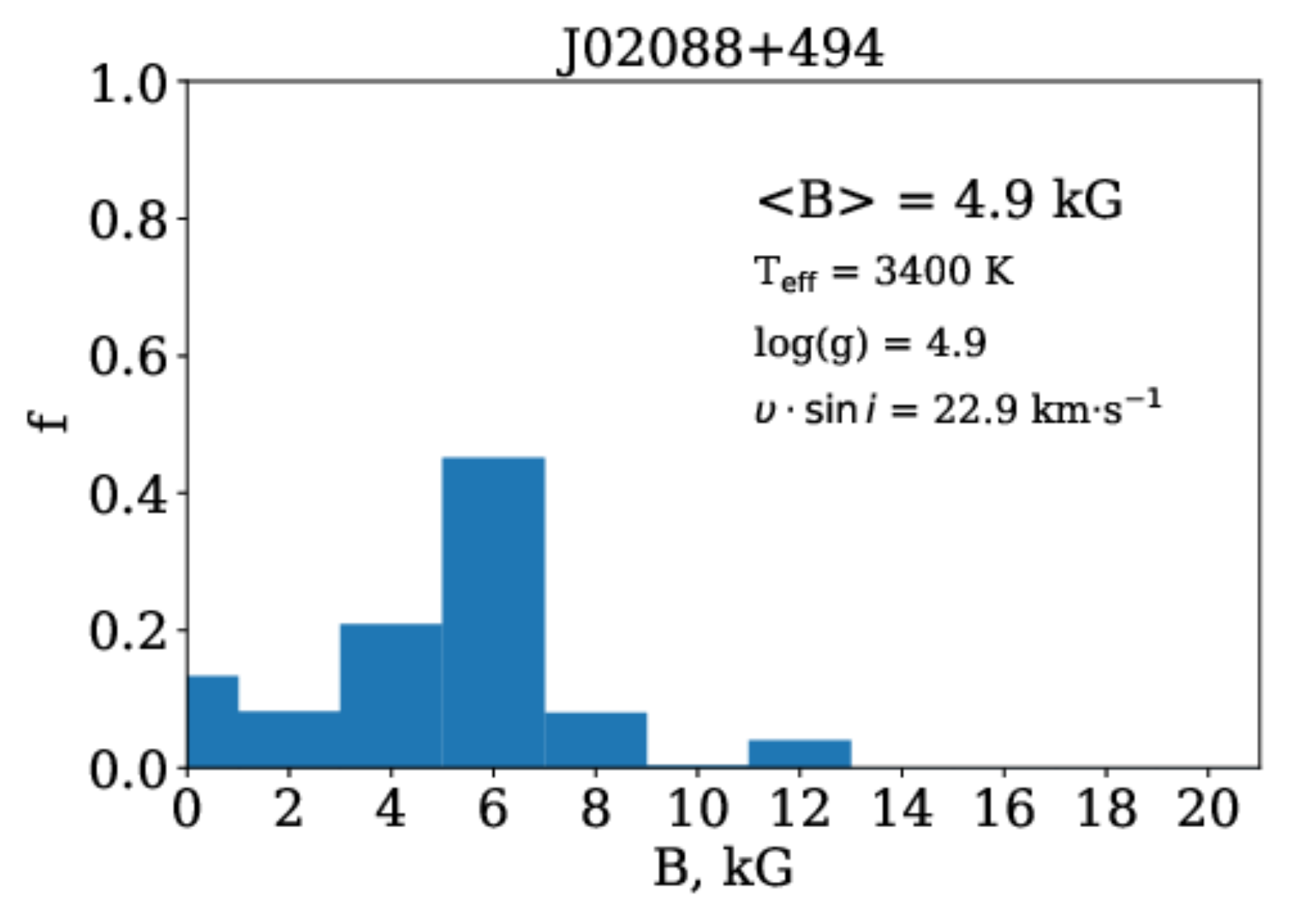}
\includegraphics[width=0.3\hsize]{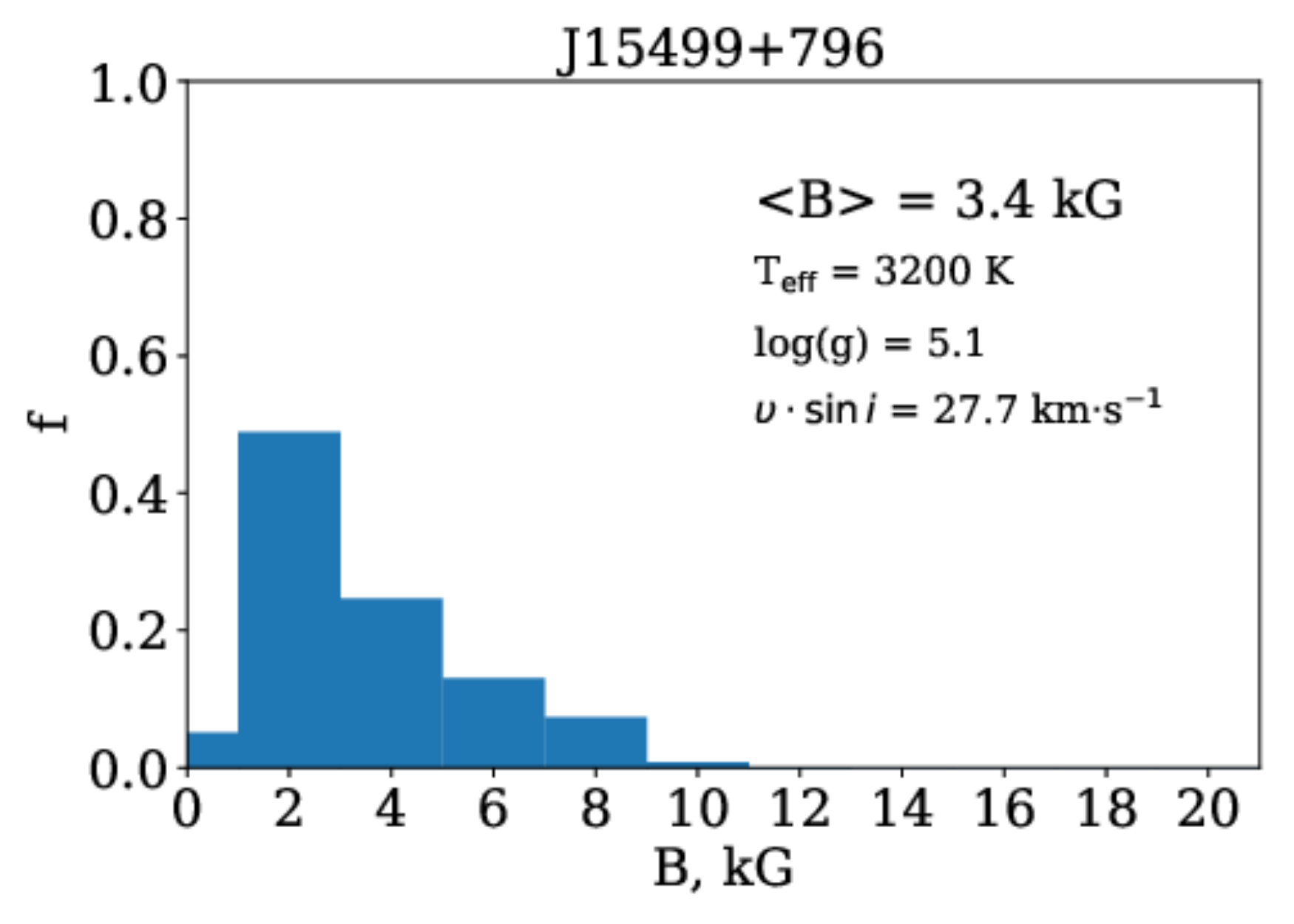}
\caption{\label{fig:f-factors-2}
Same as on Fig.~\ref{fig:f-factors-1}, continued.}
\end{figure*}

\begin{figure*}
\centerline{\hspace{0.05\hsize} \textbf{Dipole-dominant} \hspace{0.15\hsize} \textbf{Multipole-dominant}}
\centerline{
\includegraphics[width=0.3\hsize]{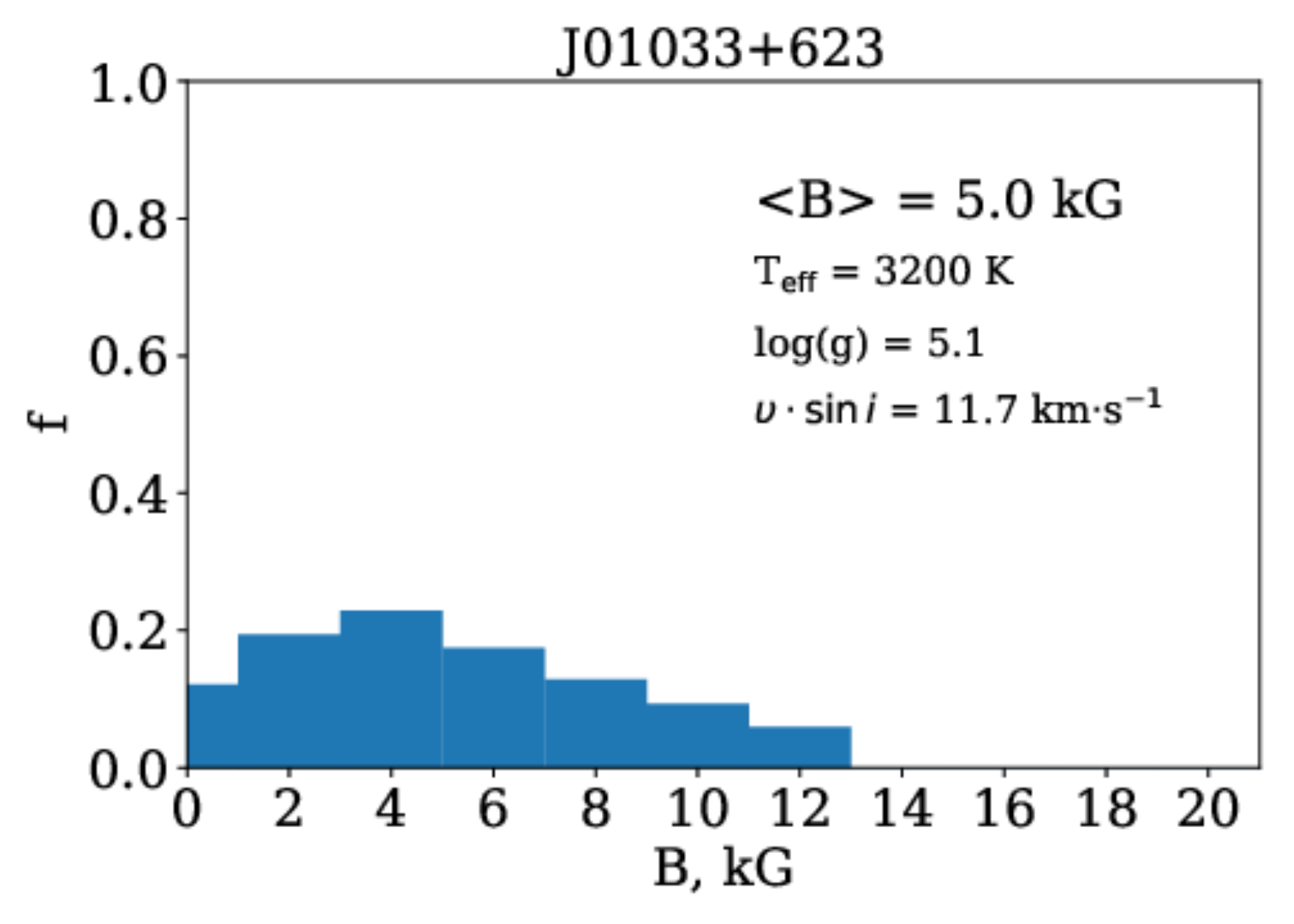}
\includegraphics[width=0.3\hsize]{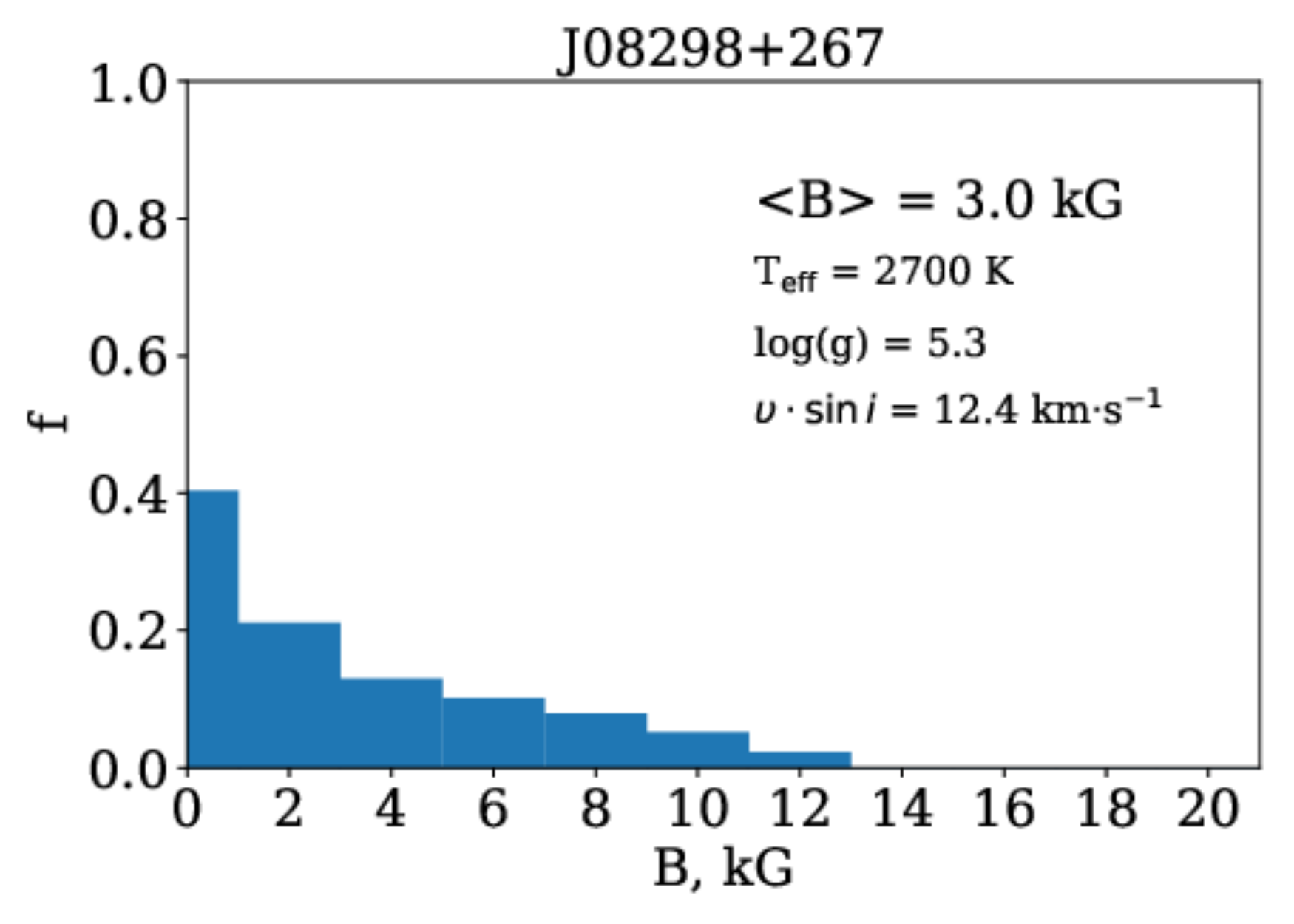}
}
\centerline{
\includegraphics[width=0.3\hsize]{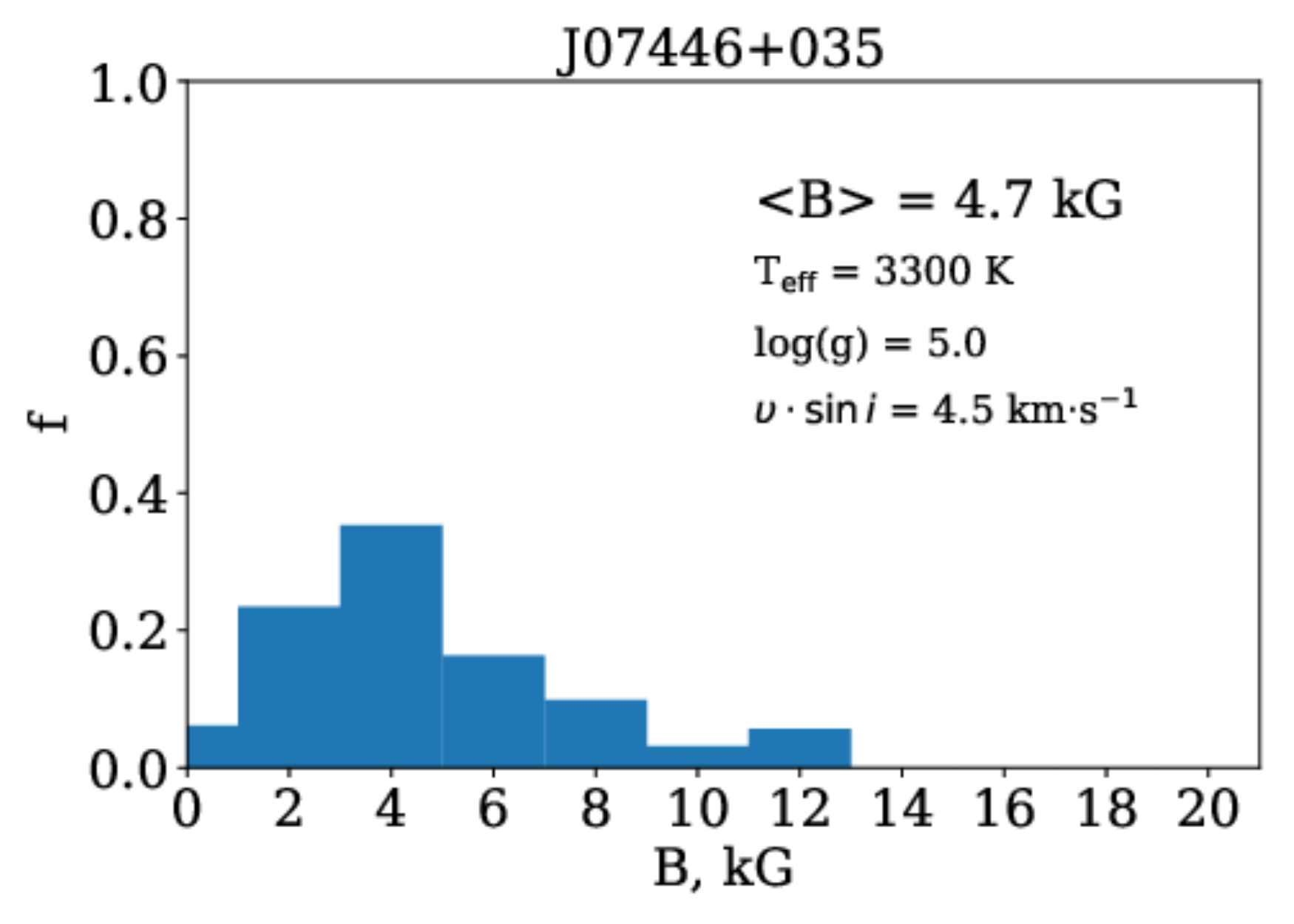}
\includegraphics[width=0.3\hsize]{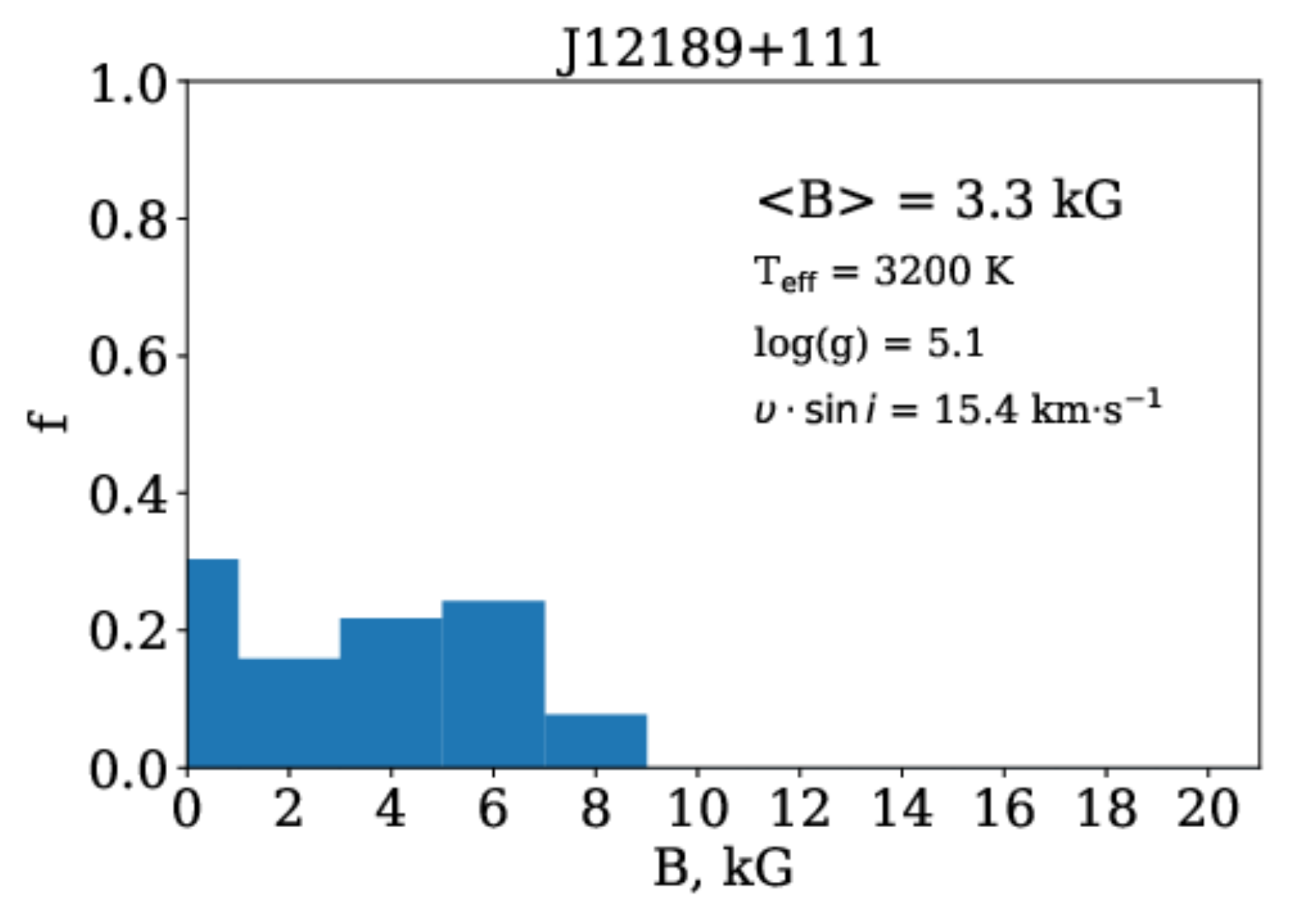}
}
\centerline{
\includegraphics[width=0.3\hsize]{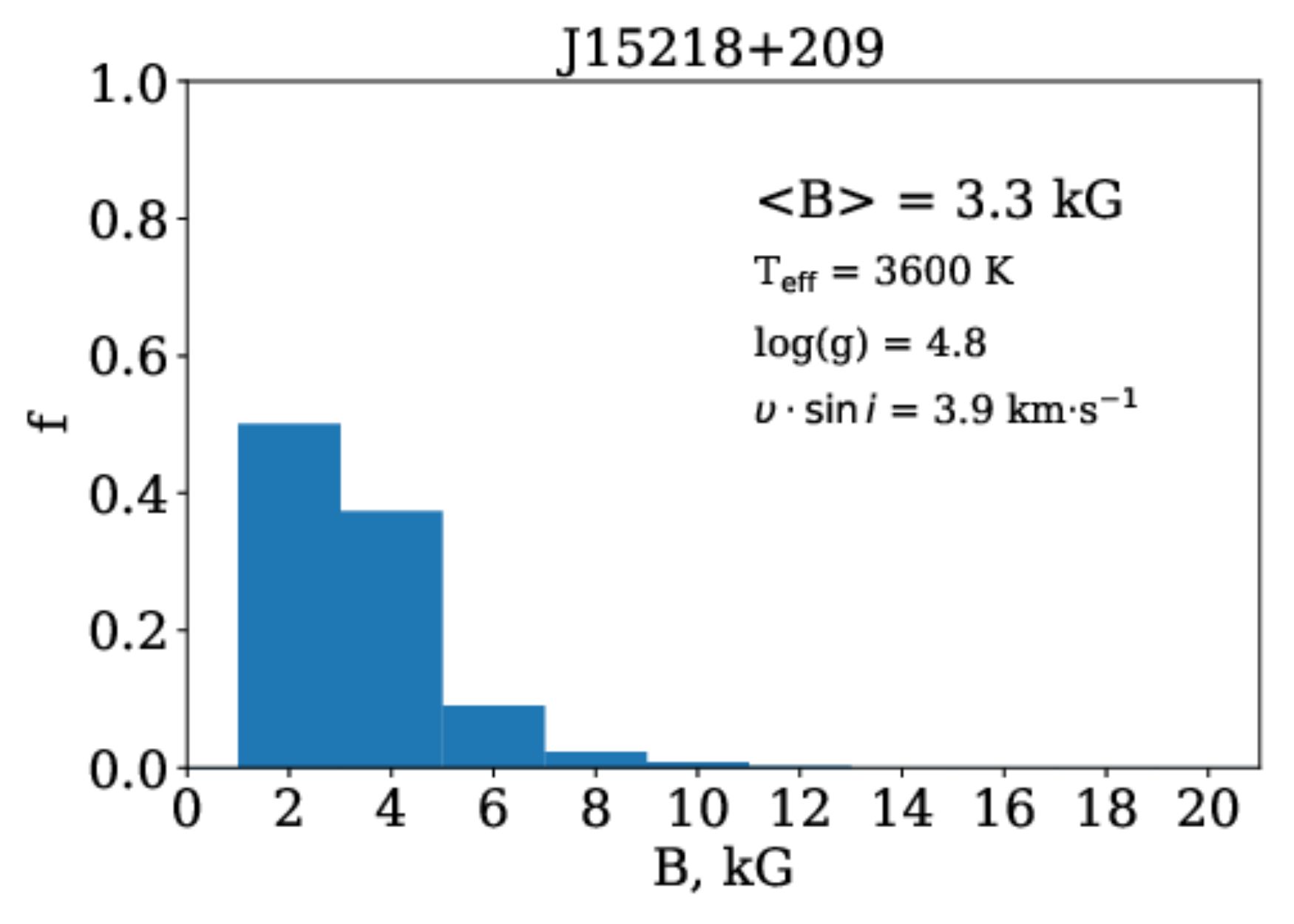}
\includegraphics[width=0.3\hsize]{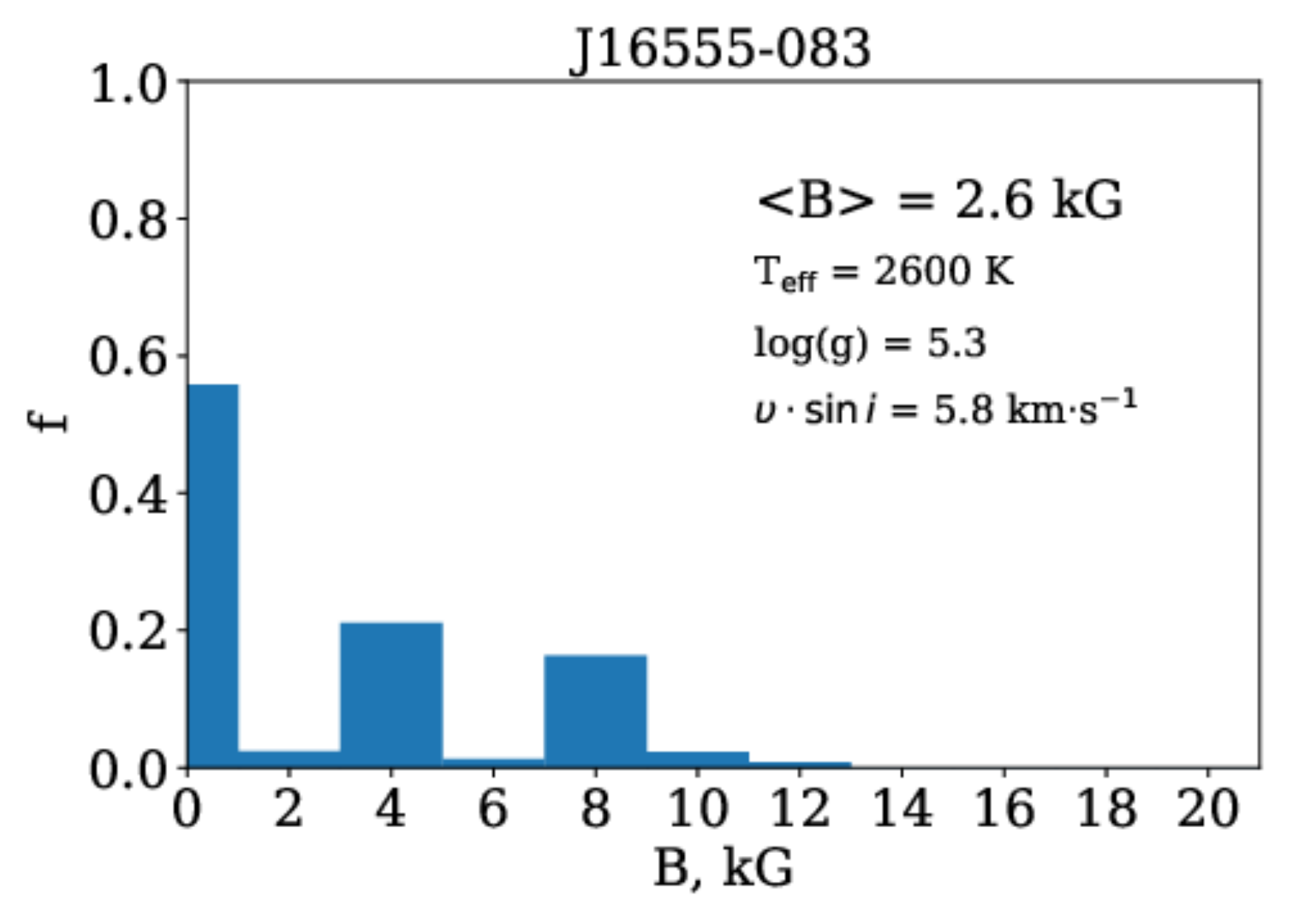}
}
\centerline{
\includegraphics[width=0.3\hsize]{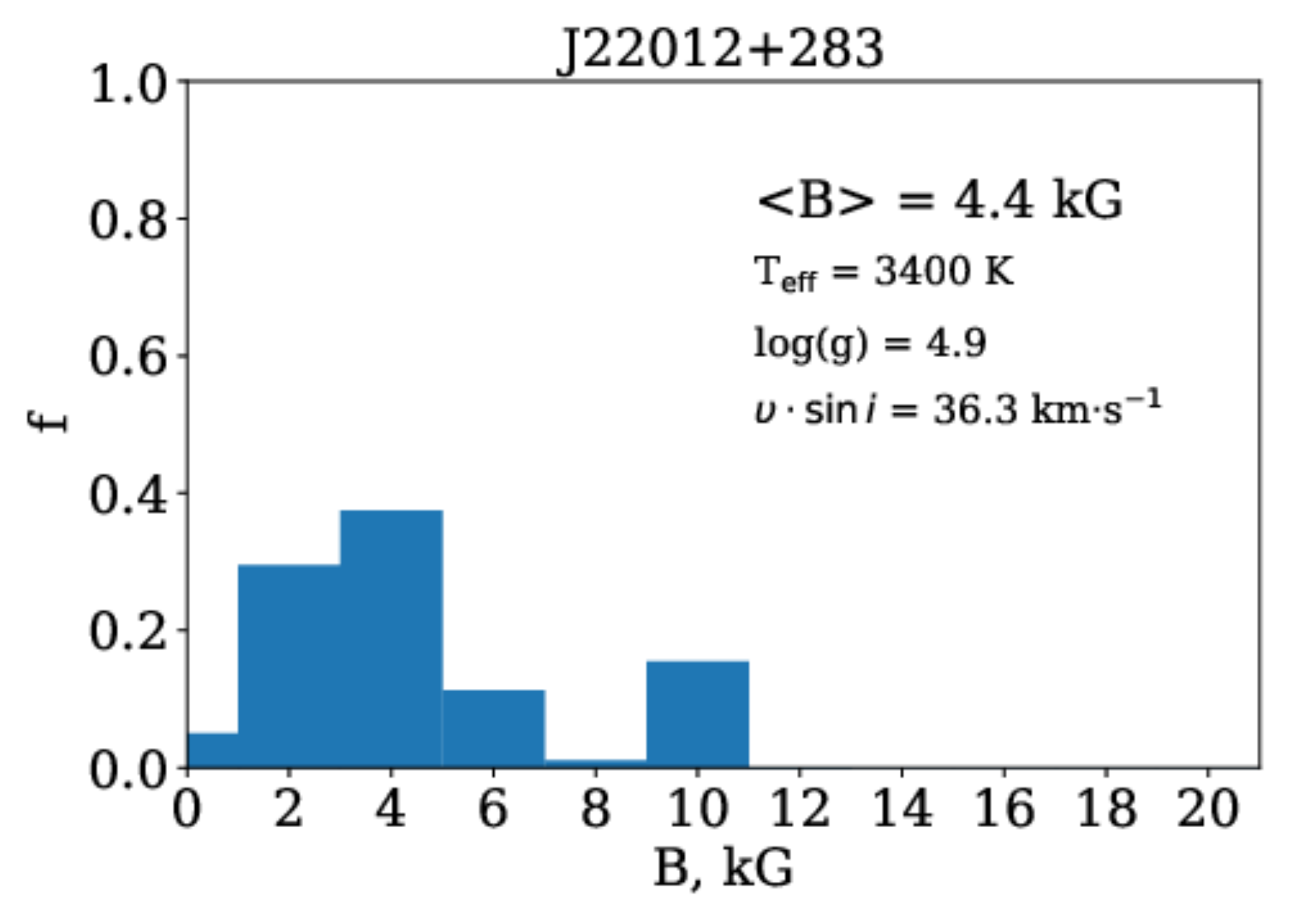}
\includegraphics[width=0.3\hsize]{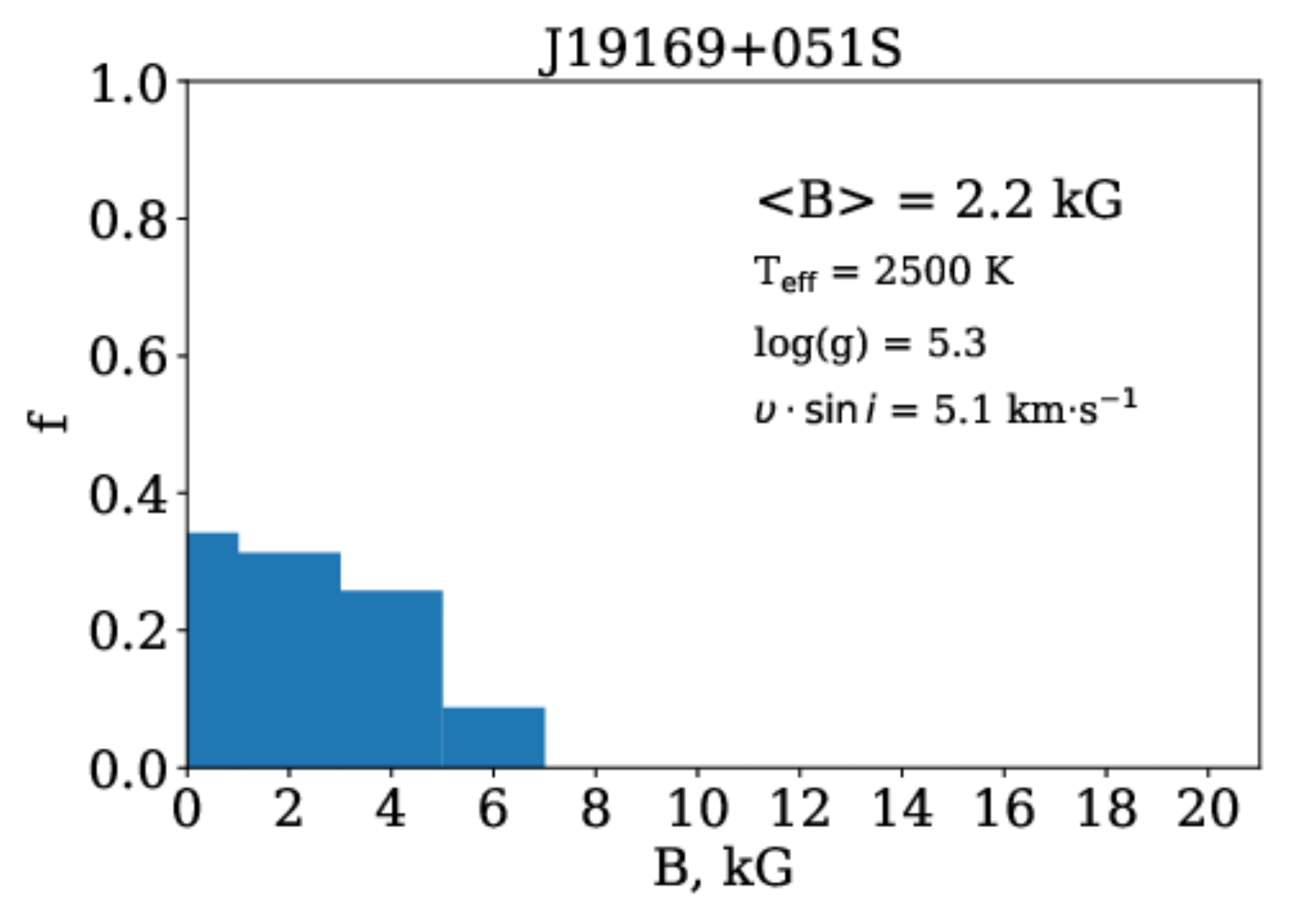}
}
\hspace{0.5\hsize}
\includegraphics[width=0.3\hsize]{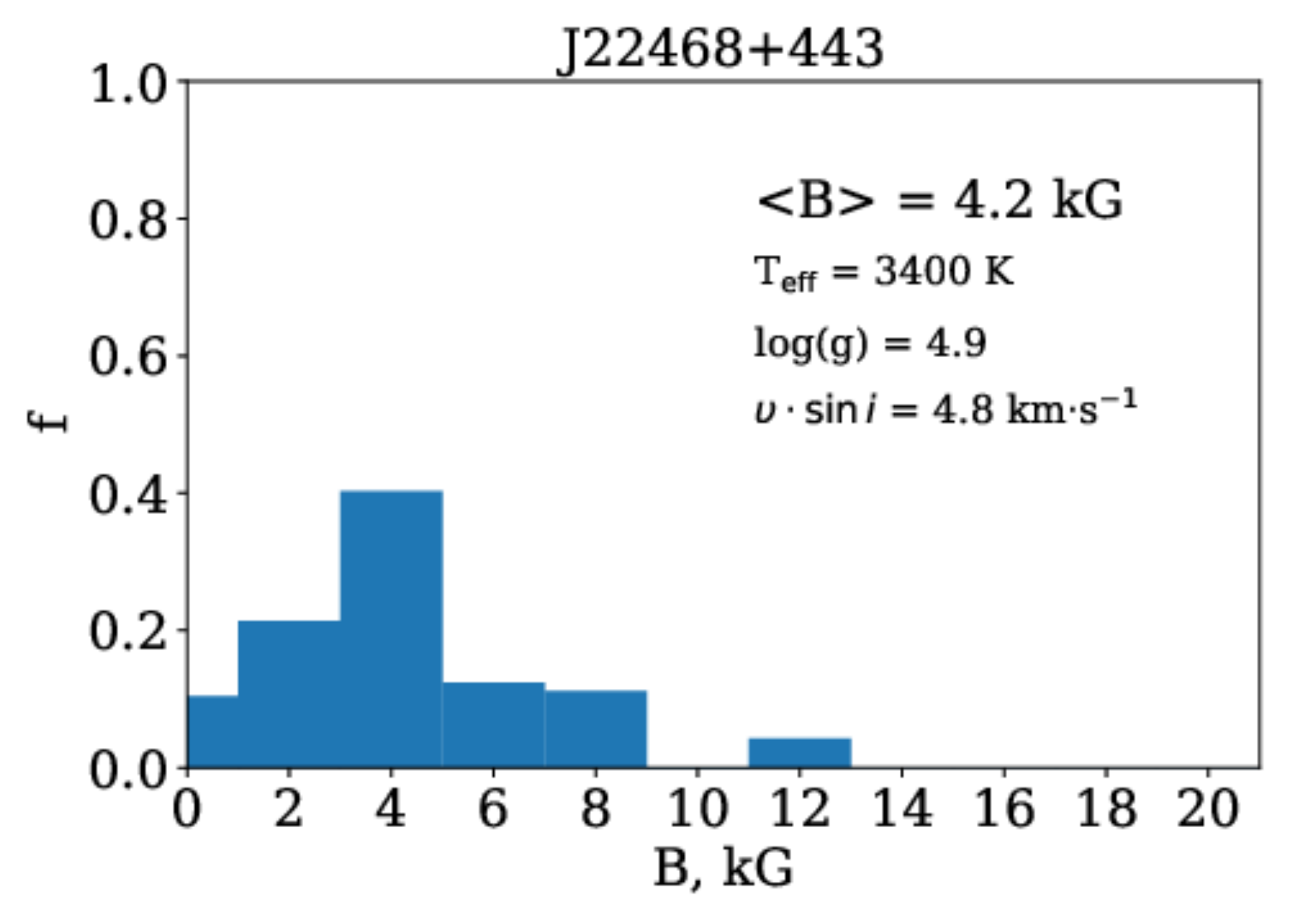}
\caption{\label{fig:f-factors-3}
Distribution of filling factors as derived from Ti lines 
for stars with known geometry of their large-scale magnetic fields.
Left column~--~dipole-dominant, right column~--~multipole-dominant.}
\end{figure*}

We use magnetic filling factors in our attempts to measure magnetic fields
because complex shapes of magnetically sensitive spectral lines in M dwarfs can not be represented
by a single magnetic field component. In order to recover accurate distributions
of filling factors data with very high SNR are required. Moreover, filling factors are sensitive
to the available magnetic information (i.e., set of spectral lines) and atmospheric parameters 
(especially elemental abundance and rotation broadening).
Therefore, the filling factors that we present in this paper 
should be taken with caution, at least for fast rotating stars.
The results are shown on Figs.~\ref{fig:f-factors-1},~\ref{fig:f-factors-2}, and~\ref{fig:f-factors-3} 
as derived from Ti lines.

Our filling factors seem to show two distinct patterns.
We either detect very smooth distributions (e.g., J01033+623, J07446+035, J08298+267 (Fig.~\ref{fig:f-factors-3}), J12156+526 (Fig.~\ref{fig:f-factors-1})),
or more patchy patterns with distinct dominant magnetic field components 
(e.g., J02088+494 (Fig.~\ref{fig:f-factors-2}), J16555-083, J22012+283 (Fig.~\ref{fig:f-factors-3}), J18022+642, J18189+661 (Fig.~\ref{fig:f-factors-1}), etc.).
Next, in our sample there are four M dwarfs that, according to their spectral types (SpT$<$M3.5), should have
partly convective envelopes (J03473-019, J05365+113, J15218+209, J22518+317), while the rest should be fully
convective objects. However, we do not observe an obvious change in the complexity of the fields between these two groups.
For instance, the distribution of filling factors in partly convective J03473-019 and J22518+317 (Fig.~\ref{fig:f-factors-1}) can be compared
to, e.g., those of fully convective J19169+051S (Fig.~\ref{fig:f-factors-3}) and J19511+464 (Fig.~\ref{fig:f-factors-1}). 
The only feature that we could notice is that
fully convective stars tend to have stronger average magnetic fields represented by magnetic components of stronger strength. 
For instance, in the hottest partly convective star J05365+113 (Fig.~\ref{fig:f-factors-1}) we derive a very weak magnetic field with only two 
magnetic components ($0$~kG and $2$~kG, respectively).

In our sample, we find several objects with remarkably very similar properties of their magnetic fields.
For the two stars with largest $\vsini$ values, J01352-072 (Fig.~\ref{fig:f-factors-1}) and J04472+206 (Fig.~\ref{fig:f-factors-2}), 
we recover the same average magnetic fields
with identical filling factors. The only difference between these stars is that J04472+206 
is one spectral type cooler.
Next, we find a twin of J07446+035 (very well studied M dwarf YZ~CMi, Fig.~\ref{fig:f-factors-3}), 
which is J23548+385 (not studied at all, Fig.~\ref{fig:f-factors-1}). 
Both  stars have very similar filling factors, average magnetic field, projected rotational velocities, 
but slightly different spectral types. In addition, another famous M dwarf J22468+443 (EV~Lac) has
filling factors similar to J07446+035 and J23548+385, although its average magnetic
field is noticeably weaker. 

We can now look for a connection between our filling factors and the geometry of large scale magnetic fields.
In our sample, we have nine stars for which the geometry of their surface magnetic fields was previously
derived from polarimetric measurements. 
In particular, \cite{2010MNRAS.407.2269M} showed that partly convective M dwarfs tend to have complex
multipole fields with strong toroidal contribution, while stars that are fully convective generate
dipole-dominant predominantly poloidal magnetic fields. Following this classification,
in Fig.~\ref{fig:f-factors-3} we group M dwarfs in columns depending 
on whether the star has dipole-dominant magnetic field (left column) or more complex
multipole-dominant field (right column). One can see that there is no obvious difference in filling factors
between these two groups. A smooth distribution is found for dipole-dominant J01033+623 on one side, 
and  multipole-dominant J08298+267 and J19169+051S on the other side. Likewise, J07446+035 and J22468+443 have
almost identical filling factors but their magnetic field geometries were found to be very different.
Nevertheless, there seems to be one common feature that is intrinsic to complex multipole fields as observed from filling factors,
which is the appearance of strong zero-field components. This zero-field component is strongest in four out of five
stars with multipole-donimant fields with an exception of J22468+443. 

\subsection{Magnetic field and rotation}

\begin{figure*}
\includegraphics[width=\hsize]{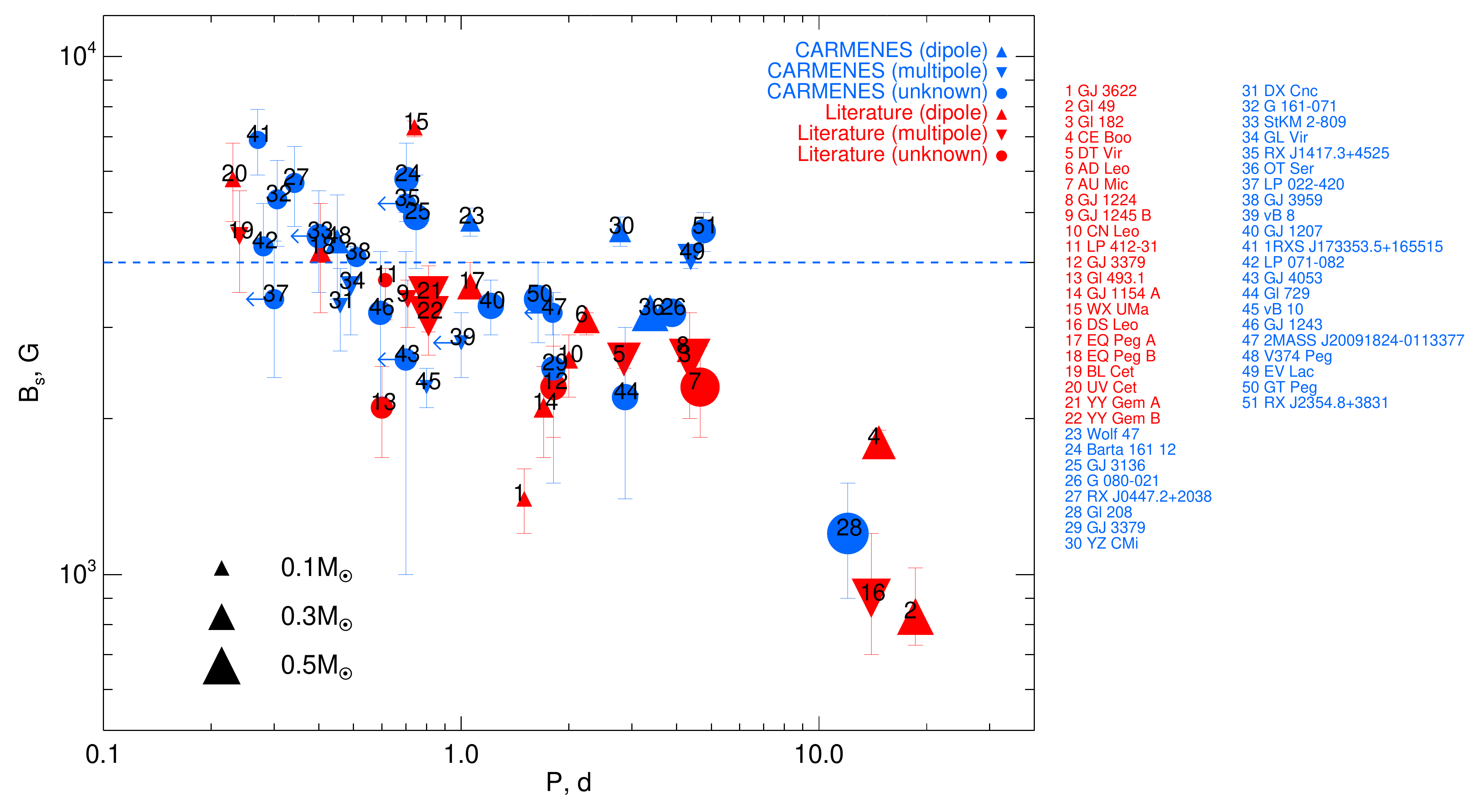}
\caption{\label{fig:bf-period}
Average magnetic fields in stars from our sample as a function of rotation period.
Measurements in stars with known dipole and multipole states 
are shown as filled upward and downward triangles, respectively. Stars with unknown dynamo states
are shown as filled blue circles. Our measurements from this work are shown with blue color
and the literature values are shown with the red one.
The symbol size scales with stellar mass (see legend on the plot).
The horizontal dashed line marks the $4$~kG threshold of saturated magnetic field assigned to stars
with multipole dynamo regime. Similar to our previous works we put a conservative 
$1$~kG error bars on measured magnetic fields in stars with $\vsini>20$~\kms.
}
\end{figure*}

The detection of very strong magnetic fields in M dwarfs needs to be understood
in terms of underlying dynamo processes. Because dynamos in these stars are powered
by convective motions subject to stellar rotation, it is essential to compare our magnetic field
measurements with rotation periods which we plot in Fig.~\ref{fig:bf-period}.
We additionally plot measurements from Sh2017 and other literature sources for stars that
are not in our sample. We also include recent measurements of the magnetic field in the eclipsing binary system
YY~Gem \citep{2019ApJ...873...69K}.
The rotation periods are from \citet[][and references therein]{2018yCat..36210126D,2010MNRAS.407.2269M}.

The general pattern of increasing magnetic field strength as period decreases down to about $P\approx4$~d
is evident from Fig.~\ref{fig:bf-period}. When M dwarfs rotate faster than that, we enter 
the regime of activity saturation in terms of X-ray fluxes \citep{1984ApJ...279..763N,2014ApJ...794..144R}. 
The magnetic field strength
in this regime show large scatter with values between $2$~kG and $7$~kG.
The magnetic field in GJ~3622 was measured in Sh2017 and its magnitude seems to be
too weak as for the rotation period of $P=1.5$~d. 
It is thus important to obtain new magnetic field measurements for this star in future studies. 

From Fig.~\ref{fig:bf-period} it is difficult to see a well defined trend of the magnetic flux density increasing with
rotation for periods below four days. We definitely lack accurate magnetic field measurements in stars with ultra-short periods
and our current results are consistent with the generally accepted conclusion that the magnetic flux density
saturates in stars with saturated activity. More work needs to be done to fully address this effect.

\section{Discussion}

\subsection{Magnetic filling factors}

From our analysis of filling factors we find two distinct pattern in their distribution.
The first one is a very smooth distribution of filling factors and the second pattern looks 
more patchy and can not be approximated by a smooth function. It is difficult to find
strict connections between patterns in our filling factors and other essential stellar parameters
such as rotation or temperature. However, it is possible to draw some preliminary conclusions.

First, we do not find a clear difference in patterns of filling factors between fully and partially convective stars.
However we observe that fully convective stars tend to have stronger average magnetic fields represented by stronger 
magnetic components. This reflects the conclusion already drawn in previous investigations that fully
convective stars generate on average stronger surface magnetic fields \citep{2007ApJ...656.1121R}.

Second, there seems to be no pronounced difference in filling factors for stars that are known to have
different geometries of their large scale fields. At the same time, our Fig.~\ref{fig:f-factors-3} suggests that
stars with multipole-dominant fields seem to have stronger zero field magnetic component compared to stars with
dipole-dominant fields.
This is an interesting observation because it may tell us that these stars have different spot
distributions and, perhaps, spot sizes on their surfaces compared to stars with dipole-dominant fields.
For instance, spots that are present in a few small localized areas on a visible disk of the star
while the rest of the photosphere is non-magnetic would result in strong zero field component. On the contrary,
a large spot or groups of spots around magnetic poles that occupy considerable fraction of the
stellar disk would have a reduced zero field component or even none if observed at small inclination. 
Recent analysis of stellar spots in a binary system Gl~65-AB seem to support this idea \citep{2017MNRAS.471..811B}. 
It would be thus interesting to combine polarimetric and photometric techniques to address this question in future. 

\begin{figure}
\includegraphics[width=\hsize]{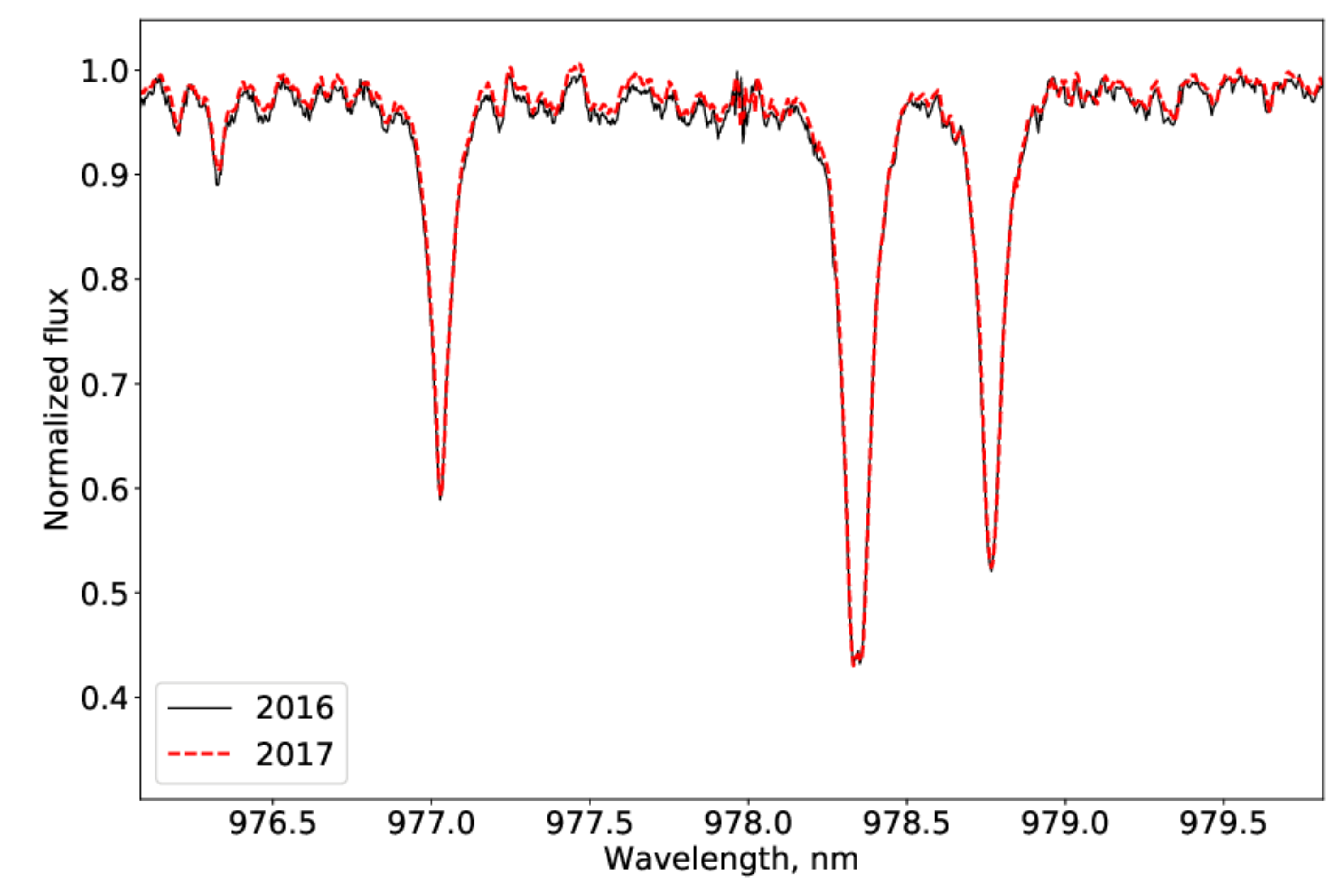}
\centerline{2016 \hspace{3.6cm} 2017}
\includegraphics[width=0.49\hsize]{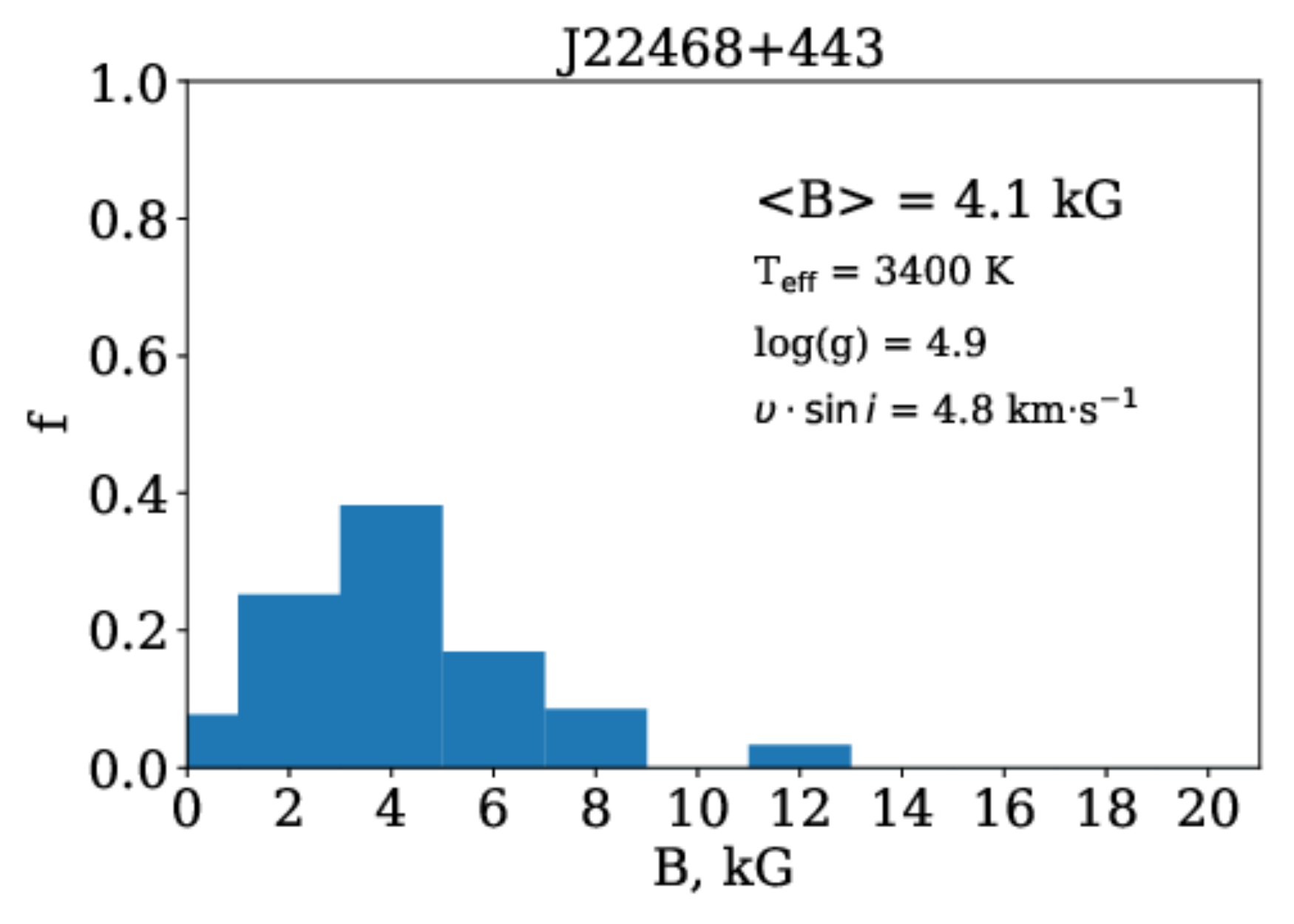}
\includegraphics[width=0.49\hsize]{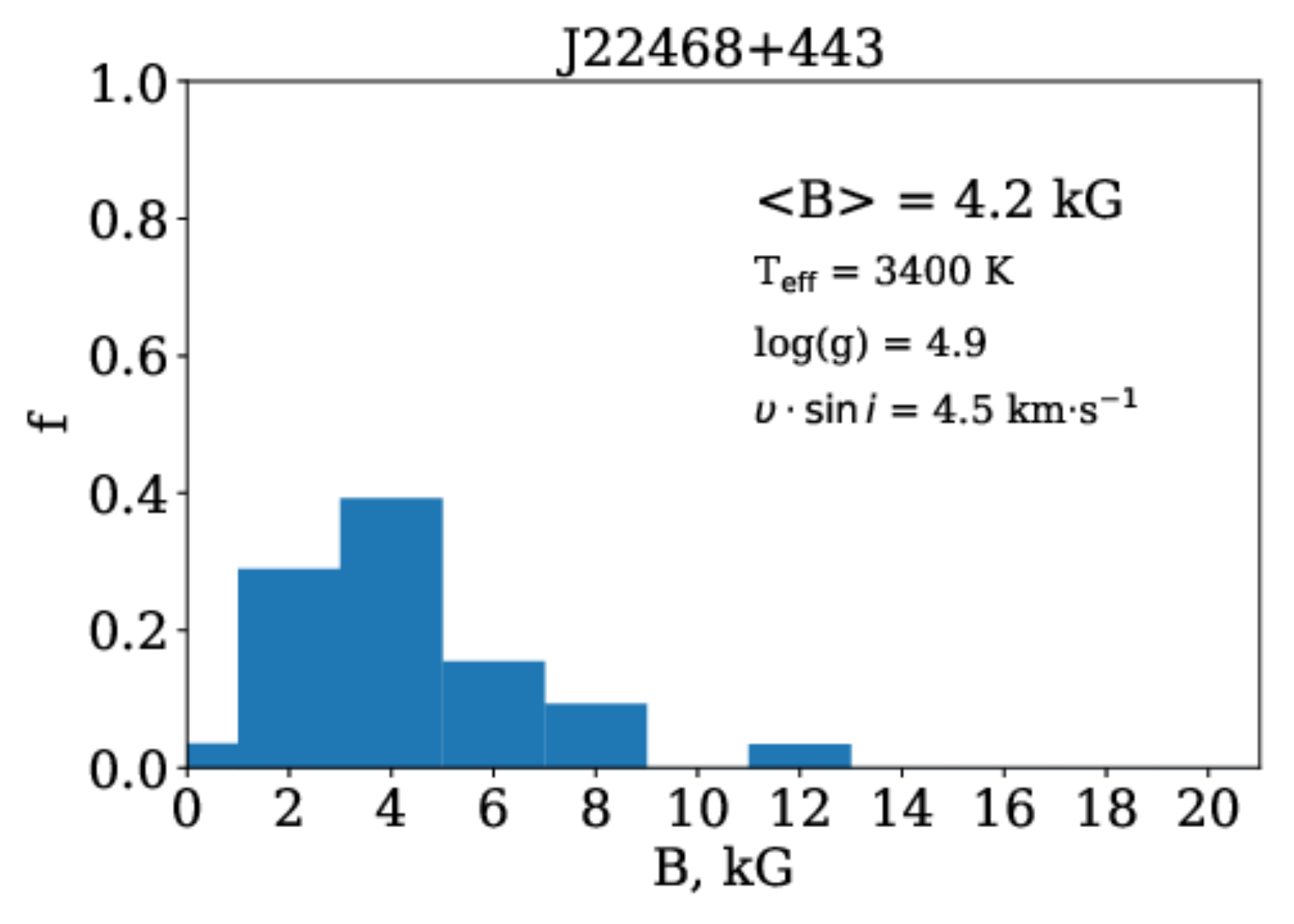}
\caption{\label{fig:evlac-2016-2017-spec-bf}
Mean CARMENES spectra (top panel) around three magnetic sensitive Ti lines obtained
in 2016 (full black line) and 2017 (dashed red line)
 and corresponding filling factors derived from these data sets (bottom panel).
}
\end{figure}

Another remarkable finding is the detection of twins in our sample.
In stars J01352-072 and J04472+206 we recover same average magnetic fields
with identical filling factors.
The other twins are J07446+035 and J23548+385. These stars have very close $\vsini$'s, spectral types,
average magnetic fields, and pattern of filling factors. Because J07446+035 has stable dipole-dominant
magnetic field geometry \citep{2008MNRAS.390..567M} we predict that this should be the same for J23548+385. 
However, for J22468+443 with a multipole-dominant field, we also
derive filling factors that look surprisingly similar to those of dipole-dominant J07446+035.
A possible explanation would be just a coincidence because J22468+443 has complex variable magnetic field
and it might just have happened that the star was observed when its surface field was simple.
Because the spectra of J22468+443 has very high SNR, we looked for a possible seasonal magnetic field
variation between spectra obtained in 2016 (June-December) and 2017 (January-October).
For each year, we co-added individual spectra to build a high SNR template as described in section Methods.
We found that line profiles of Ti did not show any significant changes, as can be seen from the top panel 
of Fig.\ref{fig:evlac-2016-2017-spec-bf} where we plot spectra obtained in 2016 and 2017 
around three magnetic sensitive Ti lines. We also derived filling factors for these two data sets and found them 
to be very similar with only marginal change in the total magnetic field strength, as illustrated in the bottom
panel of Fig.~\ref{fig:evlac-2016-2017-spec-bf}.
Note that J22468+443 was observed to have very strong variability of the large scale component of its magnetic field
with large areas of positive and negative fields located in the same hemisphere \citep[see Fig.4 in][]{2008MNRAS.390..567M}.
Such features were never detected in stars with dipole-dominant fields (e.g., WX~UMa, AD~Leo, Gl~51, etc.)
and thus we still place J22468+443 in the group of stars with complex multipole fields. Additional spectropolarimetric
measurements will surely help to address this question in more detail.

Next, with certain exceptions we can conclude that M dwarfs of spectral types around M7.0 and later
tend to have smooth distributions of their filling factors with obviously dominant zero-field magnetic component like,
e.g., J08298+267 and J19169+051S, with an exception of J16555-083 which shows more patchy filling factors pattern (Fig.~\ref{fig:f-factors-3}).
If confirmed with future studies, this would imply that magnetic dynamo in stars at the very cool end of the M type sequence
fails to generate
large spots and perhaps decays out because temperatures of this objects are too cool to support efficient
dynamo action.

In stars with large $\vsini>20$~\kms\ we usually find smooth distribution of filling factors that
peak at non-zero magnetic field component. This could mean that the stars have simple dipole-dominant magnetic fields,
i.e. consistent with what one would expect in fully convective objects with short rotation periods.

In this work we did not study the rotational variability in individual line profiles which could be caused
by variable magnetic field strength over the stellar surface. This variability could be seen in stars with largest $\vsini$
values due to a better spatial resolution provided by the Doppler broadened line profiles. 
The first evidence of such variability was reported in \citet{2017ApJ...835L...4K} 
for the B component of the binary system Gl~65. At the same time, no line profile variability
was recently detected by \citet{2019ApJ...873...69K} in another fast rotating binary YY~Gem, possibly implying that 
the variability is very weak due to a different distribution of small scale 
magnetic fields over the surface of these stars compared to the case of Gl~65~B. 
Unfortunately, the SNR of individual spectra for our stars is much lower than those used in 
both mentioned above studies and we could not detect any significant variability above the photon noise limit.
This also implies that the total magnetic field strength remains very much stable over the time spawn of our observations
and does not bias our analysis. However, it is no doubt important to study line profile variability in future studies.

\subsection{Magnetic field and rotation}

Our new measurements add nine new objects to the set of stars having very strong fields above $4$~kG
and one object (J05365+113) to the subsample of stars with long rotation period and weak fields.
From Fig.~\ref{fig:bf-period} one can still see a large scatter in magnetic fields for stars with periods
shorter than $4$~d and it remains inconclusive whether magnetic fields keep growing as periods decrease
or do they saturate to some maximum magnetic field which could be defined by the stellar dynamo state,
as suggested in Sh2017. Additional measurements are clearly needed.
In particular, it is needed to analyze stars with ultra-short rotation periods $P<0.3$~d.

Another important property of stellar magnetic fields is the geometry of their large scale components.
The current understanding of stellar magnetism predicts complex multipol-dominant magnetic fields in partly
convective M dwarfs and more simple, dipole-dominant fields in stars that are fully convective
\citep{2010MNRAS.407.2269M,2013A&A...549L...5G}. 
However, there are cases  of a fast rotating fully convective object that generate
complex mostly multipole fields (e.g., DX~Cnc, GJ~1245~B, Gl~Vir, etc.). 
Rotation alone can not explain this observation because
both types of geometries are found in stars that have similar spectral types and rotation periods.
It is believed that dynamo in M dwarfs (at least when they are fully convective) become
bi-stable and the choice of the dynamo state is somehow linked to the properties of the initial
magnetic field that existed during the star formation \citep{2013A&A...549L...5G}. Furthermore, when the rotation 
decreases (e.g., due to the magnetic braking), a fully convective star may change its dynamo state
from a stable dipole-dominant to a variable complex magnetic field thereby developing magnetic cycles, i.e. 
the magnetic field geometry will vary with time \citep{2016ApJ...833L..28Y}.

From our CARMENES data we do not have information about the geometry of the magnetic fields in our stars,
but from our measurements we find objects with very strong fields and fast rotation so we predict that
they should have dipole-dominant fields. It is therefore essential to follow up these stars with spectropolarimetric
observations. This would help to better constrain the connection between field intensity,
geometry, and the rotation of the star. 

There are many more potential application of our finding because
magnetic fields play critical role in all stages of stellar and even planetary evolution. 
For instance, knowing magnetic properties of stars
is one of the pieces in the puzzle called stellar activity and includes understanding
connections between the magnetic fields, rotation braking, stellar spots, X-ray fluxes, and finally understanding
the hazardous environment around planets orbiting these active stars \citep{2015MNRAS.449.4117V,2017ApJ...843L..33G}.
Very strong magnetic fields my have even direct impact on the planetary structure
by providing additional source of energy via the induction heating \citep{2017NatAs...1..878K,2018ApJ...858..105K}.
Especially M dwarfs with dipole dynamo states are interesting objects in this regard because
they maintain stable large scale magnetic fields whose energy decay with a separation to the star
much slower compared to stars with multipole-dominant fields.
Ultra-fast rotating M dwarfs are therefore very interesting and exotic objects in many aspects.

From our analysis we find
that in six stars with periods $P<0.3$~d we find three of them missing a significant zero field component
in their filling factor distributions.
If confirmed with future studies and it will appear that all stars with ultra-short periods show little of the non-magnetic areas, 
this would imply that most of the stellar surface is covered with active regions~--~an effect which was
proposed as one of the possible reason for activity saturation \citep{2011MNRAS.411.2099J,2014ApJ...794..144R}.
More accurate magnetic field measurements in late type M dwarfs, especially those with periods $P<0.3$~d,
are therefore essential to explore properties of stellar dynamos in this parameter range.

\section{Summary}

In this work we presented first magnetic field measurements from the high-resolution
near-infrared spectroscopic observations taken with the CARMENES instrument. We specifically concentrated
on the so-called RV-loud sample of $31$ M dwarfs presented in \citet{2018A&A...614A.122T} because
these stars are expected to have strong magnetic fields as indicated by analysis of their activity
indicators. We employed state-of-the-art radiative transfer model to measure average magnetic fields 
from the Zeeman broadening of atomic and molecular lines. Our main conclusions are summarized as follows:
\begin{itemize}
\item
We detect strong magnetic fields $\bs>1$~kG in all our targets. In $16$ of them the measurements
were done for the first time. In $12$ of them our data
indicate a presence of very strong fields above $4$~kG.
\item
We observe $17$ stars with short rotation periods $P<1$~d and our new measurements
are consistent with the effect of magnetic field saturation, however the magnetic field
may as well still grow at least in stars with dipole dynamo states.
\item
Our analysis of filling factors points towards the existence of particular
features in their patterns that may help to distinguish between the stars 
that have different dynamo states and/or spot patterns.
\item
We find two stars, J07446+035 and J23548+385, that are twins in terms of their average magnetic fields and distribution
of filling factors. It would be interesting to compare the geometry of their 
global magnetic fields and test whether they are the same as well. This would be an important additional
test for our analysis methods.
\item
Our study adds $16$ new objects to the list of stars with strong magnetic fields
and short rotation periods.
However, in order to fully characterise magnetic properties of stars and to
put our findings in the context of bi-stable dynamos
we lack information about the geometry of large scale magnetic fields
and instruments with polarimetric capabilities are needed.
Thus, the next logical step would be to follow up our targets with polarimetry
and derive maps of their photospheric magnetic fields.
\end{itemize}

\begin{acknowledgement}
CARMENES is an instrument for the Centro Astron\'omico Hispano-Alem\'an de Calar Alto (CAHA, Almeria, Spain).
CARMENES is funded by the German Max-Planck-Gesellschaft (MPG), the Spanish Consejo Superior de Investigaciones Cientificas (CSIC), 
the European Union through FEDER/ERF FICTS-2011-02 funds, and the members of the CARMENES Consortium (Max-Planck-Institut f\"ur Astronomie, 
Instituto de Astrofisica de Andalucia, Landessternwarte K\"onigstuhl, Institut de Ci\'encies de l'Espai, Insitut f\"ur Astrophysik G\"ottingen, 
Universidad Complutense de Madrid, Th\"uringer Landessternwarte Tautenburg, 
Instituto de Astrofisica de Canarias, Hamburger Sternwarte, Centro de Astrobiologia and Centro Astron\'omico Hispano-Alem\'an), 
with additional contributions by the Spanish Ministry of Economy,  the German Science Foundation through the 
Major Research Instrumentation  Programme and DFG Research Unit FOR2544 ``Blue Planets around Red Stars'', the Klaus 
Tschira Stiftung, the states of Baden-W\"urttemberg and Niedersachsen, and by the Junta de Andalucia.
L.T.-O. acknowledges support from the Israel Science Foundation (grant No.~848/16).
This work has made use of the VALD database, operated at Uppsala University, the Institute of Astronomy RAS in Moscow, and the University of Vienna.
We also acknowledge the use of electronic databases SIMBAD and NASA's ADS.
\end{acknowledgement}

\bibliographystyle{aa}
\bibliography{biblio}

\begin{appendix}

\section{Model fit to the observed spectra in FeH and Ti lines}

\begin{figure*}
\includegraphics[width=\hsize]{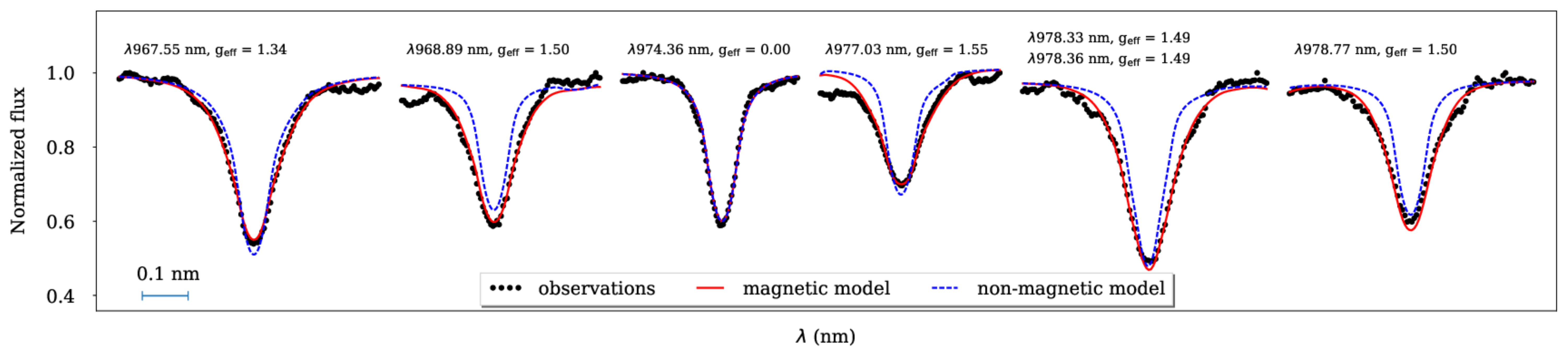}
\includegraphics[width=\hsize]{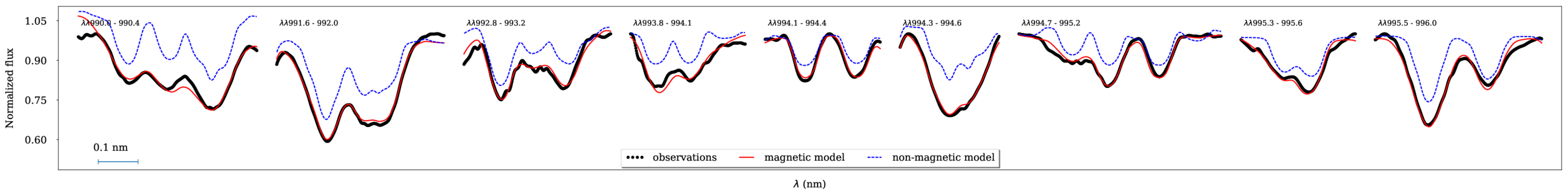}
\caption{\label{fig:J01033+623-fit}
Model fit to Ti and FeH lines in J01033+623.
We show the comparison between observed and predicted spectra
for a set of Ti (top panel) and FeH lines (bottom panel)
Black circles~--~observations; red full line~--~best fit model spectrum;
blue dashed line~--~spectrum computed assuming zero magnetic field.
}
\end{figure*}

\begin{figure*}
\includegraphics[width=\hsize]{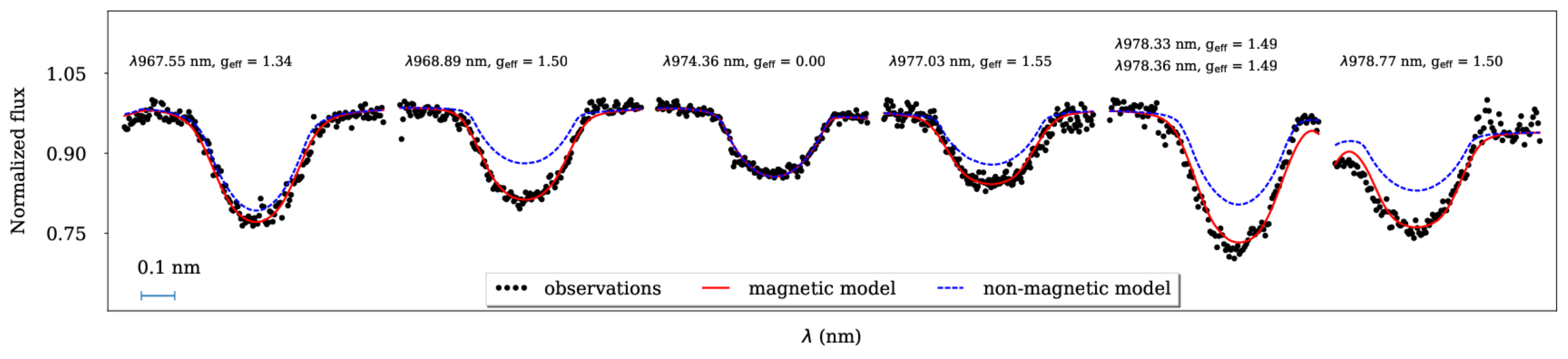}
\includegraphics[width=\hsize]{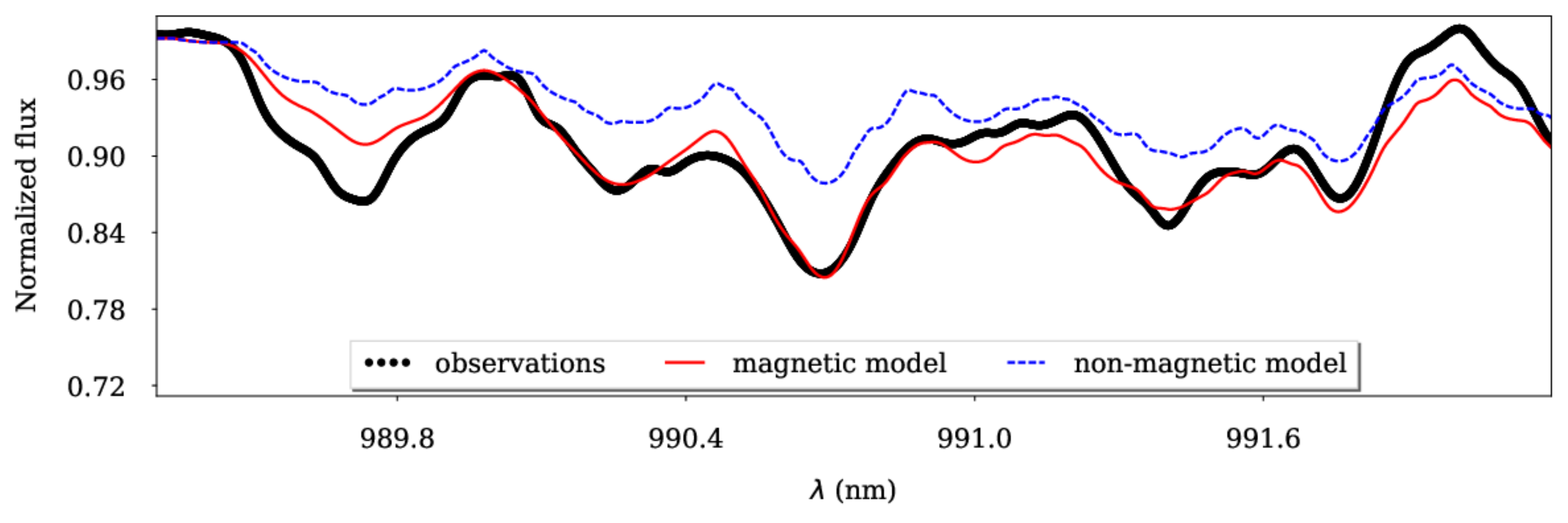}
\caption{\label{fig:J01352-072-fit}
Same as on Fig.~\ref{fig:J01033+623-fit} but for J01352-072.}
\end{figure*}

\begin{figure*}
\includegraphics[width=\hsize]{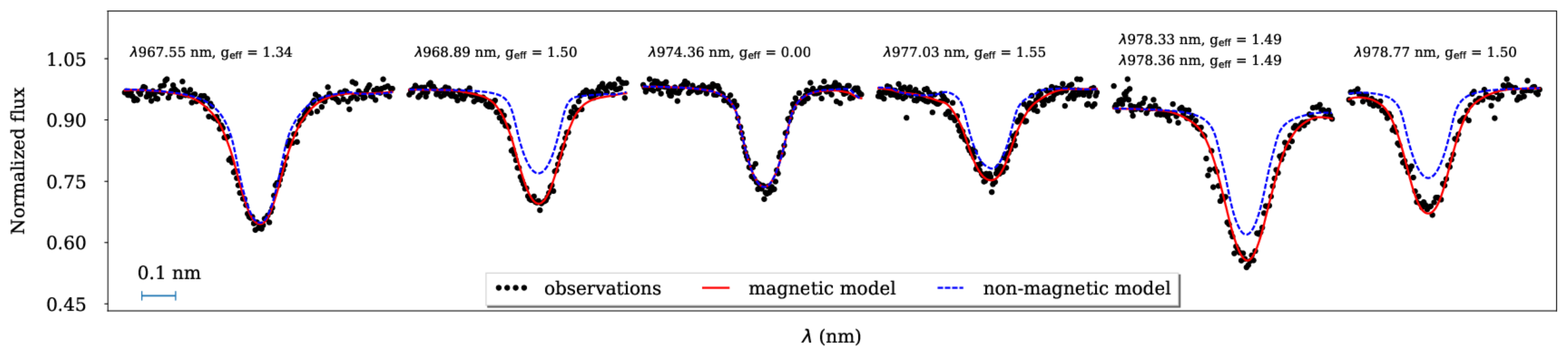}
\includegraphics[width=\hsize]{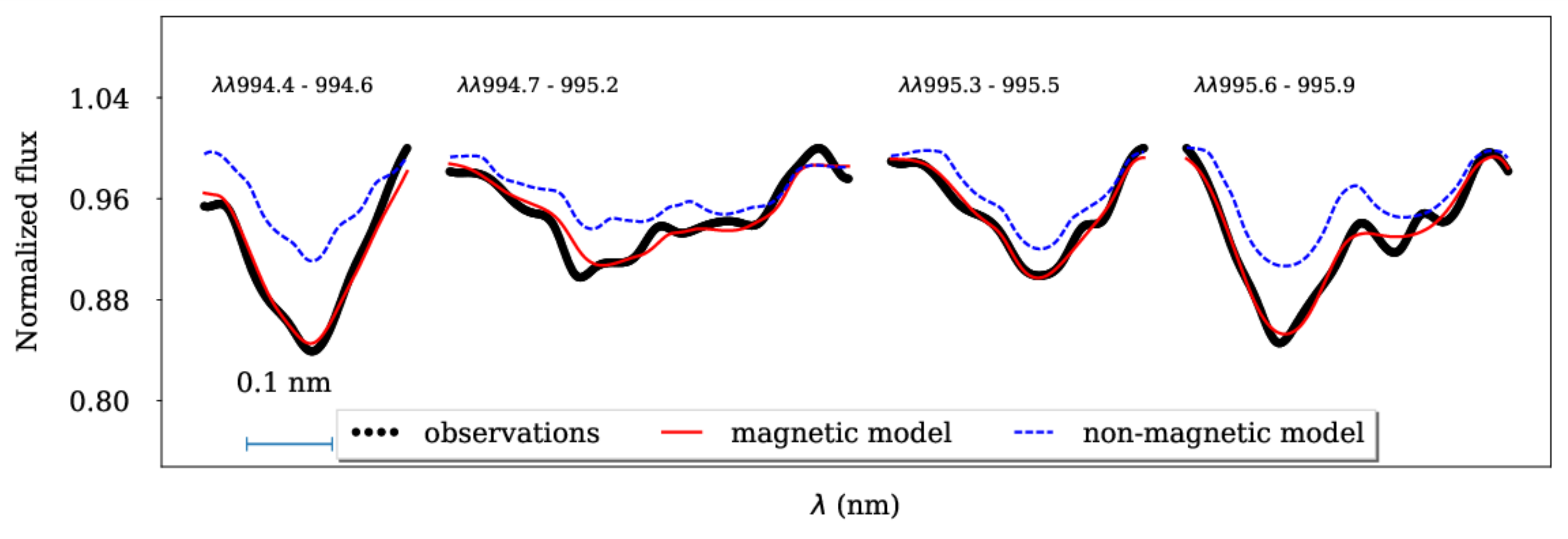}
\caption{\label{fig:J02088+494-fit}
Same as on Fig.~\ref{fig:J01033+623-fit} but for J02088+494.
}
\end{figure*}

\begin{figure*}
\includegraphics[width=\hsize]{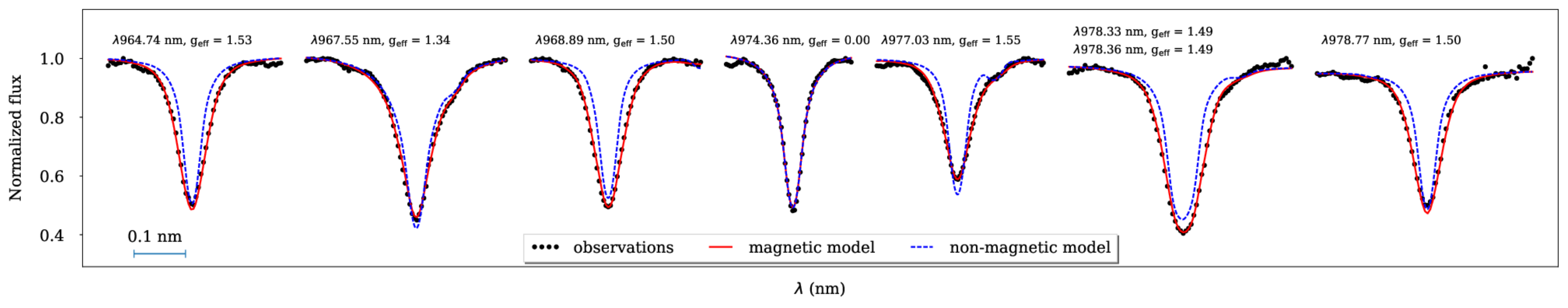}
\includegraphics[width=\hsize]{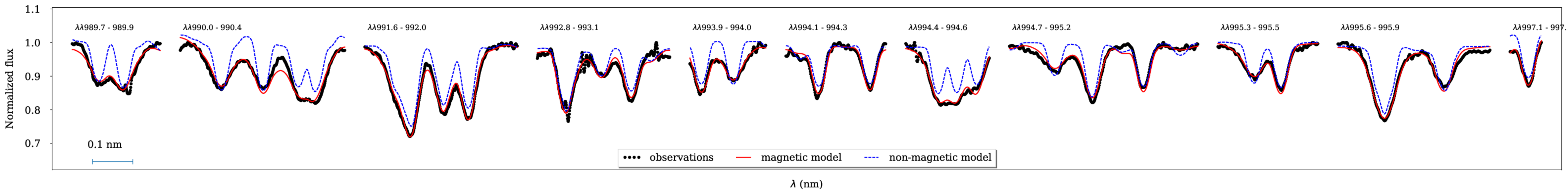}
\caption{\label{fig:J03473-019-fit}
Same as on Fig.~\ref{fig:J01033+623-fit} but for J03473-019.
}
\end{figure*}

\begin{figure*}
\includegraphics[width=\hsize]{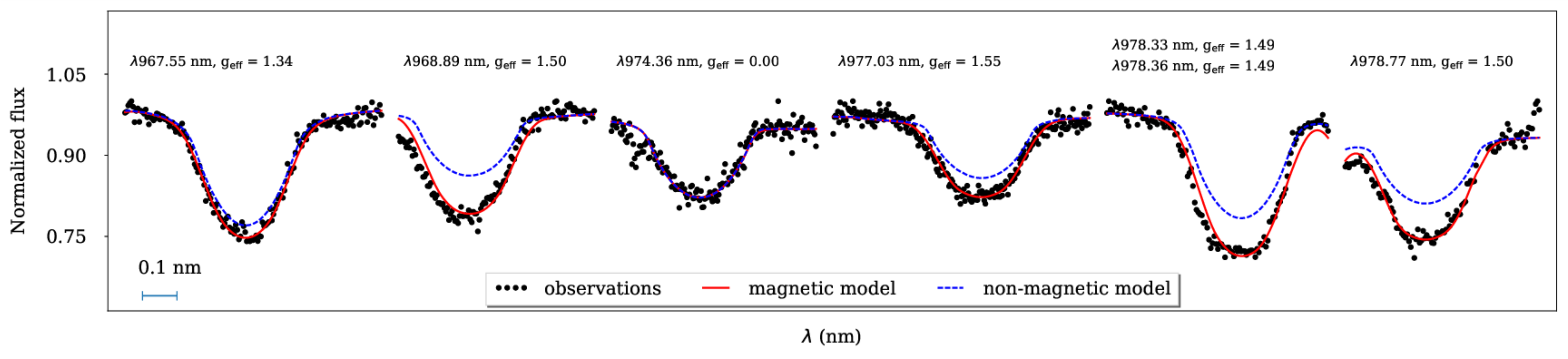}
\includegraphics[width=\hsize]{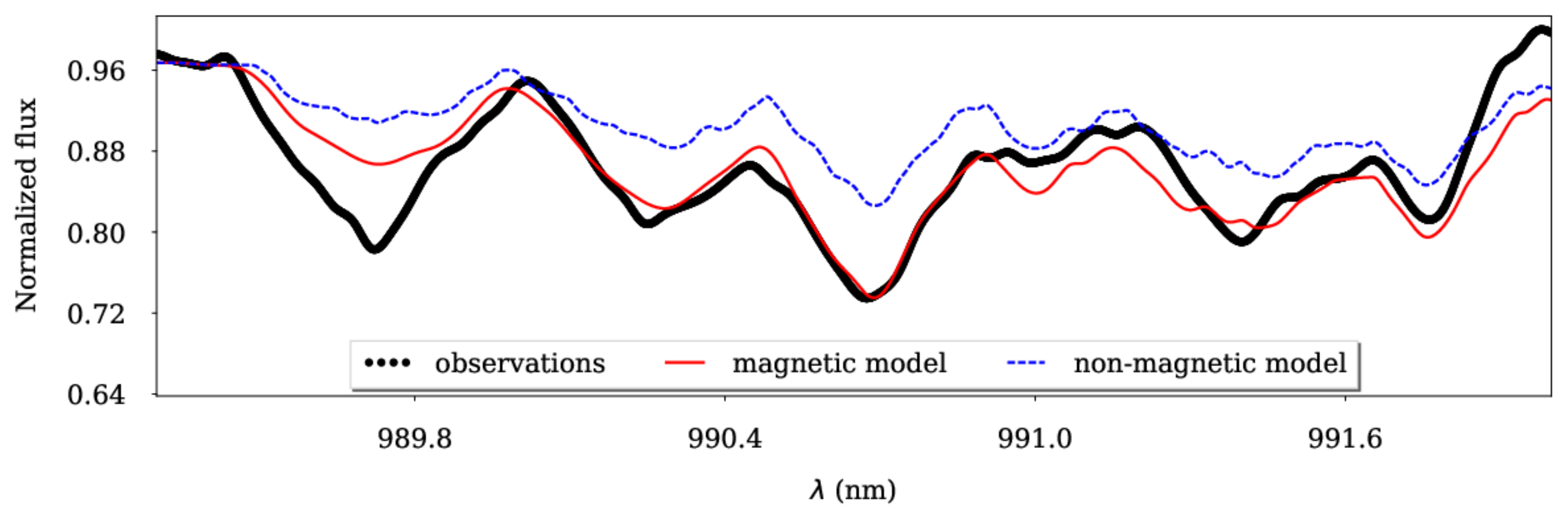}
\caption{\label{fig:J04472+206-fit}
Same as on Fig.~\ref{fig:J01033+623-fit} but for J04472+206.
}
\end{figure*}

\begin{figure*}
\includegraphics[width=\hsize]{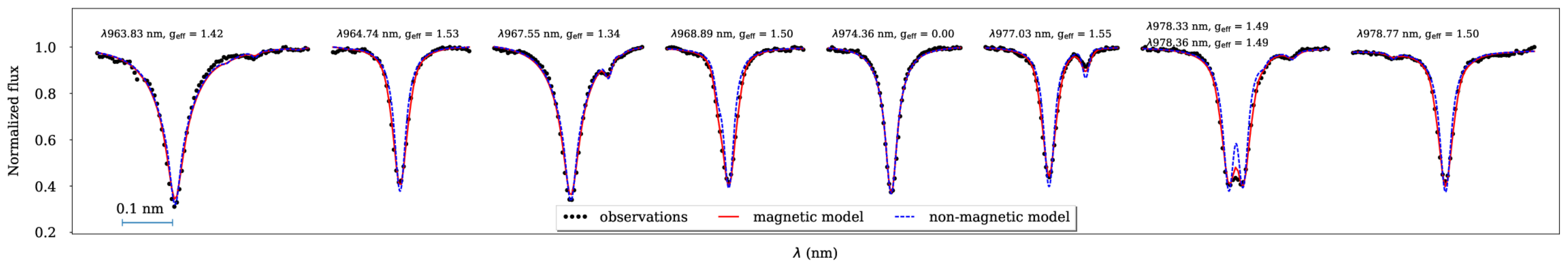}
\includegraphics[width=\hsize]{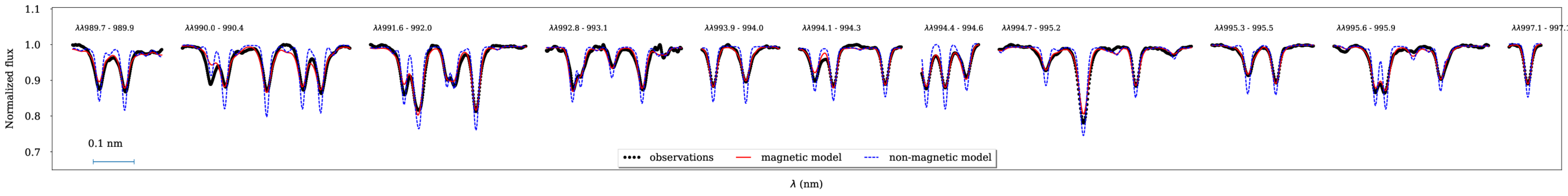}
\caption{\label{fig:J05365+113-fit}
Same as on Fig.~\ref{fig:J01033+623-fit} but for J05365+113.
}
\end{figure*}

\begin{figure*}
\includegraphics[width=\hsize]{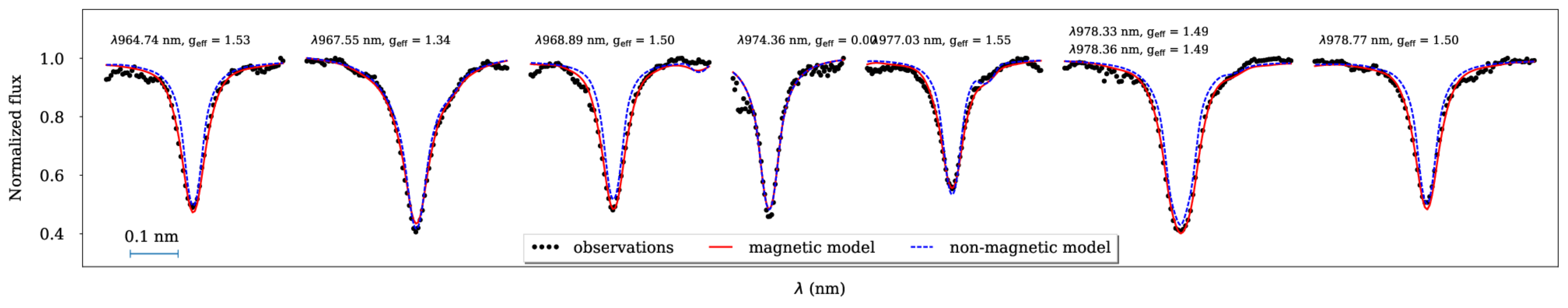}
\includegraphics[width=\hsize]{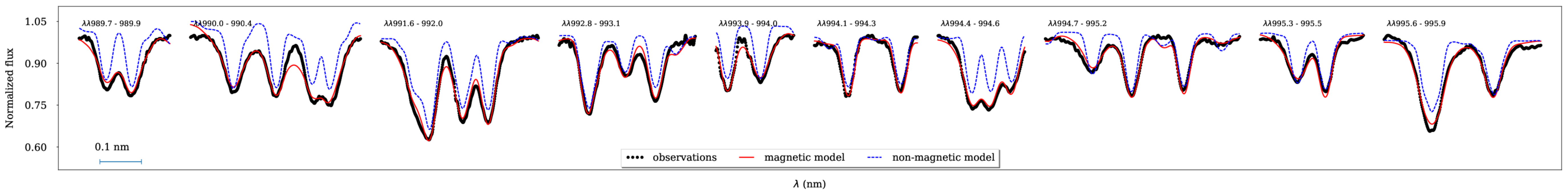}
\caption{\label{fig:J06000+027-fit}
Same as on Fig.~\ref{fig:J01033+623-fit} but for J06000+027.
}
\end{figure*}

\begin{figure*}
\includegraphics[width=\hsize]{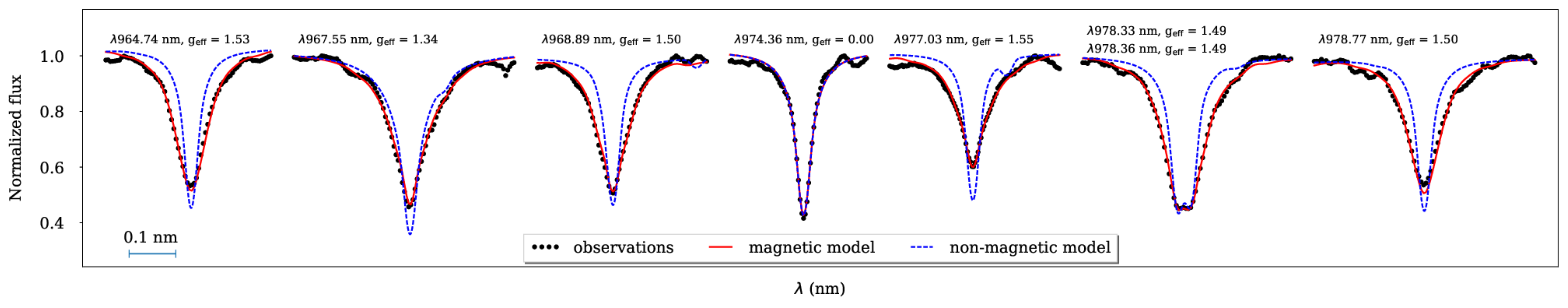}
\includegraphics[width=\hsize]{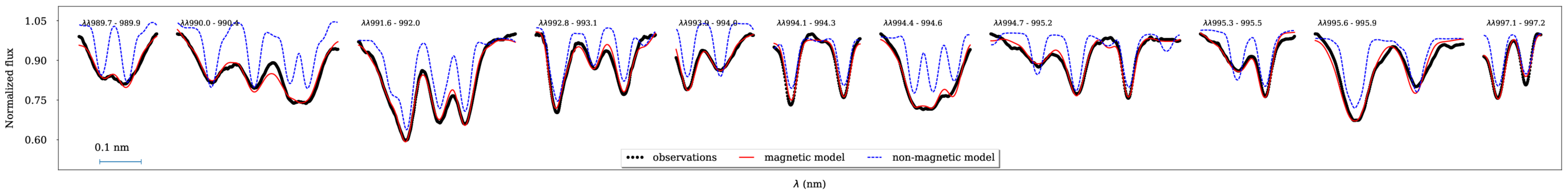}
\caption{\label{fig:J07446+035-fit}
Same as on Fig.~\ref{fig:J01033+623-fit} but for J07446+035.
}
\end{figure*}

\begin{figure*}
\includegraphics[width=\hsize]{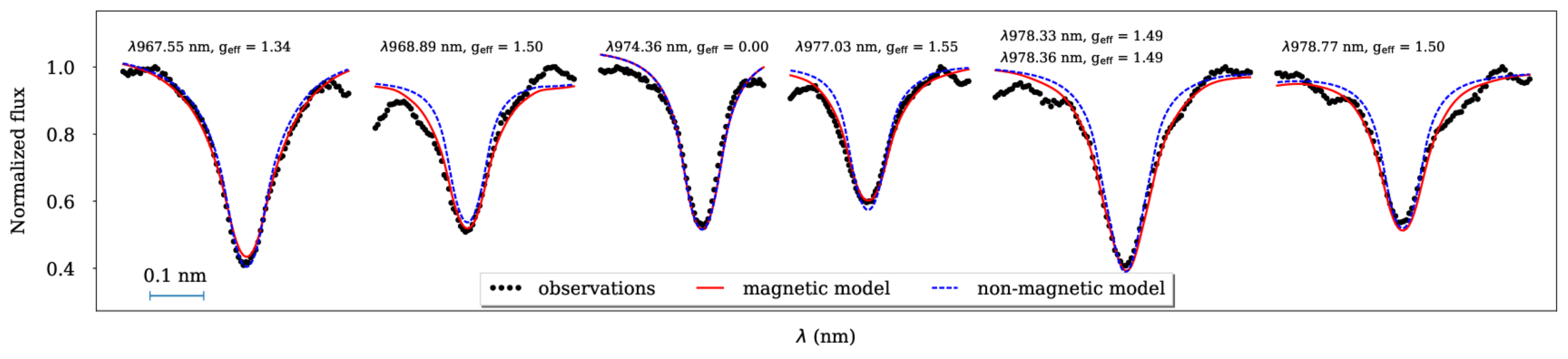}
\includegraphics[width=\hsize]{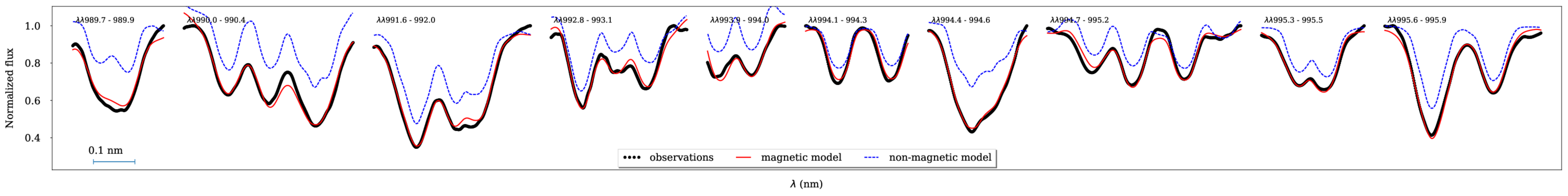}
\caption{\label{fig:J08298+267-fit}
Same as on Fig.~\ref{fig:J01033+623-fit} but for J08298+267.
}
\end{figure*}

\begin{figure*}
\includegraphics[width=\hsize]{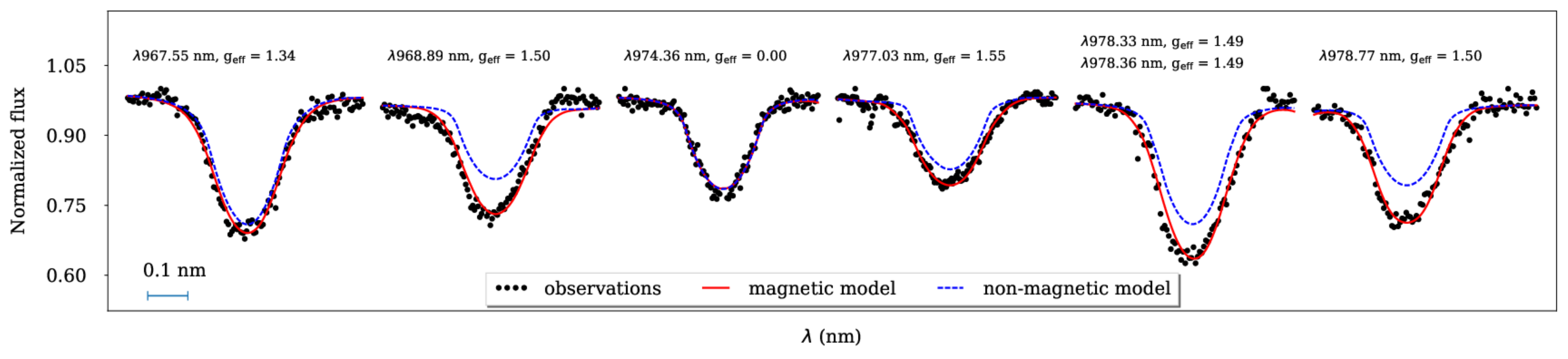}
\includegraphics[width=\hsize]{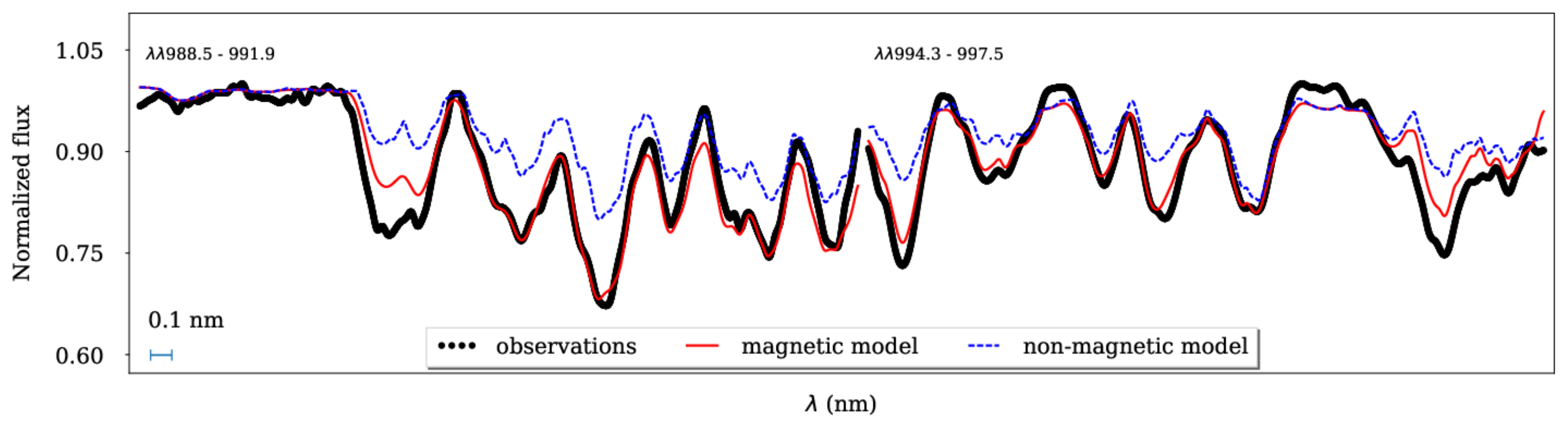}
\caption{\label{fig:J09449-123-fit}
Same as on Fig.~\ref{fig:J01033+623-fit} but for J09449-123.
}
\end{figure*}

\begin{figure*}
\includegraphics[width=\hsize]{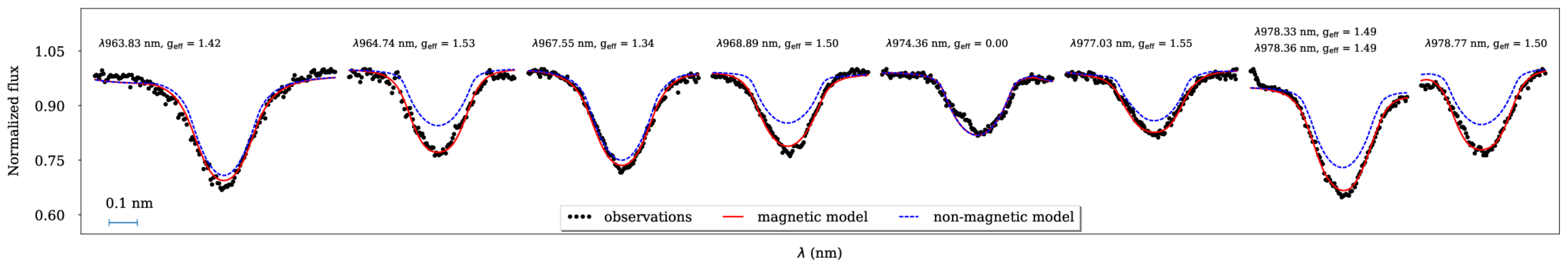}
\includegraphics[width=\hsize]{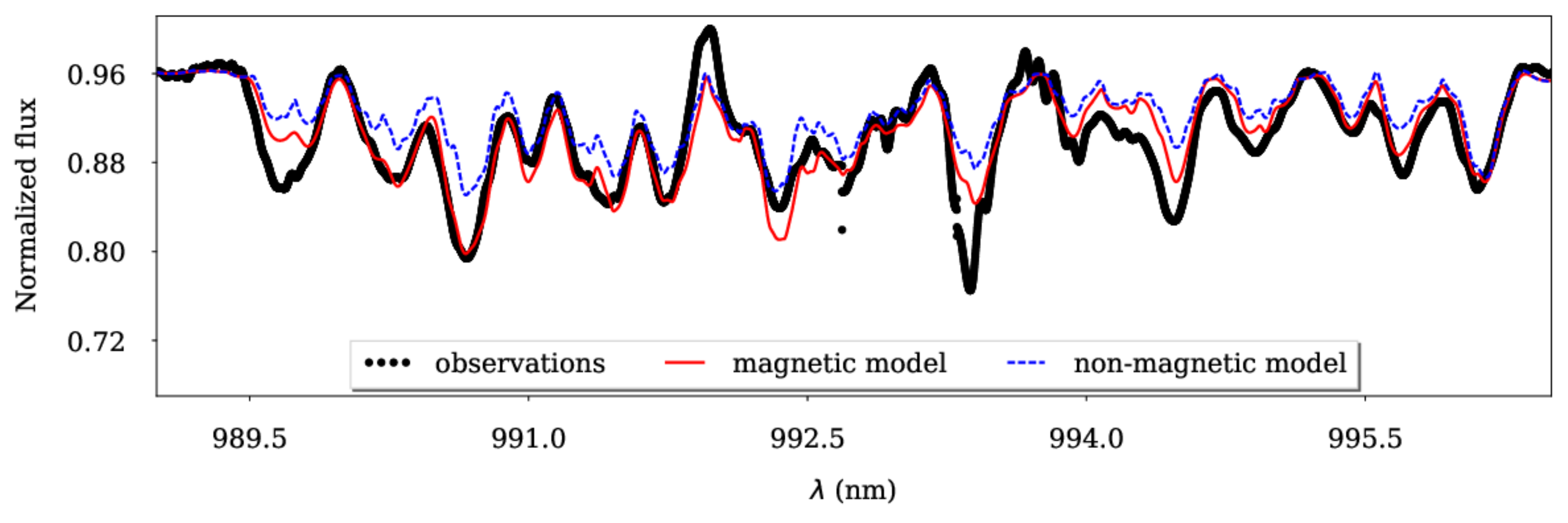}
\caption{\label{fig:J12156+526-fit}
Same as on Fig.~\ref{fig:J01033+623-fit} but for J12156+526.
}
\end{figure*}

\begin{figure*}
\includegraphics[width=\hsize]{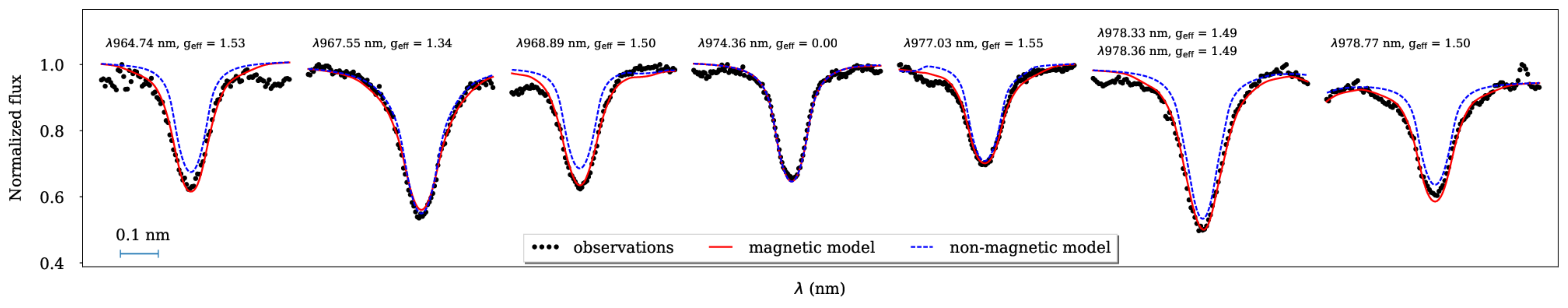}
\includegraphics[width=\hsize]{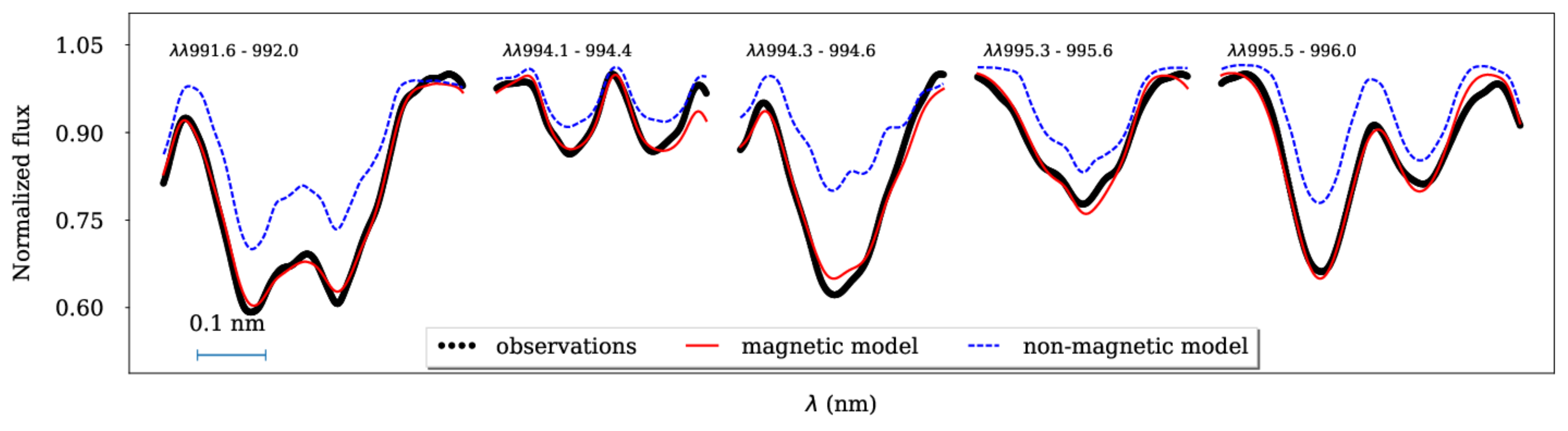}
\caption{\label{fig:J12189+111-fit}
Same as on Fig.~\ref{fig:J01033+623-fit} but for J12189+111.
}
\end{figure*}

\begin{figure*}
\includegraphics[width=\hsize]{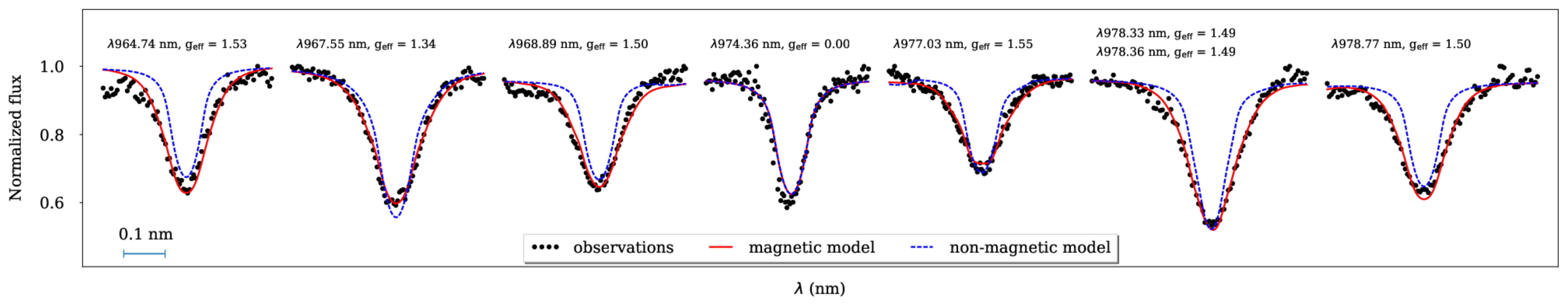}
\includegraphics[width=\hsize]{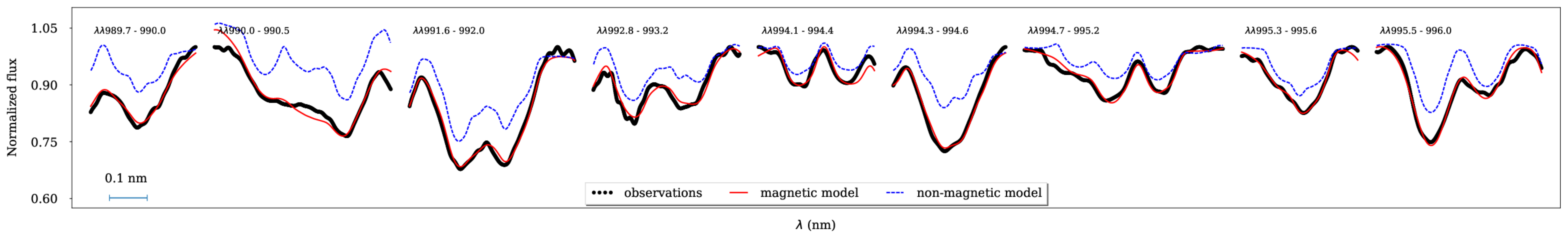}
\caption{\label{fig:J14173+454-fit}
Same as on Fig.~\ref{fig:J01033+623-fit} but for J14173+454.
}
\end{figure*}

\begin{figure*}
\includegraphics[width=\hsize]{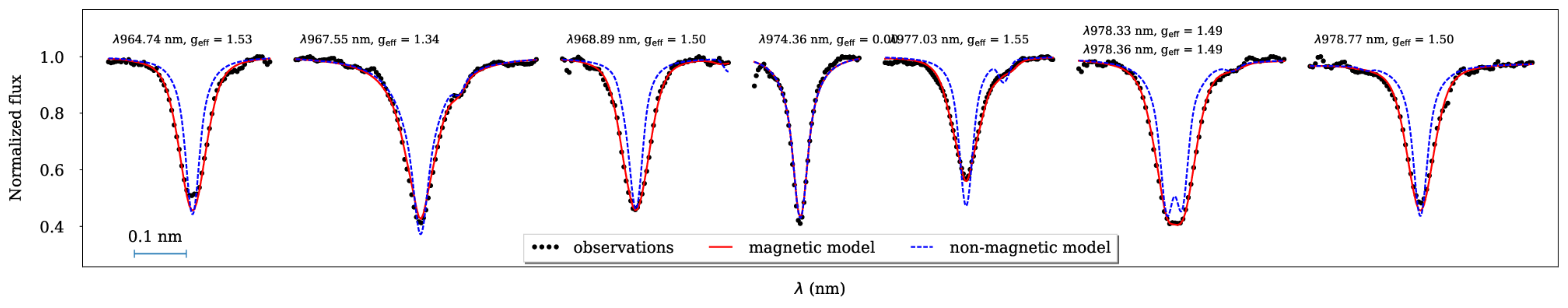}
\includegraphics[width=\hsize]{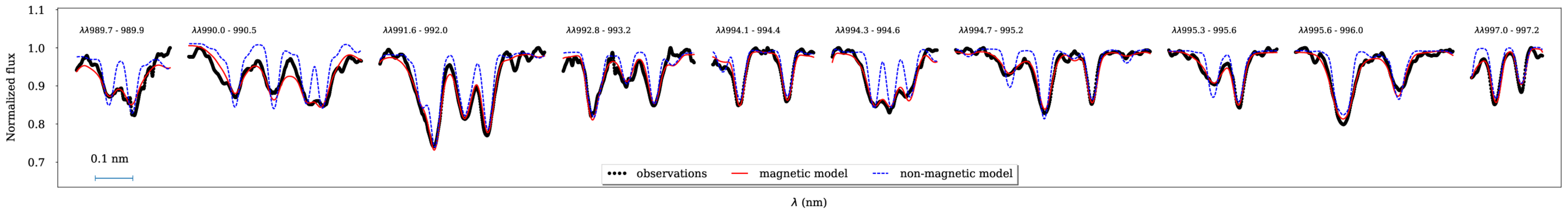}
\caption{\label{fig:J15218+209-fit}
Same as on Fig.~\ref{fig:J01033+623-fit} but for J15218+209.
}
\end{figure*}

\begin{figure*}
\includegraphics[width=\hsize]{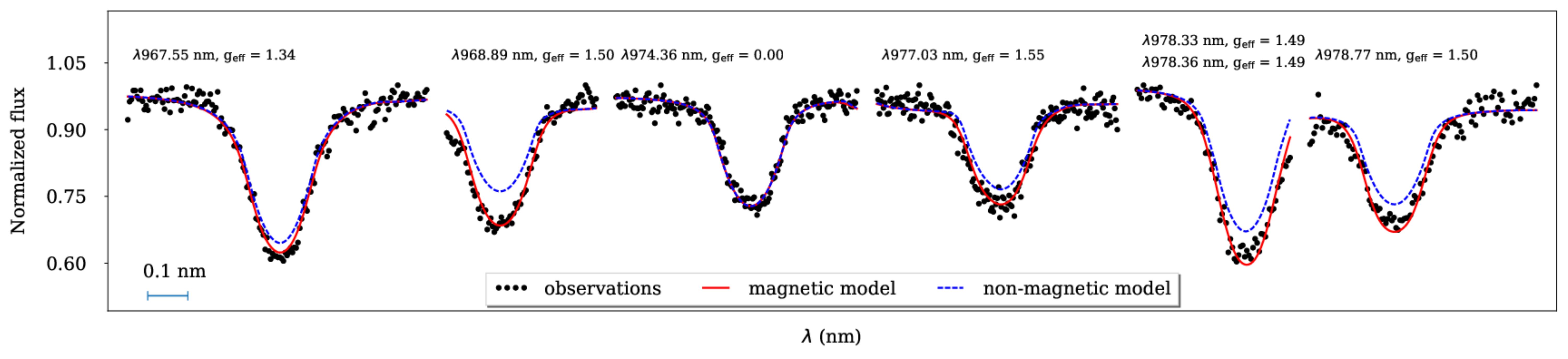}
\includegraphics[width=\hsize]{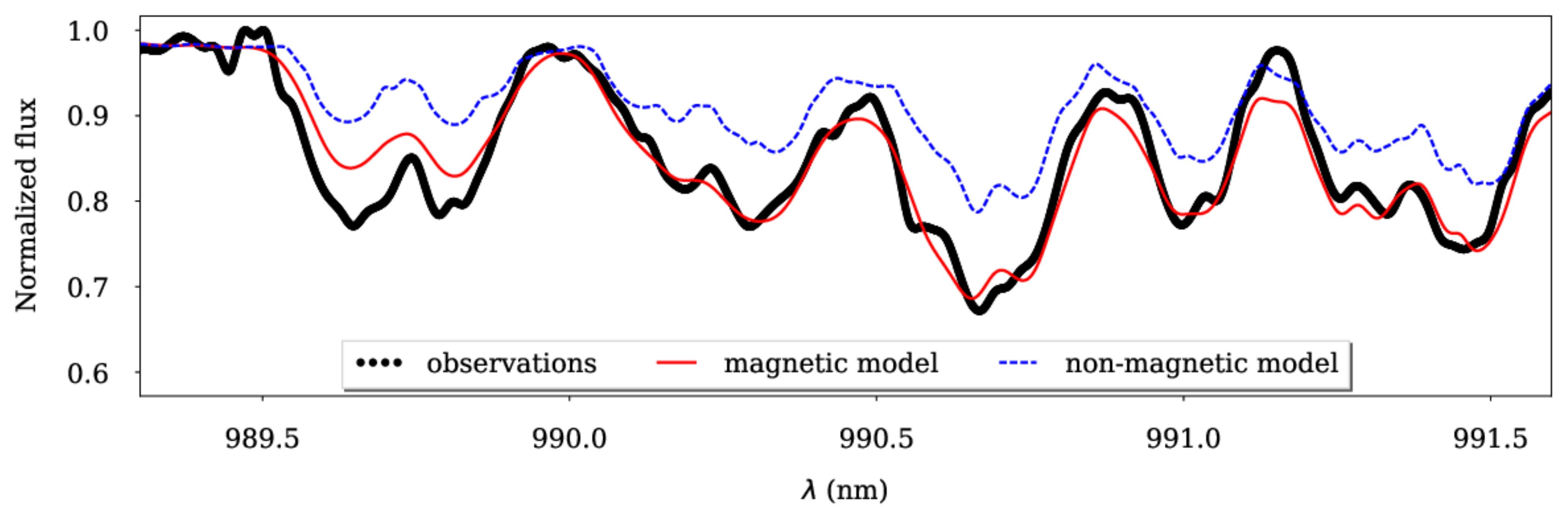}
\caption{\label{fig:J15499+796-fit}
Same as on Fig.~\ref{fig:J01033+623-fit} but for J15499+796.
}
\end{figure*}

\begin{figure*}
\includegraphics[width=\hsize]{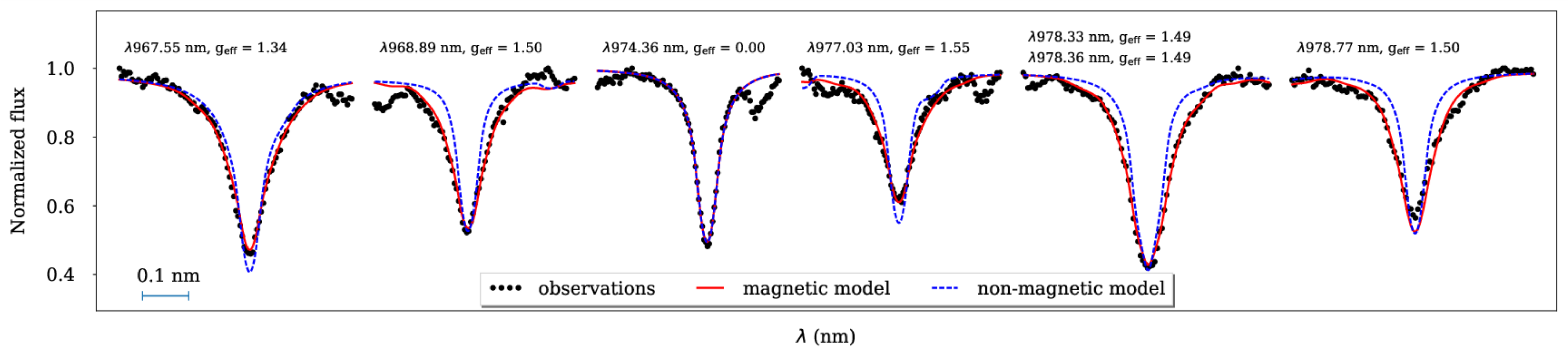}
\includegraphics[width=\hsize]{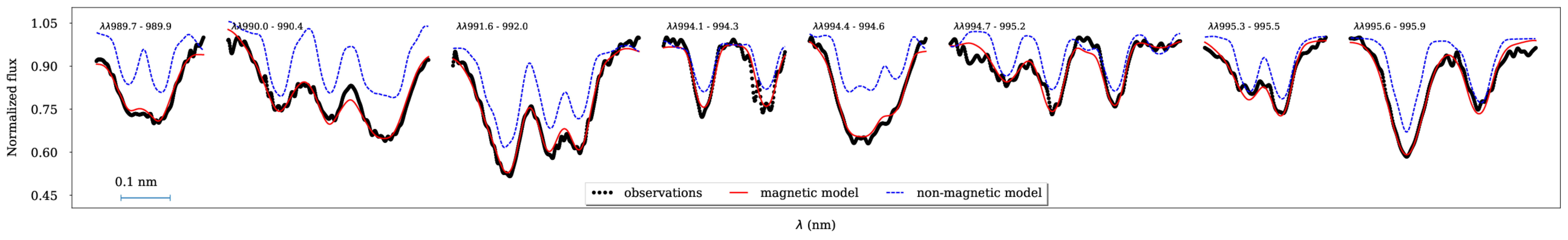}
\caption{\label{fig:J16313+408-fit}
Same as on Fig.~\ref{fig:J01033+623-fit} but for J16313+408.
}
\end{figure*}

\begin{figure*}
\includegraphics[width=\hsize]{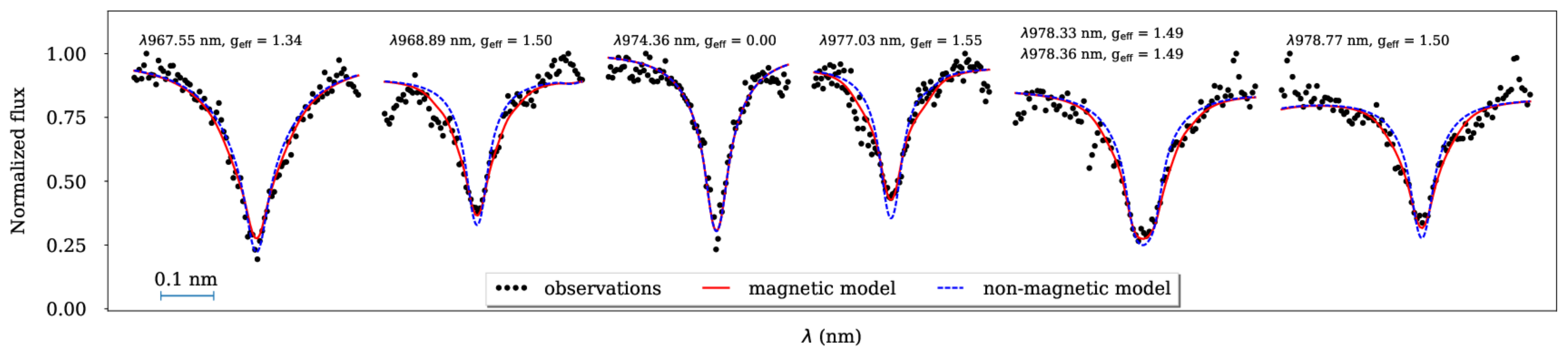}
\includegraphics[width=\hsize]{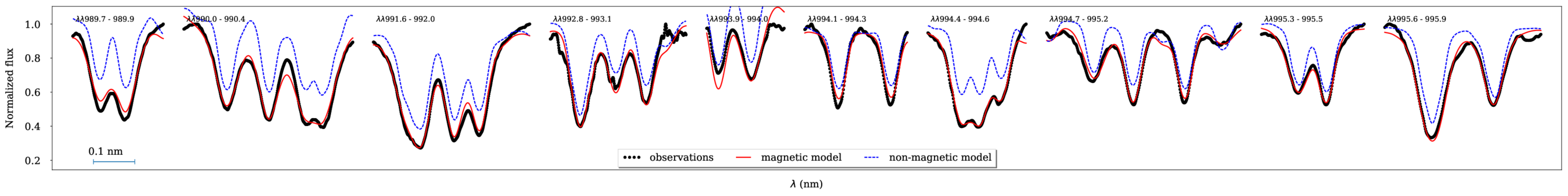}
\caption{\label{fig:J16555-083-fit}
Same as on Fig.~\ref{fig:J01033+623-fit} but for J16555-083.
}
\end{figure*}

\begin{figure*}
\includegraphics[width=\hsize]{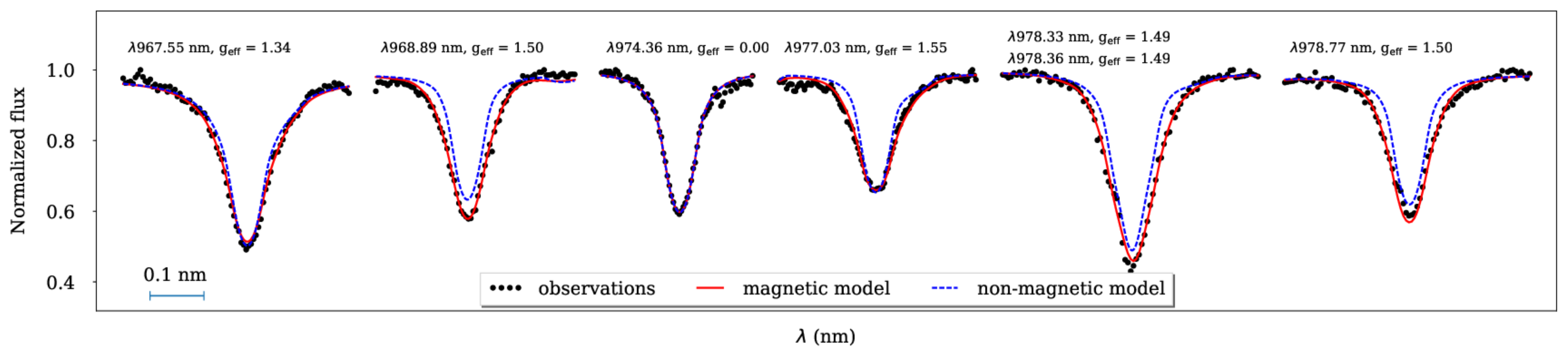}
\includegraphics[width=\hsize]{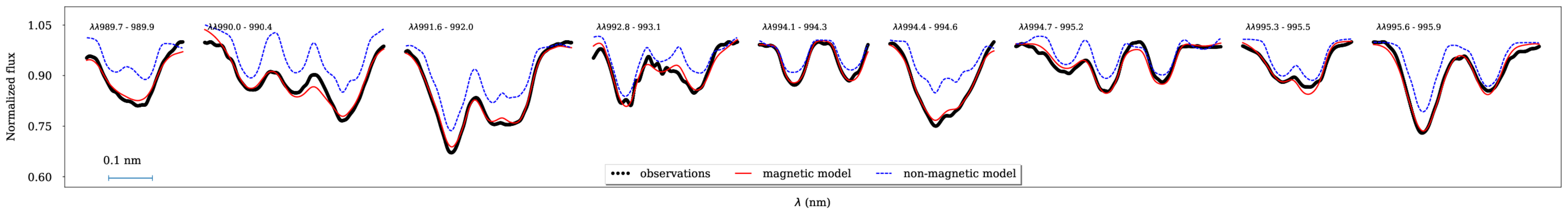}
\caption{\label{fig:J16570-043-fit}
Same as on Fig.~\ref{fig:J01033+623-fit} but for J16570-043.
}
\end{figure*}

\begin{figure*}
\includegraphics[width=\hsize]{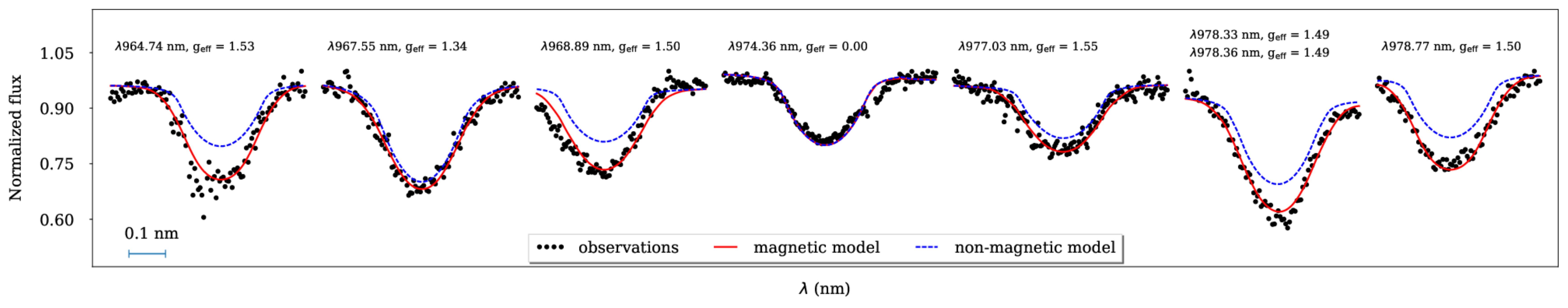}
\includegraphics[width=\hsize]{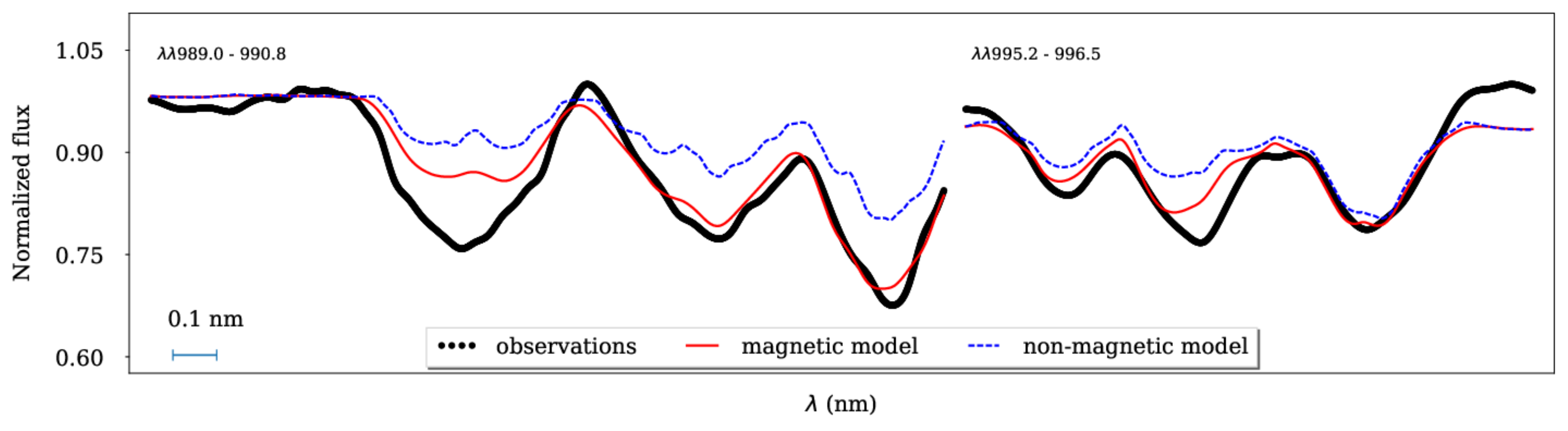}
\caption{\label{fig:J17338+169-fit}
Same as on Fig.~\ref{fig:J01033+623-fit} but for J17338+169.
}
\end{figure*}

\begin{figure*}
\includegraphics[width=\hsize]{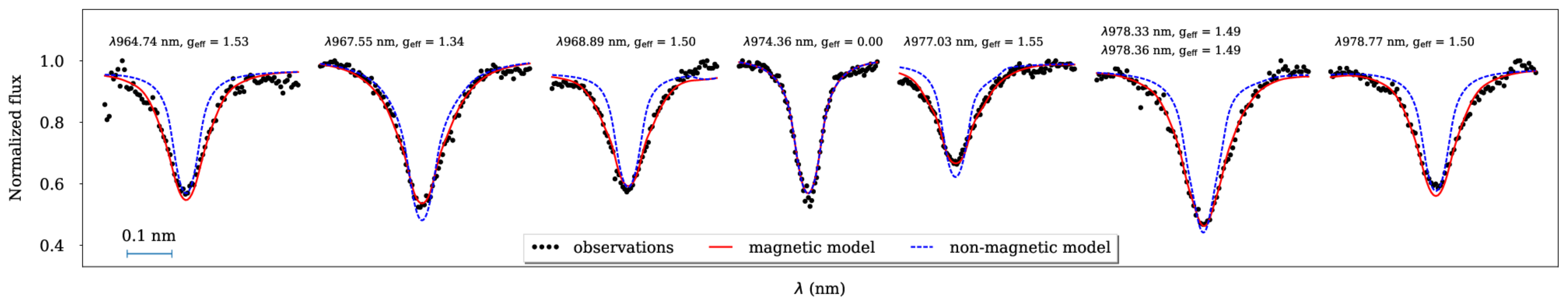}
\includegraphics[width=\hsize]{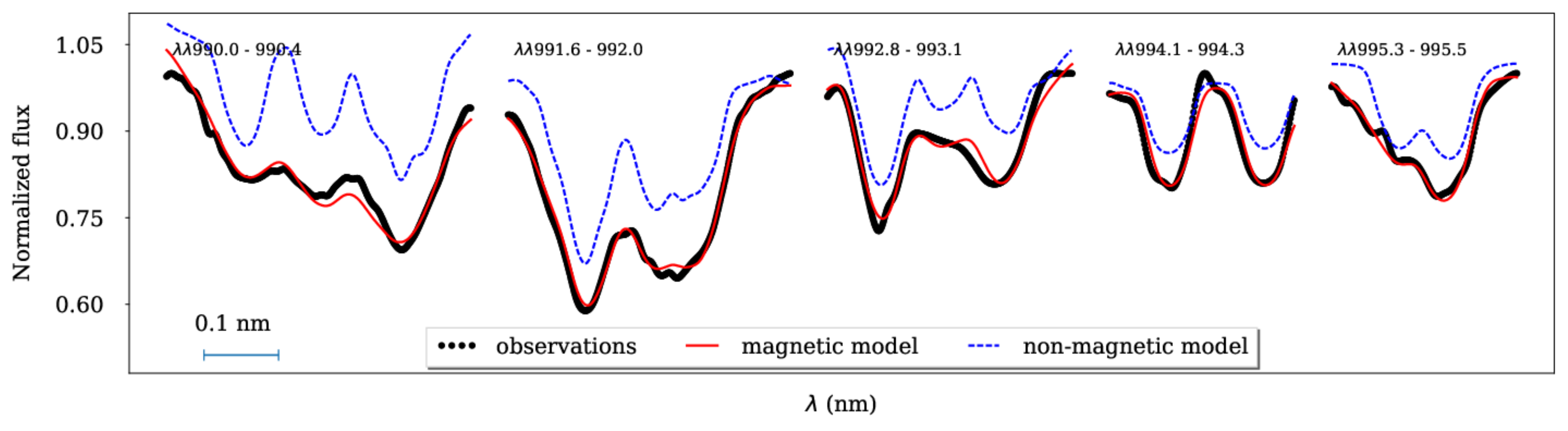}
\caption{\label{fig:J18022+642-fit}
Same as on Fig.~\ref{fig:J01033+623-fit} but for J18022+642.
}
\end{figure*}

\begin{figure*}

\includegraphics[width=\hsize]{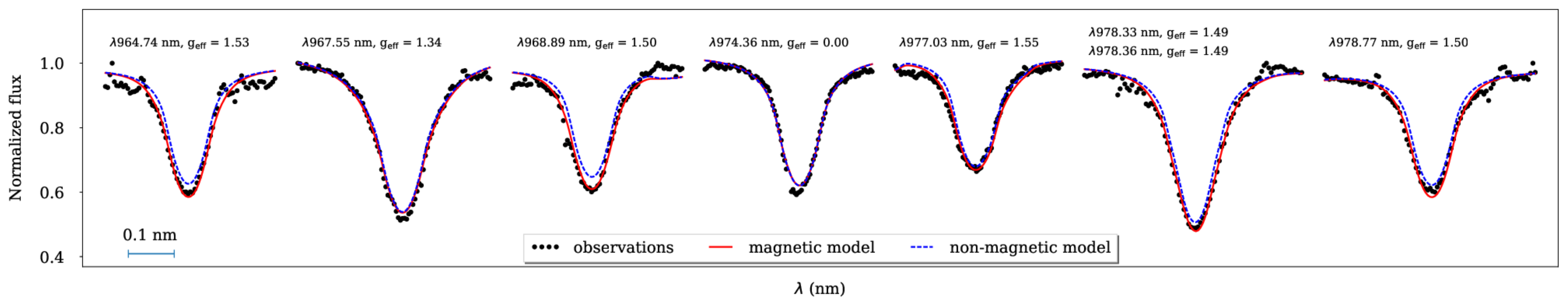}
\includegraphics[width=\hsize]{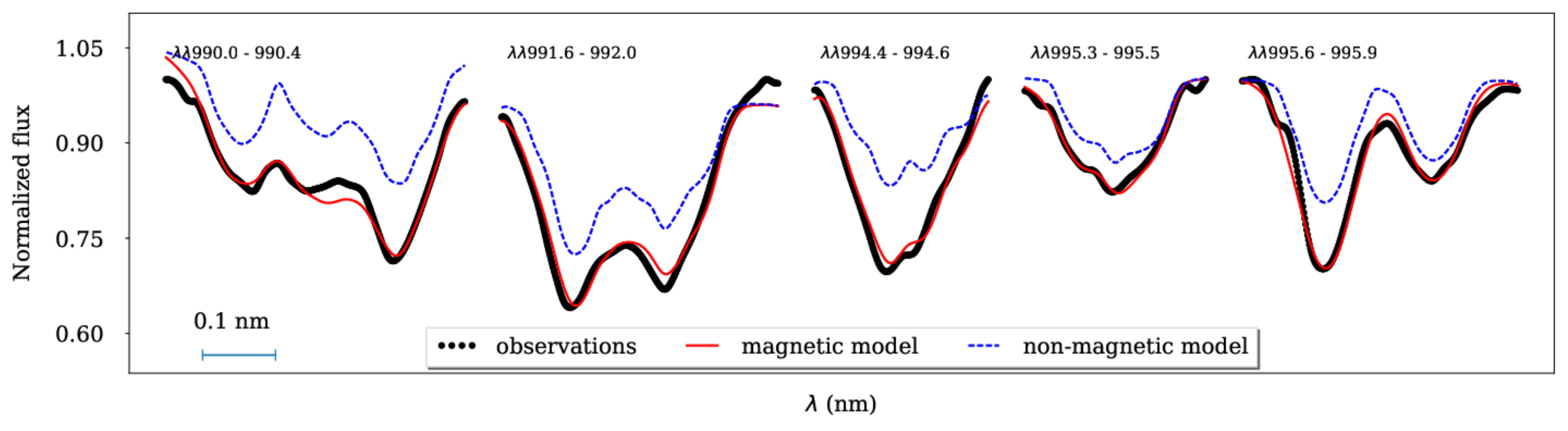}
\caption{\label{fig:J18189+661-fit}
Same as on Fig.~\ref{fig:J01033+623-fit} but for J18189+661.
}
\end{figure*}

\begin{figure*}
\includegraphics[width=\hsize]{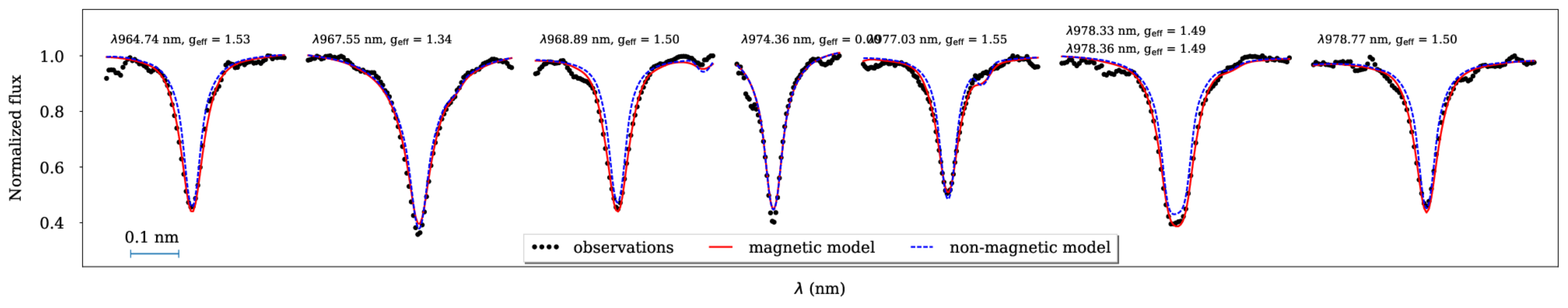}
\includegraphics[width=\hsize]{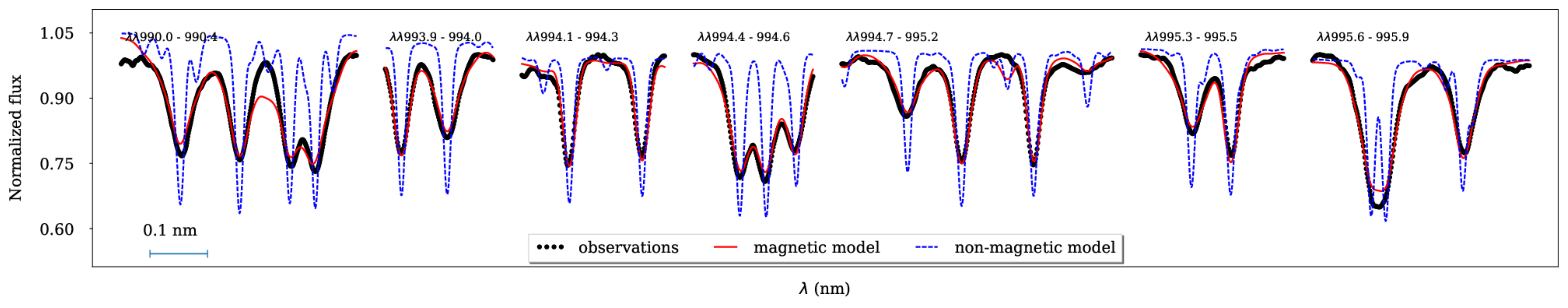}
\caption{\label{fig:J18498-238-fit}
Same as on Fig.~\ref{fig:J01033+623-fit} but for J18498-238.
}
\end{figure*}

\begin{figure*}
\includegraphics[width=\hsize]{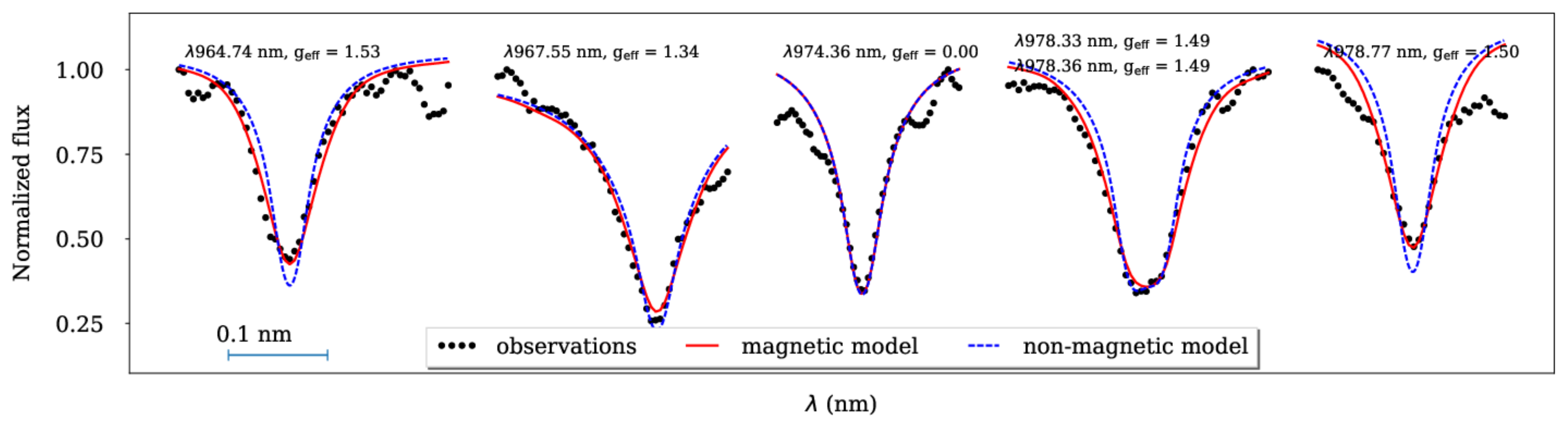}
\includegraphics[width=\hsize]{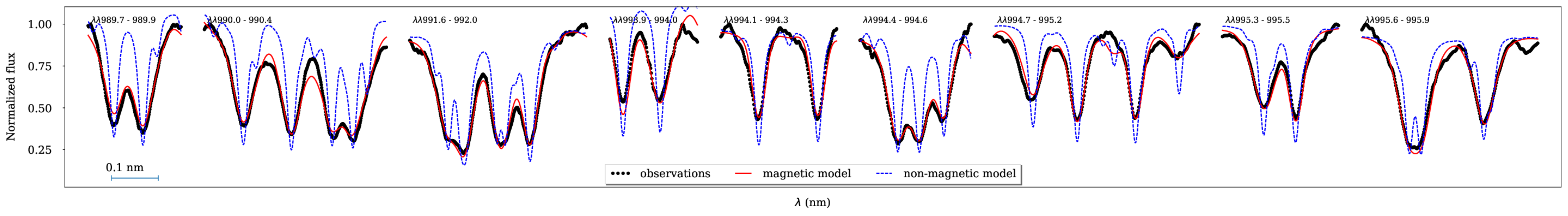}
\caption{\label{fig:J19169+051S-fit}
Same as on Fig.~\ref{fig:J01033+623-fit} but for J19169+051S.
}
\end{figure*}

\begin{figure*}
\includegraphics[width=\hsize]{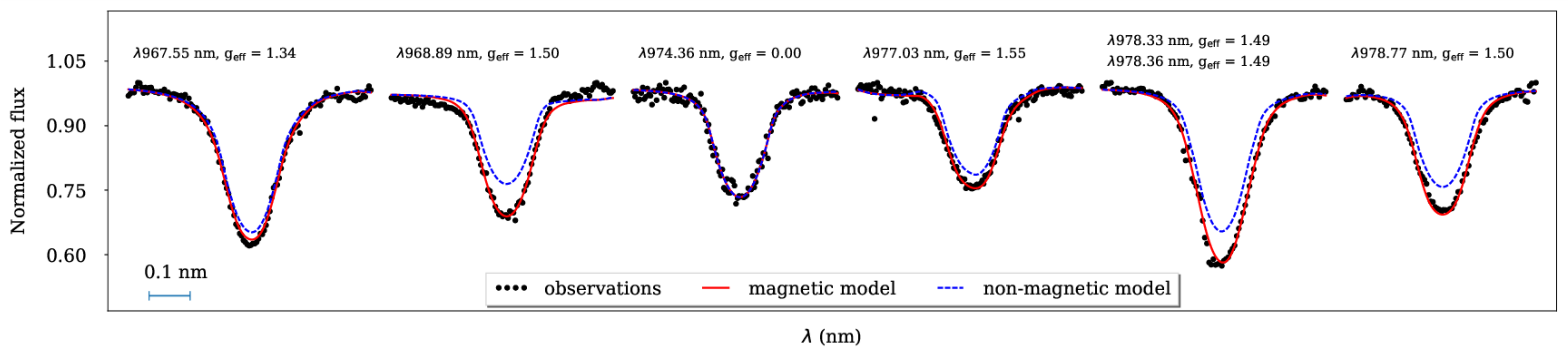}
\includegraphics[width=\hsize]{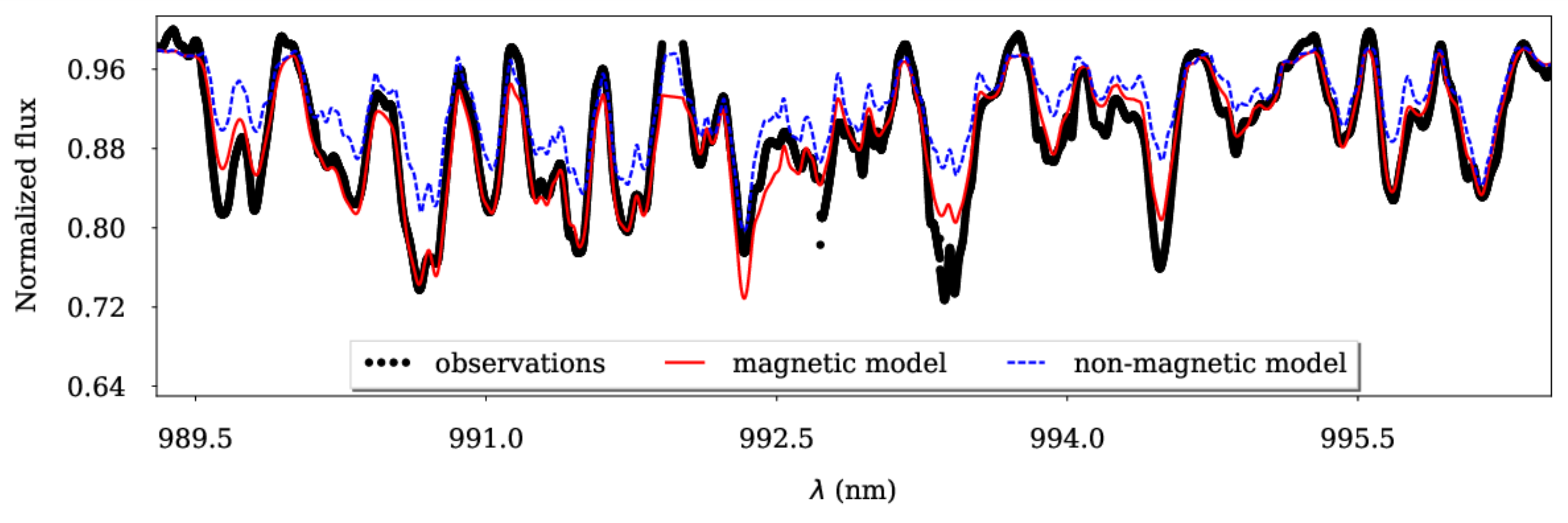}
\caption{\label{fig:J19511+464-fit}
Same as on Fig.~\ref{fig:J01033+623-fit} but for J19511+464.
}
\end{figure*}

\begin{figure*}
\includegraphics[width=\hsize]{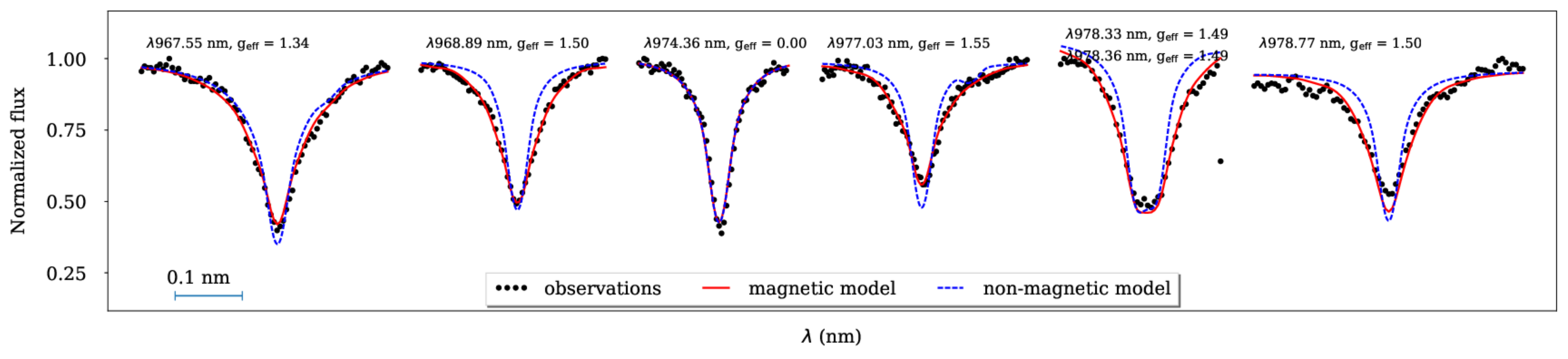}
\includegraphics[width=\hsize]{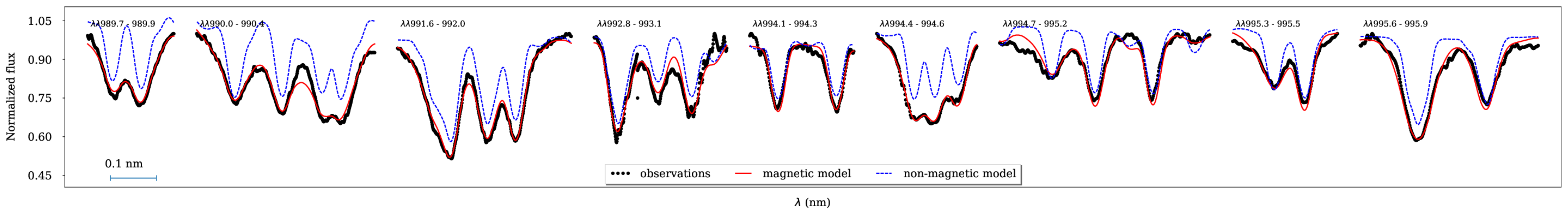}
\caption{\label{fig:J20093-012}
Same as on Fig.~\ref{fig:J01033+623-fit} but for J20093-012.
}
\end{figure*}

\begin{figure*}
\includegraphics[width=\hsize]{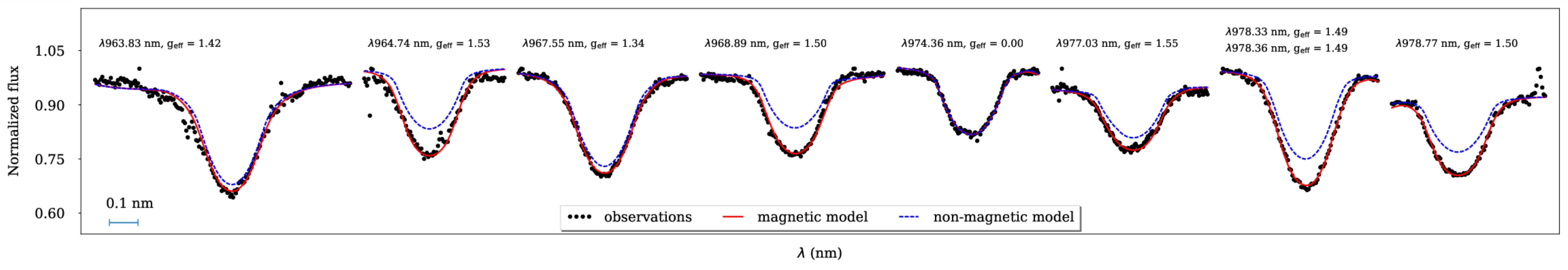}
\includegraphics[width=\hsize]{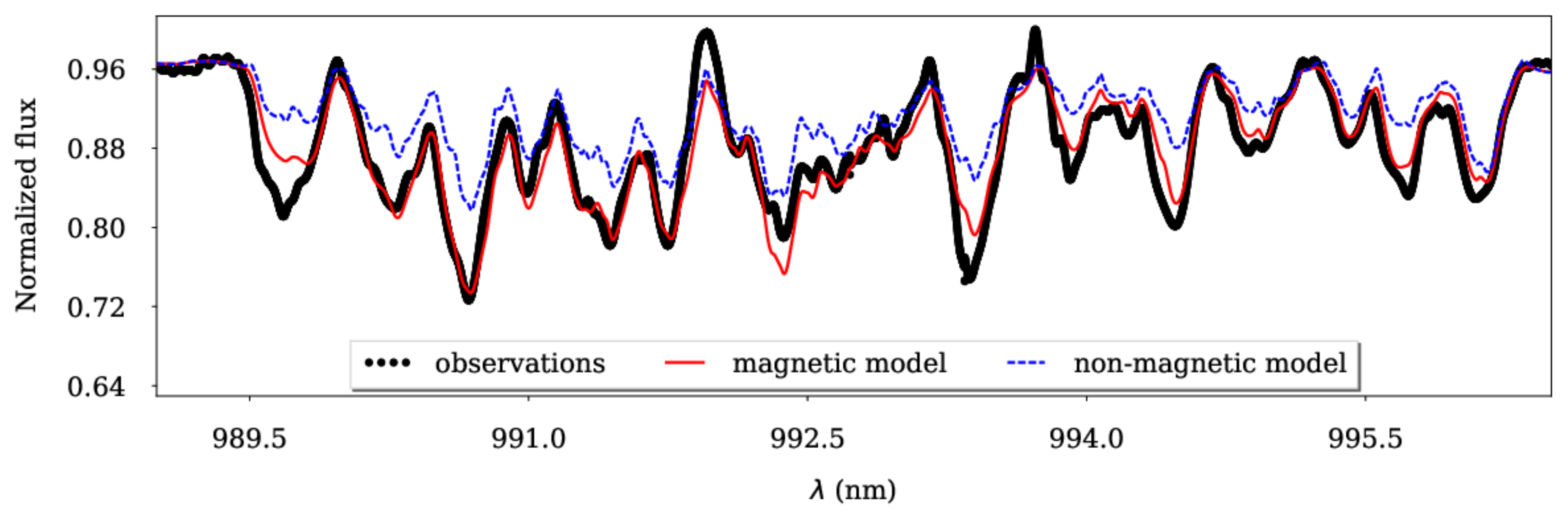}
\caption{\label{fig:J22012+283-fit}
Same as on Fig.~\ref{fig:J01033+623-fit} but for J22012+283.
}
\end{figure*}

\begin{figure*}
\includegraphics[width=\hsize]{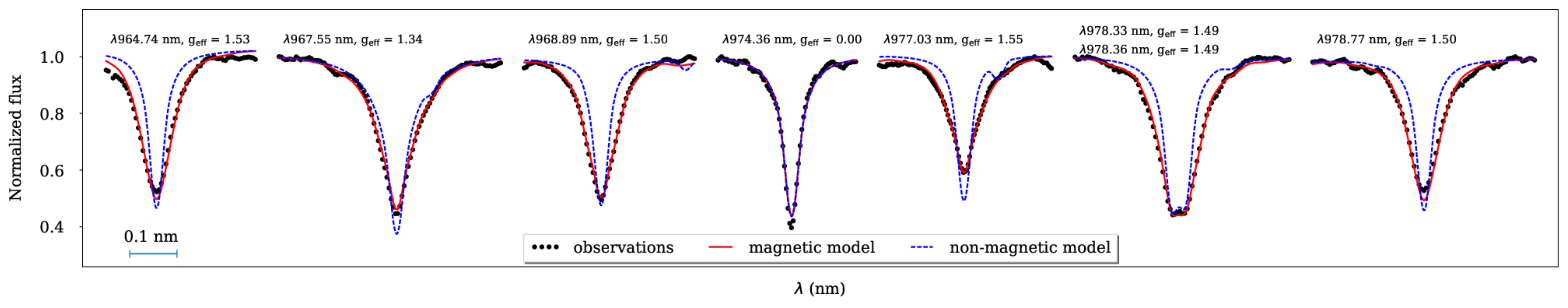}
\includegraphics[width=\hsize]{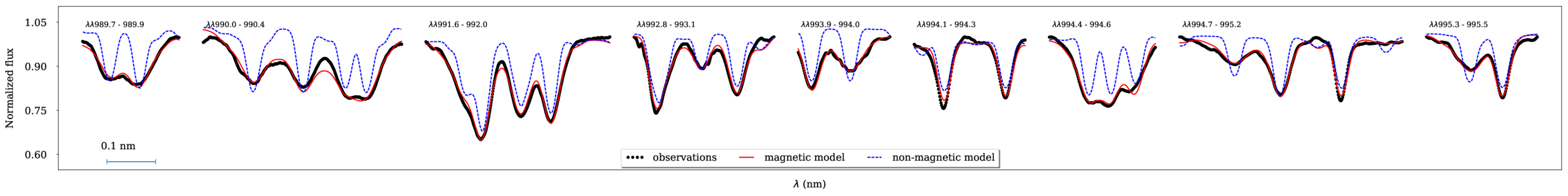}
\caption{\label{fig:J22468+443-fit}
Same as on Fig.~\ref{fig:J01033+623-fit} but for J22468+443.
}
\end{figure*}

\begin{figure*}
\includegraphics[width=\hsize]{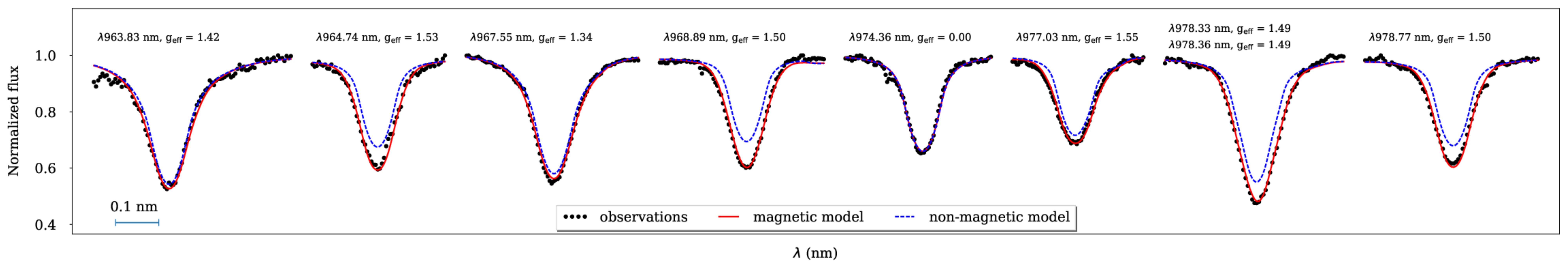}
\includegraphics[width=\hsize]{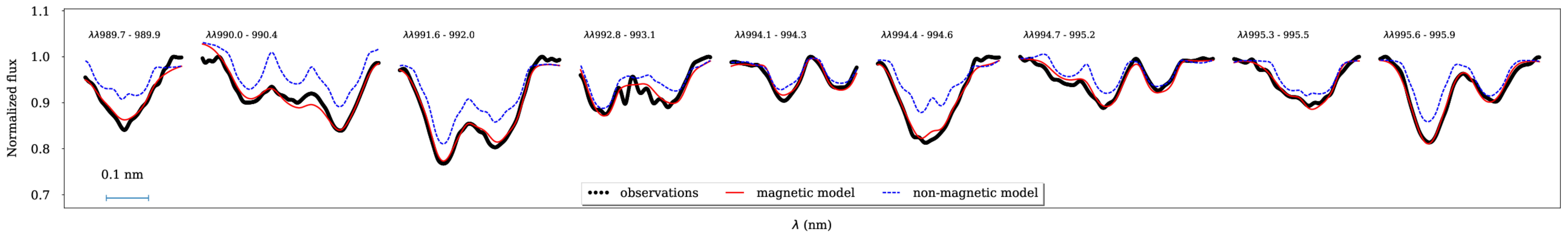}
\caption{\label{fig:J22518+317-fit}
Same as on Fig.~\ref{fig:J01033+623-fit} but for J22518+317.
}
\end{figure*}

\begin{figure*}
\includegraphics[width=\hsize]{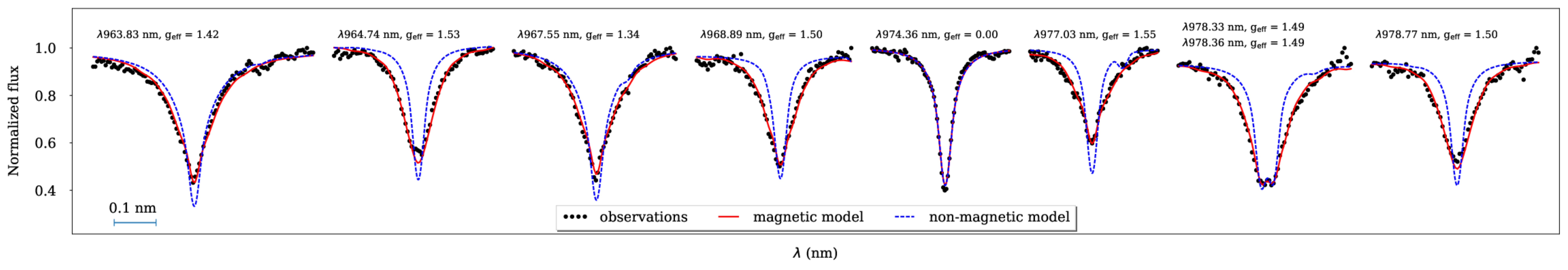}
\includegraphics[width=\hsize]{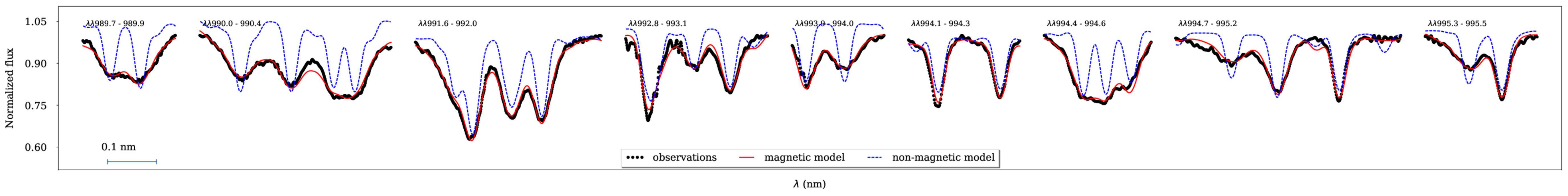}
\caption{\label{fig:J23548+385-fit}
Same as on Fig.~\ref{fig:J01033+623-fit} but for J23548+385.
}
\end{figure*}

\end{appendix}

\listofobjects

\end{document}